\documentclass[pdflatex,prd,twocolumn,showpacs,superscriptaddress,nofootinbib]{revtex4-2}
\usepackage{bm,amsmath,amssymb}
\usepackage{graphicx}
\usepackage{subfigure}
\usepackage{color}
\usepackage{xcolor}
\usepackage{soul}
\usepackage{hyperref}
\usepackage{multirow}
\usepackage{lineno}
\usepackage{lipsum}
\usepackage{physics}
\hypersetup{
    colorlinks = false,
    urlcolor = red
    }

\newcommand \beqa {\begin{eqnarray}}
\newcommand \eeqa {\end{eqnarray}}
\newcommand \hmu {\hat{\mu}}
\newcommand \hs {\hat{s}}

\newcommand \he {\hat{\epsilon}}
\newcommand \hsi {\hat{\sigma}}

\newcommand \hp {\hat{p}}

\newcommand \hn {\hat{n}}
\newcommand \pade {Pad\'e~}
\newcommand \xb {\bar{x}}

\newcommand{\nth}{{}^{\rm th}} % nth
\newcommand{\simulat}{\texttt{SIMULATeQCD}}
\newcommand{\Tc}{T_{\rm c}}               % critical temp
\newcommand{\Tpc}{T_{\rm pc}}   % pseudo-critical temp

\usepackage{graphicx} 
\graphicspath{{figs_r05/}}

\begin{document}
%\linenumbers
\title{Equation of state and
speed of sound of (2+1)-flavor QCD
in strangeness-neutral matter at non-vanishing net baryon-number density.}

%----------------------------------
%     authors & affiliations
%-----------------------------------------------------------------------------

\author{D. Bollweg}
\affiliation{Physics Department, Brookhaven National Laboratory, Upton, New York 11973, USA}
\affiliation{RIKEN-BNL Research Center, Brookhaven National Laboratory, Upton, New York 11973}
\author{D. A. Clarke}
\affiliation{Fakult\"at f\"ur Physik, Universit\"at Bielefeld, D-33615 Bielefeld, Germany}
\affiliation{Department of Physics and Astronomy, University of Utah, 
Salt Lake City, Utah 84112, United States}
\author{J. Goswami}
\affiliation{RIKEN Center for Computational Science, Kobe 650-0047, Japan}
\author{O. Kaczmarek}
\affiliation{Fakult\"at f\"ur Physik, Universit\"at Bielefeld, D-33615 Bielefeld, Germany}
\author{F. Karsch}
\affiliation{Fakult\"at f\"ur Physik, Universit\"at Bielefeld, D-33615 Bielefeld, Germany}
\author{Swagato Mukherjee}
\affiliation{Physics Department, Brookhaven National Laboratory, Upton, New York 11973, USA}
\author{P. Petreczky}
\affiliation{Physics Department, Brookhaven National Laboratory, Upton, New York 11973, USA}
\author{C. Schmidt}
\affiliation{Fakult\"at f\"ur Physik, Universit\"at Bielefeld, D-33615 Bielefeld, Germany}
\author{Sipaz Sharma}
\affiliation{Fakult\"at f\"ur Physik, Universit\"at Bielefeld, D-33615 Bielefeld, Germany}
\collaboration{HotQCD collaboration}
\date{\today}

%-----------------------------------------------------------------------------
%     abstract
%-----------------------------------------------------------------------------
\begin{abstract}
We update results on the QCD equation of state in (2+1)-flavor QCD with non-zero conserved charge chemical potentials obtained from an eighth-order Taylor
series.To construct bulk thermodynamic observables, we use continuum extrapolated results for the second and fourth order expansion coefficients of pressure, while using spline interpolated result based on high statistics $N_{\tau}=8$ data for the sixth and eighth order expansion coefficients. We present results for basic bulk
thermodynamic observables of strangeness-neutral 
strong-interaction matter, i.e. pressure,
number densities, energy and entropy 
density, and resum Taylor series results
using \pade approximants.
Furthermore, we calculate the
speed of sound as well
as the adiabatic compression factor of the
strangeness-neutral matter
on lines of constant entropy per net 
baryon number. 
We show that the equation of state
($P(n_B), \epsilon (n_B)$)
is already well described by the $4^{\rm th}$-order Taylor series in almost the entire range of temperatures
accessible with the beam energy scan
in collider mode at the Relativistic Heavy Ion Collider.
\end{abstract}


\maketitle

\section{Introduction}

Fundamental properties of hot and dense matter controlled by the strong force are encoded in the 
equation of state (EoS). The EoS is sensitive to the
change of degrees of freedom that dominate properties
of strong-interaction matter at low temperatures and densities on the one hand and high temperature and densities on the other hand. It describes fluctuations
of e.g. energy and particle densities that signal the
occurrence of phase transitions. The EoS finds application 
in the hydrodynamic modelling of hot and dense matter
created in heavy ion collisions 
\cite{Braun-Munzinger:2015hba}
and in constraining the ``cosmic trajectory'' of 
matter in the expanding early universe \cite{Middeldorf-Wygas:2020glx}. 

The EoS of strong-interaction matter is obtained  
from quantum chromodynamics (QCD) at finite temperature and non-zero values of the conserved charge 
chemical potentials. It is  well studied at vanishing values of the chemical potentials  \cite{Karsch:2000ps,Borsanyi:2013bia,HotQCD:2014kol}. 
For non-vanishing values of the chemical potentials
that couple to net baryon number ($\mu_B$), net electric charge ($\mu_Q$) and net strangeness 
($\mu_S$), lattice QCD calculations have to face
the well-known sign problem. Currently, this renders 
direct numerical calculations with non-zero,
real-valued chemical potentials 
impossible. For this reason numerical
calculations at non-zero $(\mu_B,\mu_Q,\mu_S)$ 
have been performed by either using 
Taylor expansions in terms of the chemical potentials, {e.g.} \cite{Gavai:2001fr,Allton:2002zi}
or by performing numerical simulations at 
imaginary values of the chemical potential, 
\cite{DElia:2002tig,deForcrand:2002hgr}.
Using the latter and analyzing results with
various ans\"atze for an analytic continuation
to real values of the chemical potential, the
EoS of (2+1)-flavor QCD has been determined
\textcolor{black}{
\cite{Borsanyi:2012cr,Guenther:2017hnx,Borsanyi:2022qlh}.
}
Following the Taylor expansion
approach, results for the QCD EoS 
have been obtained from up to sixth-order expansions \cite{Ejiri:2005uv,Bazavov:2017dus}.

We extend here the Taylor expansion approach for the EoS in (2+1)-flavor QCD using recent high-statistics results
for Taylor expansion coefficients up to eighth-order
\cite{Bollweg:2022rps} in ($\mu_B, \mu_Q, \mu_S)$. We
will make use of the Pad\'e-resummation of Taylor series
discussed in \cite{Bollweg:2022rps} and determine
direct Taylor series results as well as Pad\'e-resummed
approximants for pressure, net baryon-number, energy and
entropy densities. Making use of our high-statistics
data at temperatures in the vicinity of the 
pseudo-critical temperature we will,
for the first time, present contributions to higher-order derivatives of bulk thermodynamic observables, {\it i.e.}
the speed of sound and adiabatic compressibility of 
strangeness-neutral strong-interaction matter.
These quantities have been 
calculated previously at 
vanishing values of the chemical
potentials \cite{Gavai:2004se,Borsanyi:2013bia,HotQCD:2014kol}.

We focus on a situation most relevant in the context of 
heavy ion collision experiments, 
in which overall net strangeness
number density ($n_S$) vanishes and the ratio
of net electric charge ($n_Q$) to net
baryon-number ($n_B$) densities is close to 
the isospin-symmetric limit of strangeness 
neutral matter, $n_Q=n_B/2$.
In collisions of heavy nuclei,
e.g. gold or lead, the ratio
$n_Q/n_B$ is approximately 0.4. Earlier 
studies of the QCD EoS with a strange quark
mass tuned to its physical value,
and degenerate light quark masses
tuned to reproduce the physical, light hadron spectrum, 
have been performed for
different values of $r\equiv n_Q/n_B$.
They have shown
that differences in bulk thermodynamic observables, arising
from the deviation of $r$
from the isospin-symmetric value,
$r=0.5$, are small \cite{Bazavov:2017dus}. We therefore
stick here to an analysis of
the thermodynamics of strangeness-neutral,
isospin-symmetric systems, which is 
equivalent to setting the 
electric charge chemical potential
$\mu_Q$ to zero.

This paper is organized as follows.
In Section~\ref{sec:thermConstraint} we summarize basic
thermodynamic relations and outline
calculations in QCD with non-vanishing
chemical potentials using the Taylor series approach and the resummation of Taylor series using \pade approximants. 
In Section~\ref{sec:taylorCoeff} we present our results on Taylor expansion coefficients 
for bulk thermodynamic observables and discuss their properties at low and high 
temperature by comparing with HRG model calculations and $\order{g^2}$ high-temperature
perturbation theory, respectively. We furthermore discuss the structure of
expansion coefficients in the vicinity
of the pseudo-critical temperature.
In Section~\ref{sec:bulkTherm} we analyze thermodynamic 
observables as function of the baryon
chemical potential. We determine the baryon chemical potential as function of net 
baryon-number density to arrive at the 
QCD equation of state, $P(n_B)$. We furthermore present results for basic thermodynamic
observables on the pseudo-critical line as well as on lines
of constant ratio of entropy to net baryon number. The latter are discussed in Section~\ref{sec:constantS}. 
We finally give our conclusions in Section~\ref{sec:conclusion}. In three appendices we give some details on
directional partial derivatives (\ref{app:partial}) and
discuss $\order{g^2}$ perturbative results for the EoS of strangeness-neutral matter (\ref{app:IG})
as well as the calculation of the isentropic speed of sound (\ref{app:sound}).

\section{Thermodynamics in (2+1)-flavor QCD and global
constraints}\label{sec:thermConstraint}

We present here the basic thermodynamic
observables that will be analyzed by us
in later sections using results from
up to eighth-order Taylor expansions of the
pressure of QCD with a strange quark 
mass tuned to its physical value and 
two degenerate light quark masses 
that correspond in the continuum limit
to a physical pion mass value of about
$135$~MeV.

We also briefly summarize basic relations used in  
the Taylor expansion approach to the thermodynamics of
QCD at non-zero values of the conserved charge chemical potentials and discuss our approach to the utilization of
Pad\'e-resummed Taylor series for thermodynamic observables. Further details on these three topics 
can be found in \cite{Bollweg:2022rps}.

\subsection{Thermodynamic observables}
\label{sec:observables}

In a grand canonical ensemble the pressure, $p$, 
is a function of temperature $T$ and a set of chemical potentials $\vec{\mu}=(\mu_B,\mu_Q,\mu_S)$,
which couple to the currents of conserved charges for net baryon number ($B$), electric 
charge ($Q$) and strangeness ($S$). The
pressure is given in terms of the logarithm of the grand canonical partition function
\begin{eqnarray}
\frac{p}{T^4} &=& \frac{1}{VT^3}\ln\mathcal{Z}(T,V,\hmu_B,\hmu_Q,\hmu_S) \; .
\label{pressure}
\end{eqnarray}
We will often use dimensionless variables and observables obtained by 
rescaling the dimensionful observables with appropriate 
powers of the temperature, e.g.
$\hat{\mu}_X\equiv \mu_X/T$ or $\hat{p}\equiv p/T^4$. With this we obtain for the
number densities,
\begin{equation}
    \hn_X =\frac{1}{VT^3} \frac{\partial \ln\mathcal{Z}(T,V,\hmu_B,\hmu_Q,\hmu_S)}{\partial \hmu_X} \;,\; X=B,\ Q,\ S\; ,
    \label{energy0}
\end{equation}
and the energy density,
\begin{eqnarray}
    \hat{\epsilon} &=&\frac{1}{VT^2} \frac{\partial \ln\mathcal{Z}(T,V,\hmu_B,\hmu_Q,\hmu_S)}{\partial T} 
    \nonumber \\
    &=& 3 \hat{p} +T \frac{\partial \hat{p}}{\partial T}
    \; .
    \label{energy}
\end{eqnarray}
The entropy density reads
\begin{eqnarray}
    \hs &=& 
\he +\hp -\hmu_B \hn_B - \hmu_Q \hn_Q -\hmu_S \hn_S \; .
\label{entropy}
\end{eqnarray}
Enforcing the constraint for isospin-symmetric systems, $n_Q=n_B/2$ in strangeness-neutral matter ($n_S=0$) is equivalent to demanding $\mu_Q=0$ \cite{Bazavov:2017dus}. Thermodynamic observables thus become
functions of $T$ and $\hmu_B$ only,
i.e. $\hp\equiv p(T,\hmu_B)/T^4$
and $\hs =\he +\hp -\hmu_B \hn_B $.

In addition to the basic bulk thermodynamic observables introduced above,
we also will calculate their temperature 
derivatives, imposing external
constraints. We calculate the isentropic speed
of sound in a strangeness-neutral medium with fixed $n_Q/n_B$,
\cite{Floerchinger:2015efa},
\begin{eqnarray}
    c_s^2 \equiv
      c^2_{\vec{X}}
      &=&\left(
    \frac{\partial p}{\partial \epsilon}\right)_{\vec{X}} 
    =\frac{\left(
    \partial p/\partial T\right)_{\vec{X}}}{\left(\partial \epsilon/\partial T\right)_{\vec{X}}}
    \; .
    \label{dpdesn}
\end{eqnarray}
with $\vec{X}=(s/n_B,n_Q/n_B,n_S)$.
This also gives the adiabatic compressibility 
\begin{eqnarray}
\kappa_s &=& \frac{1}{n_B} \left( \frac{\partial n_B}{\partial p} \right)_{\vec{X}}
   = \frac{1}{c_s^2 (\epsilon +p-\mu_S n_S)}
\; .
\label{kappaS}
\end{eqnarray}
The calculation of $c_s^2$ and $\kappa_s$
requires taking directional derivatives,
which can be done numerically on lines of
fixed constraints $\vec{X}$ in the parameter
space $(T,\vec{\mu})$, or using analytic 
relations.
Some details on partial derivatives at 
fixed constraints are given in 
Appendix~\ref{app:partial}.

\subsection{Taylor expansions}
\label{sec:Taylor}

We start with the expansion of the pressure, $\hp=p/T^4$, in terms of
the three conserved charge chemical
potentials $(\hmu_B,\hmu_Q,\hmu_S)$,
\begin{eqnarray}
\hp &=& \frac{1}{VT^3}\ln\mathcal{Z}(T,V,\hmu_B,\hmu_Q,\hmu_S) \nonumber \\
&=& \sum_{i,j,k=0}^\infty
\frac{\chi_{ijk}^{BQS}}{i!j!\,k!} \hmu_B^i \hmu_Q^j \hmu_S^k \; ,
\label{Pdefinition}
\end{eqnarray}
with $\chi_{000}^{BQS}\equiv p(T,0)/T^4$
and expansion coefficients $\chi_{ijk}^{BQS}$, 
\begin{equation}
\chi_{ijk}^{BQS}\equiv \chi_{ijk}^{BQS}(T) =\left. 
\frac{\partial \hp}{\partial\hmu_B^i \partial\hmu_Q^j \partial\hmu_S^k}\right|_{\vec{\mu}=0} \; .
\label{suscept}
\end{equation}
These expansion coefficients are 
cumulants of conserved charge fluctuations
evaluated at vanishing chemical
potential\footnote{We often suppress the argument ($T$) of these generalized
susceptibilities and also suppress superscripts and subscripts of
$\chi_{ijk}^{BQS}$ whenever one of the subscripts vanishes, e.g.
$\chi_{i0k}^{BQS}\equiv \chi_{ik}^{BS}$.}.

From this expansion and the constraints, we obtain Taylor expansions for
pressure and number densities,
\begin{eqnarray}
\Delta \hat{p}&\equiv& \frac{p(T,\mu_B)}{T^4} -\frac{p(T,0)}{T^4} =
\sum_{k=1}^{\infty} P_{2k}(T) \hmu_B^{2k} \; ,
\label{Pn} \\
\hn_X &=& \sum_{k=1}^{\infty} N_{2k-1}^X(T) \hmu_B^{2k-1} \;\; ,\;\;
X=B,\ Q,\ S \;\; .
\label{nX}
\end{eqnarray}
The construction of the expansion coefficients\footnote{See \cite{Bazavov:2020bjn} for
this notation and explicit definitions of $\bar{\chi}_0^{B,n}$.}, 
$P_{2k}\equiv \bar{\chi}^{B,2k}_0/(2k)!$ 
and $N_{2k-1}$
has been discussed previously for the general case of
strangeness-neutral matter and a fixed ratio
$n_Q/n_B=r$ \cite{Bazavov:2017dus,Bollweg:2022rps}.
This gives the chemical potentials
$\hmu_Q$ and $\hmu_S$ as Taylor series in terms of $\hmu_B$,
\begin{eqnarray}
\hmu_Q(T,\mu_B) &=& q_1(T)\hmu_B + q_3(T) \hmu_B^3+q_5(T) \hmu_B^5 +\dots
\\
\hmu_S(T,\mu_B) &=& s_1(T)\hmu_B + s_3(T) \hmu_B^3+s_5(T) \hmu_B^5 +\dots \; . 
\label{qs}
\end{eqnarray}
Here we only consider the 
isospin-symmetric case, $n_Q/n_B=1/2$, which fixes $\mu_Q=0$. In that case 
one finds, for instance, a simple relation between
the expansion coefficients of the net
baryon-number density and the pressure,
$N_{2k-1}^B=2k P_{2k}$.

Taking derivatives of the expansion
coefficients with respect to temperature
we also obtain Taylor series for the energy
and entropy densities, 
\begin{eqnarray}
\Delta\hat{\epsilon}&\equiv&\frac{\epsilon(T,\hmu_B)}{T^4} - \frac{\epsilon(T,0)}{T^4} = 
\sum_{k=1}^{\infty} \epsilon_{2k}(T) \hmu_B^{2k}\; ,
\label{energyc} \\
\Delta \hat{s}&\equiv&\frac{s(T,\hmu_B)}{T^3} - \frac{s(T,0)}{T^3} = \sum_{k=1}^{\infty} \sigma_{2k}(T) \hmu_B^{2k}\; .
\label{entropyc}
\end{eqnarray}
We note that to obtain Eqs.~\ref{energyc} and 
\ref{entropyc} one must take the temperature derivatives before
applying the constraints. 
In the case $\mu_Q=0,\ n_S=0$,
as well as for $\mu_Q=0,\ \mu_S=0$ the
expansion coefficients are directly related 
to the expansion coefficients of the Taylor
series of the pressure,  
\begin{eqnarray}
    N_{2k-1}^B(T) &=& 2k P_{2k}(T) \;\; ,
    \label{N2k}
    \\
    \epsilon_{2k}(T) &=&3 P_{2k}(T)+T\frac{{\rm d}P_{2k}(T)}{{\rm d}T}\;\; ,
    \label{e2k}
    \\
    \sigma_{2k}(T) &=& 
(4-2k) P_{2k}(T)+ T\frac{{\rm d}P_{2k}(T)}{{\rm d}T} \;\; .
\label{e-s-coeff}
\end{eqnarray}
We also introduce the notation
$\he_0\equiv \epsilon(T,0)/T^4$ and $\hsi_0\equiv s(T,0)/T^3$.

\subsection{Resummation with \pade approximants}
\label{sec:Pade}

In \cite{Bollweg:2022rps} we analyzed
the resummation of Taylor series
for the pressure and net baryon-number density using $[n,m]$ \pade approximants. We showed that 
in the temperature range [135~MeV:175~MeV], 
the [4,4] \pade approximants for the pressure have poles only in the complex plane. The 
location of these poles provide an estimate 
for the range of validity of the Taylor series.
For chemical potentials in this range, one finds
good agreement between the \pade approximants 
and the eighth-order Taylor series for 
pressure and net baryon-number density\footnote{We note that the recently developed multi-\pade approach \cite{Dimopoulos:2021vrk} 
provides similar information on the location of poles in the complex plane and may, in the future, also be used to calculate thermodynamic observables.}.
We will use here 
the \pade approximants for the Taylor
series of the pressure also to determine energy and entropy densities
as well as other thermodynamic observables introduced in Sec.~\ref{sec:observables}.

As has been discussed in \cite{Bollweg:2022rps}, the Taylor series for the pressure of 
isospin-symmetric matter with either $\mu_S=0$ or $n_S=0$ may conveniently be rewritten as
\begin{eqnarray}
\frac{\Delta p(T,\mu_B)}{T^4} &=&\frac{P_2^2}{P_4}
\sum_{k=1}^{\infty} c_{2k,2} \xb^{2k}\; ,
\\
&=& \frac{P_2^2}{P_4}\left(\xb^2+\xb^4+ c_{6,2} \xb^6
+ c_{8,2} \xb^8 + ...\right) 
\nonumber \; \; ,
\end{eqnarray}
with $\xb =\sqrt{P_4/P_2}\ \hmu_B$ and
\begin{eqnarray}
    c_{6,2}=\frac{P_6 P_2}{P_4^2}  \;\; , \;\;
    c_{8,2}= \frac{P_8P_2^2}{P_4^3} 
    \;\; .
    \label{parameter}
\end{eqnarray}

\begin{figure*}[ht]
\includegraphics[width=0.48\linewidth]{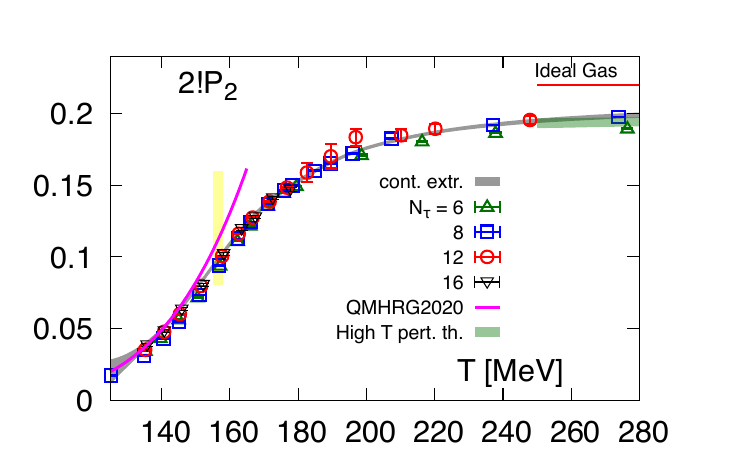}
\includegraphics[width=0.48\linewidth]{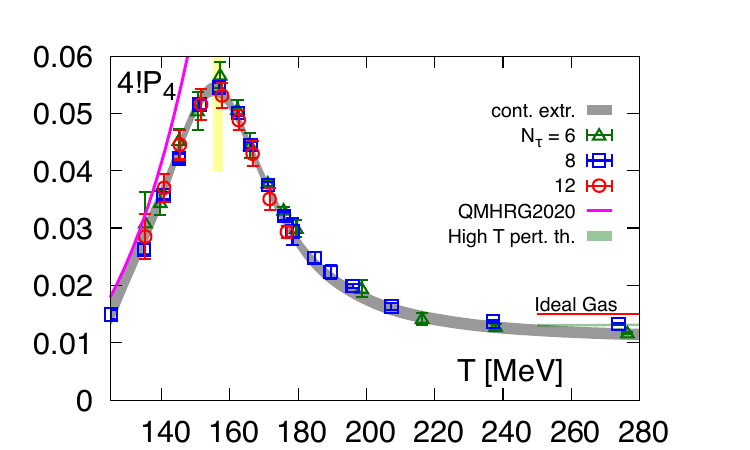}
\includegraphics[width=0.48\linewidth]{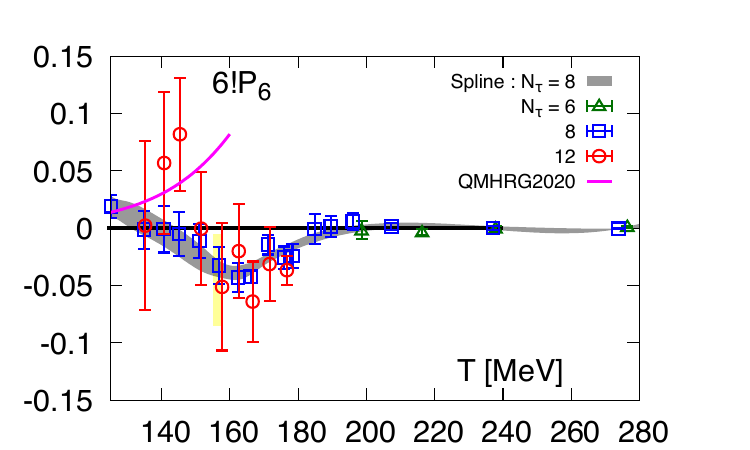}
\includegraphics[width=0.48\linewidth]{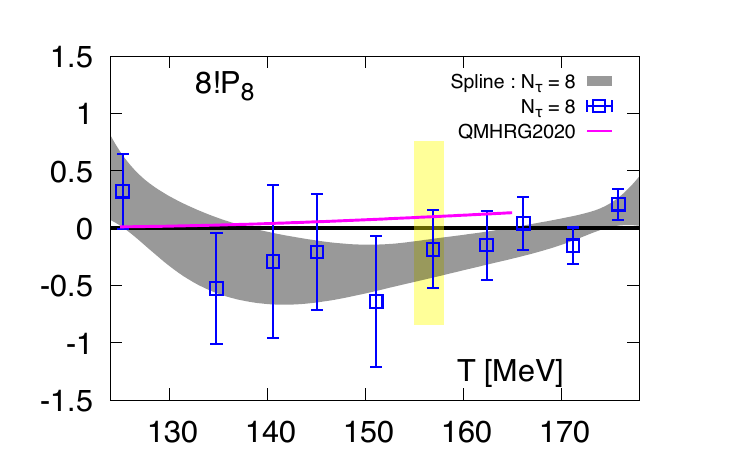}
\caption{The $n\nth$-order Taylor
expansion coefficients, $P_n(T)$, for the Taylor series of the pressure of
(2+1)-flavor QCD as function of $\hmu_B=\mu_B/T$ versus temperature.
Shown are the expansion coefficients $n! P_{n}$ for isospin-symmetric,
strangeness-neutral matter ($\mu_Q=0$, $n_S=0$). The solid magenta lines 
show results from a corresponding 
HRG model calculation using the 
QMHRG2020 list of hadron resonances
\cite{Bollweg:2022rps}.
Yellow bands show the location of 
the pseudo-critical temperature $\Tpc(0)=156.5(1.5)$~MeV \cite{HotQCD:2018pds}. The green
bands at high temperature show the 
$\order{g^2}$ perturbative result using a renormalization scale in the range $[4\pi T,8\pi T]$ \textcolor{black}{(for details see Appendix B)}.}
\label{fig:taylor8}
\end{figure*}

Using expansion coefficients from
fourth, sixth and eighth-order
Taylor series for the pressure we construct 
several $[n,m]$ \pade approximants, 
\begin{eqnarray}
P[2,4] &=& \frac{\xb^2}{1-\xb^2 + (1-c_{6,2}) \xb^4}\;\; ,  \label{pd24}\\
P[4,2] &=& \frac{\xb^2+(1-c_{6,2})\xb^4}{1 -c_{6,2} \xb^2}\;\; ,  \label{pd42}\\
P[4,4] &=&  \frac{(1-c_{6,2}) \xb^2 +
\left(1 - 2 c_{6,2} +c_{8,2} \right) \xb^4}{(1-c_{6,2})+
(c_{8,2}-c_{6,2}) \xb^2 + (c_{6,2}^2 - c_{8,2}) \xb^4} .
\nonumber \\
&& ~
\label{pd44}
\end{eqnarray}
The resulting approximants for the 
pressure are then given by
\begin{equation}
    \left( \frac{\Delta p(T,\mu_B)}{T^4}\right)_{[n,m]} =\frac{P_2^2}{P_4} P_{[n,m]} \; .
    \label{nmPade}
\end{equation}
For the 
number density we construct the $[3,4]$ \pade approximant.

Similar \pade approximants can be obtained
for the energy density and entropy densities
by replacing the expansion coefficients
$P_{2k}$ with the corresponding expansion coefficients
$\epsilon_{2k}$ and $\sigma_{2k}$. 
Rather then constructing in this way \pade
approximants for various thermodynamic
observables, we also directly 
determine Pad\'e-based approximants from Eq.~\ref{nmPade},
using the relevant thermodynamic relations, 
e.g. Eqs.~\ref{energy}, \ref{entropy} and
\ref{nX}, for the energy, entropy and net baryon-number densities. 
The Pad\'e-based approximants,
called P-\pade in the following,
ensure thermodynamic 
consistency among different observables and, for 
instance, ensure that the singularities in the 
approximations for, $n_B/T^3$, $\epsilon /T^4$ and $s/T^3$  
coincide with those of the \pade approximation
for the pressure. The Pad\'e-based (P-\pade) result for
the energy and entropy densities are obtained
by taking appropriate partial derivatives of the $[n,m]$ \pade approximation for 
the pressure with respect to temperature and chemical potential, and using 
Eqs.~\ref{energy} and \ref{entropy},
\begin{eqnarray}
    \left( \frac{\Delta \epsilon}{T^4}\right)_{[n,m]} 
    &=&3 \left( \frac{\Delta p}{T^4}\right)_{[n,m]} 
    +T\frac{\rm d}{{\rm d}T}\left( \frac{P_2^2}{P_4} P_{[n,m]}\right) \; ,
    \label{energyPade}
    \\
    \left( \frac{\Delta s}{T^3}\right)_{[n,m]} 
    &=&\left( \frac{\Delta \epsilon}{T^4}\right)_{[n,m]} + \left( \frac{\Delta p}{T^4}\right)_{[n,m]} 
    \nonumber \\
    &&- \hmu_B \frac{P_2^2}{P_4} \frac{{\rm d}P_{[n,m]}}{{\rm d}\hmu_B} \; .
    \label{entropyPade}
\end{eqnarray}

\section{Taylor expansion coefficients of the pressure in (2+1)-flavor QCD}\label{sec:taylorCoeff}

We update here our previous analysis of the 
EoS of (2+1)-flavor QCD performed in 
lattice QCD calculations with the highly improved staggered quark (HISQ) action
and a tree-level improved gauge action
\cite{Bazavov:2017dus}.

\begin{figure*}[ht]
\includegraphics[width=0.48\linewidth,page=1]{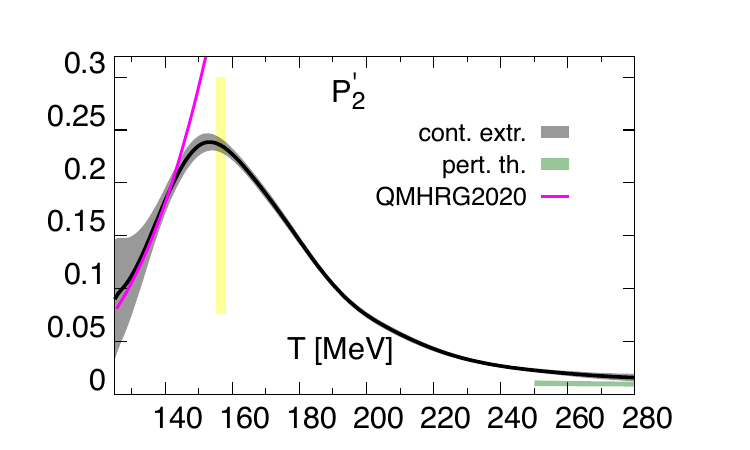}
\includegraphics[width=0.48\linewidth,page=2]{/fig2.pdf}
\includegraphics[width=0.48\linewidth,page=3]{/fig2.pdf}
\includegraphics[width=0.48\linewidth,page=4]{/fig2.pdf}
\caption{Derivatives of Taylor expansion coefficients $P_2$, $P_4$, $P_6$, and $P_8$
with respect to temperature. 
The solid \textcolor{black}{magenta} lines at low temperatures 
show results from a corresponding 
HRG model calculation using the 
QMHRG2020 list of hadron resonances
\cite{Bollweg:2022rps}.
Yellow bands show the location of the
pseudo-critical temperature $\Tpc$ \cite{HotQCD:2018pds} and
the green
bands at high temperature show the 
$\order{g^2}$ perturbative result using a renormalization scale in the range $[4\pi T,8\pi T]$.
}
\label{fig:dPdT}
\end{figure*}

\subsection{Data sets and analysis details}

For our calculation of bulk 
thermodynamic observables we use
high-statistics data sets for
$(2+1)$-flavor QCD with light and
strange quark masses tuned to physical values ($m_\ell/m_s=1/27$). These data
sets have been generated using the
HISQ action with tree-level coefficients
and a tree-level improved gauge action
\cite{Bazavov:2017dus}. 
Gauge configurations were generated using $\simulat$ code~\cite{mazur2021,Bollweg:2021cvl,Mazur:2023lvn}.
New data sets exist for the temperature interval $[125~{\rm MeV}:175~{\rm MeV}]$
and have been presented in
\cite{Bollweg:2022rps}. Additional updates
exist for the lowest temperature, where we doubled the number of gauge field 
configuration to 2.2M. 
\textcolor{black}{It is only in this temperature range that $8^{th}$ order expansion coefficients have been calculated.}
 
Compared to our previous calculation of the 
EoS \cite{Bazavov:2017dus}, the data sets 
for $N_\tau=8$ and $12$ contain more than
an order of magnitude larger statistics and,
moreover, include entirely new data sets on
lattices with temporal extent $N_\tau=16$.

At temperatures $T> 180$~MeV, we added to our analysis data from calculations 
with a slightly larger light quark mass
($m_\ell/m_s=1/20$), which had been used by HotQCD
previously \cite{Bazavov:2017dus}.
In all cases results have been obtained on
lattices with spatial extent $N_\sigma=4N_\tau$. 

Continuum extrapolations of $P_2$ and $P_4$ 
have been performed in three different temperature intervals, using different ans\"atze.
In the temperature interval 
$[135~{\rm MeV}:175~{\rm MeV}]$ we use rational polynomial functions
as described in  \cite{Bollweg:2022rps} and \cite{Bollweg:2021vqf}. 
\textcolor{black}{
The continuum extrapolations for $4^{th}$ order cumulants follows the same approach as that used for the $2^{nd}$ order cumulants
\cite{Bollweg:2022rps}, except it is based on only three different values of the cut-off, i.e. $N_{\tau} = 6,8$ and $12$.
Given our current statistical error for $N_{\tau}=12$ data on $4^{th}$ order cumulants, continuum extrapolations based on $1/N_{\tau}^2$ as well as $1/N_{\tau}^2 + 1/N_{\tau}^4$ corrections agree well within errors.  However, since there are no degrees of freedom left when including $1/N_{\tau}^4$ corrections, we choose to use the $1/N_{\tau}^2$ extrapolations when quoting errors.
}

For the high-temperature part $T \geq 175~{\rm MeV}$, 
we use a polynomial ansatz to interpolate the data to the continuum limit. 
Moreover, for low temperatures in the range 
$[125~{\rm MeV}:145~{\rm MeV}]$ 
we consider an HRG-motivated ansatz, which
takes into account the exponential drop
of thermodynamic observables at low temperatures and incorporates cutoff-dependent corrections,
\begin{eqnarray}
\hspace{-0.7cm}f(T) = A \left(\frac{B}{ T}\right)^{3/2} {\rm e}^{- B / T} 
\left(1 + d_1 T + d_2 T^2 + \frac{c}{N_{\tau}^2}\right)\; .
\end{eqnarray}
Finally, we match all the three functions and their first derivative at $T = 135~{\rm MeV}$ and at $T = 175~{\rm MeV}$.
For the interpolation of $P_6$ and $P_8$ we only used spline interpolations of the $N_{\tau} = 8$ data set. 
Much of the data analysis in this study was facilitated through the \texttt{AnalysisToolbox}~\cite{toolbox}.
Statistical uncertainty represented by bands in all figures is calculated
through bootstrap resampling. 
Central values are returned as the median with lower and
upper error bounds given by the 32\% and 68\% quantiles, respectively. 
Spline interpolations, when needed, are cubic with evenly spaced knots. They are
calculated using the \texttt{LSQUnivariateSpline} method of
SciPy \cite{2020SciPy-NMeth}. 
Central values and error bands 
shown for the lattice QCD results
are smoothed using splines.
Temperature derivatives of lattice QCD data
are calculated by fitting the temperature dependence with a spline, 
then calculating the derivative of the spline numerically.
\textcolor{black}{We have checked that our spline interpolations and the resulting derivatives
are stable under variation of the number of knots used for the interpolation.
In general we use splines with 12 knots.}

\subsection{Taylor expansion coefficients for the pressure and the trace anomaly}

The Taylor series in terms of the conserved charge chemical potentials 
$\vec{\mu}=(\mu_B, \mu_Q, \mu_S)$
have been reorganized to obtain the
Taylor series for the pressure in strangeness-neutral, isospin-symmetric matter given in Eq.~\ref{Pn}. 
Results for the expansion coefficients
$(2k)! P_{2k}\equiv \bar{\chi}_0^{B,n}$
are shown in Fig.~\ref{fig:taylor8}.
For the fits and interpolations of
different order expansion coefficients
shown in this figure we use the same approach as discussed in \cite{Bollweg:2022rps}.
{\it I.e.} for the $\order{\hmu_B^2}$ expansion coefficient the band shows 
a continuum extrapolation based on
$N_\tau=6,\ 8,\ 12$ and $16$ datasets;
for the $\order{\hmu_B^4}$ expansion coefficient we show a continuum extrapolation based on
$N_\tau=6,\ 8$ and $12$ datasets and for 
the higher-order expansion coefficients we only use results
from our high-statistics calculations on lattices
with temporal extent $N_\tau=8$, where more than
1.5 million gauge field configurations have been generated at each temperature value.
In this case the curves show spline 
interpolations of the $N_\tau=8$ data.
Here we note that the $6^{\rm th}$ and $8^{\rm th}$ order expansion coefficients of the pressure have been also estimated using lattice
QCD calculations with imaginary chemical potentials in 2+1 flavor case using $N_{\tau}=8$ lattices and stout2 action 
\cite{DElia:2016jqh},
and in 2+1+1 flavor case on $N_{\tau}=12$
lattices using stout4 action
\cite{Borsanyi:2018grb}.
The qualitative features of the expansion
coefficients obtained in these calculations
are similar to the ones obtained here.

In the following we discuss several features 
of the expansion coefficients in the 
high, low and intermediate 
temperatures ranges, respectively.

\subsubsection{High-temperature region}
It is apparent that the expansion coefficients $P_{2k}$ approach
the high-temperature ideal gas limit rapidly. In fact, for 
temperatures $T\gtrsim 250$~MeV, 
deviations from the ideal gas values are only about $12$\% for
$P_2$ and $P_4$. Moreover, at these temperature values the expansion coefficients $P_{2k}$ start being consistent with zero for $k\ge 3$, as is the case in 
a massless, ideal gas. For $k=1$ and 2, deviations from the ideal
gas limit are well described by $\order{g^2}$ high-temperature perturbation theory, although
for a detailed quantitative comparison, refined
resummed approaches such as hard thermal loop 
(HTL) \cite{Haque:2014rua} or 3-$d$ effective
theory (EQCD) \cite{Mogliacci:2013mca} may be necessary, as discussed in Refs. \cite{Bazavov:2013uja,Ding:2015fca,Bellwied:2015lba}.
In Fig.~\ref{fig:taylor8} we
show the ideal gas limit result (red line) and $\order{g^2}$ corrections (green band), using 
a 2-loop running coupling \cite{Prosperi:2006hx} $g^2(T)$ with a
renormalization scale $k_T\pi T$ with $4\le k_T\le 8$ as discussed in Appendix~\ref{app:IG}.

For the higher-order expansion
coefficients, $k\ge 3$, the ideal gas limit
as well as $\order{g^2}$ corrections vanish,
which also is consistent with the results
shown in Fig.~\ref{fig:taylor8}.

We also note that
for vanishing quark masses, the strange quark sector in strangeness-neutral, isospin-symmetric matter does
not contribute to the perturbative
expansion of the pressure up to $\order{g^2}$, {\it i.e.} although 
in this case the strangeness chemical potential is non-zero, $\mu_S=\mu_B/3$, 
the flavor chemical potential for strange quarks vanishes, $\mu_s=0$.
The rapid approach to perturbative behavior 
of the Taylor expansion coefficients of pressure, and as such also to the $\hmu_B$-dependent contribution to the pressure,
is in contrast to the behavior seen at $\hmu_B=0$, where it was found that even at
$T\simeq 300$~MeV deviations from the ideal gas limit amount to almost 50\% for the pressure and 25\% for the energy density \cite{HotQCD:2014kol}. 

\subsubsection{Low-temperature region}
At low temperatures lattice QCD results for 
$P_2$ and $P_4$ approach results
obtained  in hadron resonance gas
(HRG) model calculations using
non-interacting, point-like resonances. 
In Fig.~\ref{fig:taylor8} QCD results 
for the pressure coefficients are compared
to HRG model calculations that use the QMHRG2020 list of hadron resonances \cite{Bollweg:2021vqf}.
It is apparent that differences between QCD
and HRG model calculations show up earlier
and are more pronounced with increasing order of the 
expansion coefficients. While $P_2/P_2^{{\rm HRG}}$ deviates from unity by about 10\% at $T\simeq\Tpc$, this 
deviation reaches already 50\% for
$P_4/P_4^{{\rm HRG}}$. Moreover, it is 
evident from the comparison of the 
temperature dependence of $P_4$ calculated
in QCD and the QMHRG model, respectively, that
the slope of $P_4(T)$ differs already
significantly for $T>140$~MeV.

Deviations from HRG model calculations thus
are even more apparent in the slope of $P_{2k}(T)$. 
The $T$-derivatives, $P'_{2k}=T{\rm d}P_{2k}/{\rm d}T$, of the expansion 
coefficients are shown in Fig.~\ref{fig:dPdT}. As expected 
$P'_2(T)$ starts deviating from HRG model
calculations already at $T\simeq 140$~MeV and differences are about $30$\%  in the vicinity of the pseudo-critical temperature, $\Tpc$. For
$P'_4(T)$ differences are significant in
the entire low-temperature region, $T\ge 125$~MeV.

The functions $P'_{2k}(T)$ are
the Taylor expansion coefficients of the $\hmu_B$-dependent part of the trace anomaly,
\begin{equation}
    \frac{\Delta(\epsilon-3p)}{T^4} =\sum_k P'_{2k} \hmu_B^{2k}\; .
    \label{e3p}
\end{equation}
We thus can expect that differences between
HRG and QCD results will be larger for
energy and entropy densities than for pressure
and net baryon-number density as the former two observables
receive contributions from $P'_{2k}(T)$.
This will be discussed in Sec.~\ref{sec:taylorEnergyEntropy}.

\subsubsection{Intermediate temperature range: Pseudo-critical region}
The expansion coefficients shown in
Fig.~\ref{fig:taylor8}, as well as their $T$-derivatives shown in  Fig.~\ref{fig:dPdT}, make it clear that at temperatures in the 
vicinity of $\Tpc$, deviations from the asymptotic behavior at low and
high temperature are large. 
As has been noted before, the temperature dependence and the relation between subsequent expansion coefficients in the vicinity of 
$\Tpc$ resemble many features expected from universal scaling in the
vicinity of a second-order phase
transition. The maximum in
$T{\rm d}P_2/{\rm d}T$ is close to 
$\Tpc$. In fact, as $P_2(T)$ is an energy-like observable
\cite{Clarke:2020htu}, its derivative with respect to $T$,
\begin{eqnarray}
P'_2 =T\frac{{\rm d}P_2}{{\rm d}T}\equiv \left. \frac{T}{2} \frac{\partial^3 \hp}{\partial T \partial\hmu_B^2} \right|_{\hmu_B=0}\; ,
\end{eqnarray}
behaves like a specific heat, which in the chiral limit will develop a pronounced peak 
at the chiral phase transition temperature $\Tc$. 
This can be used as one of the
definitions of a pseudo-critical temperature. However, unlike the magnetization-like susceptibilities, which diverge at $\Tc$, the energy-like 
susceptibility will only lead to a local
maximum at $\Tc$. Using the maximum of
$P'_2$ as a definition for $\Tpc$ thus
may be more strongly affected by regular
contributions to the partition function.
Its maximum, on the other hand, 
characterizes the temperature at which 
$P''_2(T)$ vanishes. As the maximum
and minimum of $P''_2(T)$ on the left
and right of this crossing point approach each other and will diverge
in the chiral limit, the crossing point 
itself is a well defined estimator for a
pseudo-critical temperature. The temperature at
this crossing point will 
converge to the uniquely defined $\Tc$
in the chiral limit.
We find for the location of the maximum 
of $P'_2$,
\begin{eqnarray}
{\rm d}P'_2/{\rm d}T = 0 \;\; \Rightarrow \;\; T_{{\rm pc},P_2} = 153.8 (7) (5) \ {\rm MeV} \; , \;~ 
\label{Tpcp2}
\end{eqnarray}
where the first error is statistical and the 
second error reflects the uncertainty in 
defining the $T$-scale used in all our 
calculations. 

\begin{figure}[t]
\includegraphics[scale=0.64]{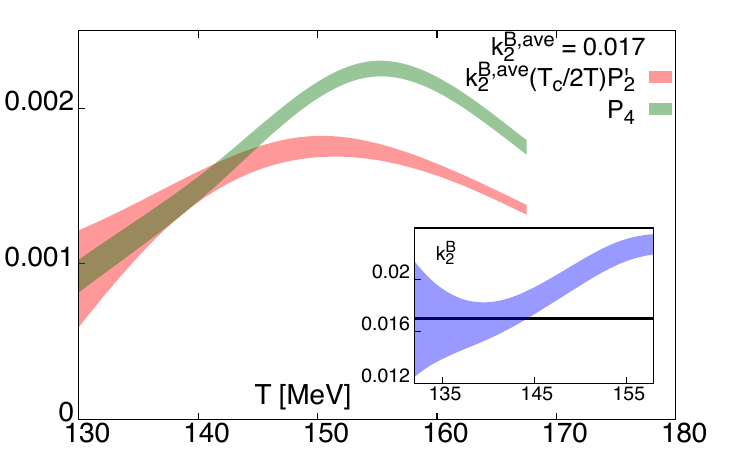}
\caption{Comparison of the fourth-order Taylor expansion coefficient, $P_4$, 
of the pressure and the temperature derivative of the second-order
expansion coefficient, $P_2$. The red and green bands reflect the errors on $P'_2$ and $P_4$, respectively. The inset shows results for 
$\kappa_2^B$ obtained from an appropriately scaled ratio of $P'_2$ and $P_4$ as discussed in
the text. From this we find 
$0.012 \le \kappa_2^B\le 0.022$.
}
\label{fig:P4-e2}
\end{figure}

Similarly, the maximum and minimum, visible in $P'_4$ (Fig.~\ref{fig:dPdT}~(top, right)) will
diverge in the chiral limit. 
The temperature at which $P'_4$ changes sign thus
also is a good observable to define a pseudo-critical temperature.
From the continuum-extrapolated fits shown
in Fig.~\ref{fig:dPdT} we thus deduce
another estimator for $\Tpc$,
\begin{eqnarray}
P'_4 = 0 \;\; \Rightarrow \;\; T_{{\rm pc},P_4} = 155.3 (2) (5) \ {\rm MeV} \; . \;~
\label{Tpcp4}
\end{eqnarray}
Both estimators for $\Tpc$, based on temperature-like derivatives of $p/T$, are
in good agreement with results obtained previously by the HotQCD
Collaboration using

\textcolor{black}{
the average over 5 different observables
that can be used to define a pseudo-critical temperature\footnote{Note
that pseudo-critical temperatures are not unique and depend on the
observable used to define them.} i.e. $T_{pc} = 156.5(1.5)$ MeV
\cite{HotQCD:2018pds}.
Other determinations of pseudo-critical temperature,
using the zero temperature subtracted chiral susceptibility only
\cite{Borsanyi:2020fev}, gave $T_{pc}=158.0(6)$ MeV, and a
recent, not yet continuum extrapolated, calculation with
twisted mass fermions in (2+1+1)-flavor QCD examined three different
definitions of pseudo-critical temperature giving results ranging
from $146.2(21)(1)$~MeV to $157.8(7)(10)$~MeV \cite{Kotov:2021rah}.
}
.

The fact that the temperature derivative
of $P_4$, on the one hand, and the  temperature derivative of the mixed
observable $P'_2$ (involving $T$- and
$\hmu_B$ derivatives), on the other hand, lead  
to quite similar results for the value of a pseudo-critical
temperature is naturally understood in terms
of universal behavior in the vicinity of a
critical point. Close to $\Tc$ higher-order 
derivatives of the pressure are dominated
by the contribution of the so-called singular 
part of $p/T$, 
\begin{equation}
    p/T \sim f_f(z)  ,
\end{equation}
where $f_f$ is the universal scaling function corresponding to the universality
class of the chiral phase transition\footnote{The chiral phase transition generally is expected to belong to the $3$-$d$, $O(4)$ universality class, although larger symmetry groups may become relevant, if the anomalous $U(1)$ symmetry of QCD gets effectively resorted \cite{Pelissetto:2013hqa}.}.
The argument $z$ of the scaling function $f_f(z)$ and its pre\-factor generally depend on the quark masses. However, as we are interested here only
in derivatives with respect to 
$T$ and $\hmu_B$, keeping the quark masses fixed, we may think of
the argument of the scaling function as just 
being proportional to the reduced 
temperature $t$, {\it i.e.} $z\equiv c_m t$, with $t$ given by a leading-order Taylor expansion,
\begin{eqnarray}
    t= \frac{1}{t_0}&& \left( \frac{T-\Tc}{\Tc} + \kappa_2^B \hmu_B^2 \right.
    \nonumber \\
    &&\left. +\order{\hmu_B^4,\hmu_B^2(T-\Tc),(T-\Tc)^2}\right) \; .
\end{eqnarray}
Here $t_0$, as well as the chiral phase
transition temperature $\Tc$ and the curvature coefficient $\kappa_2^B$, are non-universal parameters. 
This suggests that up to
a proportionality factor, taking two
derivatives with respect to $\hmu_B$
is equivalent to taking one derivative
with respect to $T$
\cite{Kaczmarek:2011zz,Karsch:2019mbv}.

The anticipated relation between derivatives 
with respect to $T$ and $\hmu_B$, respectively,
is apparent in the structure of the expansion coefficients shown in Figs.~\ref{fig:taylor8} and \ref{fig:dPdT}; $P'_2$ and $P_4$ receive 
singular contributions from $f''_f(z)$, 
\begin{eqnarray}
{P_2^s}' &=& -\frac{2 T}{\Tc} \kappa_2^B A f_f''(z) + {\rm subleading}
\nonumber \\
P_4^s &=& - (\kappa_2^B)^2 A f''_f(z) + {\rm subleading}\; ,
\end{eqnarray}
where we summarized common factors in the
factor $A=(h_0/2t_0^2T^3)~h^{-\alpha/\beta\delta}$.

At temperatures in 
the region between $\Tc$ and the pseudo-critical temperature for physical quark mass values,
the relative magnitude of 
$P'_{2}$ and $P_{4}$ 
is well described by scaling relations.
Matching values of $P'_2$ and $P_4$ in the
vicinity of $\Tc$ thus ensures that their 
$T$-derivatives, which
will diverge in the chiral limit,
will become identical when approaching this 
limit. 

In Fig.~\ref{fig:P4-e2} we show the
expansion coefficient $P_4$ and
compare it with an appropriately 
rescaled temperature derivative, $P'_2$, of the expansion coefficient $P_2$. To determine the rescaling factor $\kappa_2^B$ we calculate the
ratio $2T P_4/\Tc P'_2$ in the 
temperature interval $[130~{\rm MeV}:156.5~{\rm MeV}]$ and average the
results in this interval. As can be seen in the
inset of Fig.~\ref{fig:P4-e2} these estimates
increase with increasing $T$, indicating the
relevance of regular contributions to the
expansion coefficients $P_{2k}$, which lead 
to deviations from a $T$-independent result
for $\kappa_2$ that one would expect to 
find in the scaling regime. For the 
curvature coefficient we thus find 
$0.012 \le \kappa_2^B\le 0.022$. This result
is in good agreement with other determinations of
the curvature coefficient $\kappa_2^B$
\cite{HotQCD:2018pds,Bellwied:2015rza,Bonati:2018nut} that are based on properties of magnetization-like observables.

\subsection{Taylor expansion coefficients for energy and entropy densities}\label{sec:taylorEnergyEntropy}

\begin{figure*}[t]
\includegraphics[scale=0.65]{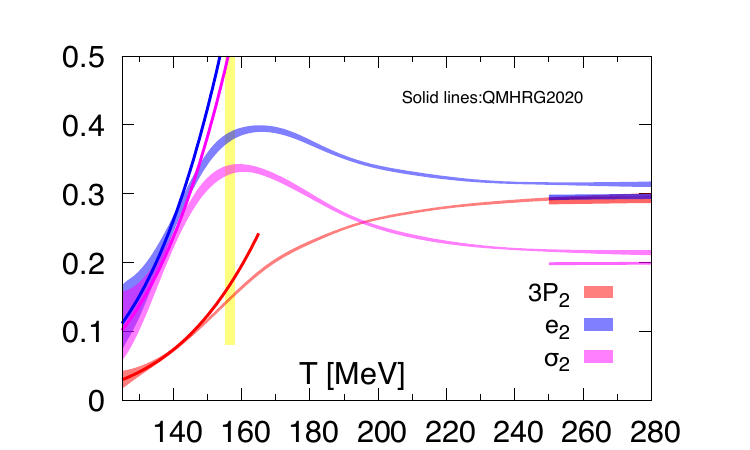}
\includegraphics[scale=0.65]{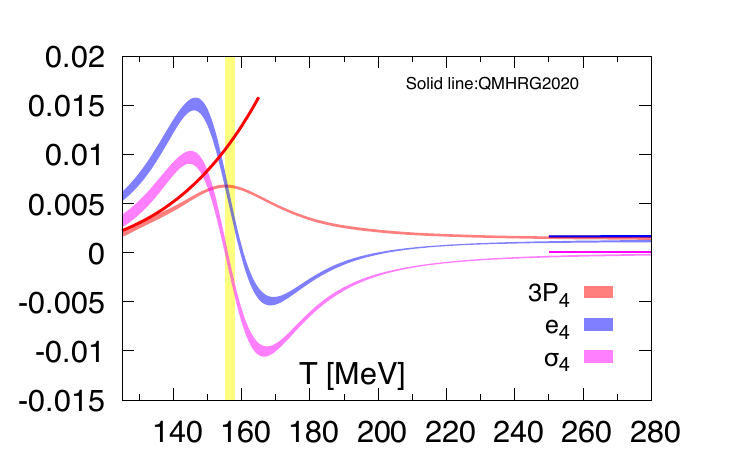}
\includegraphics[scale=0.65]{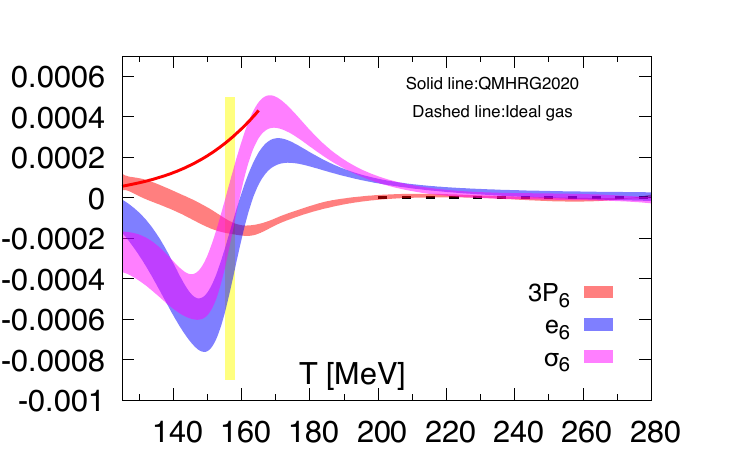}
\includegraphics[scale=0.65]{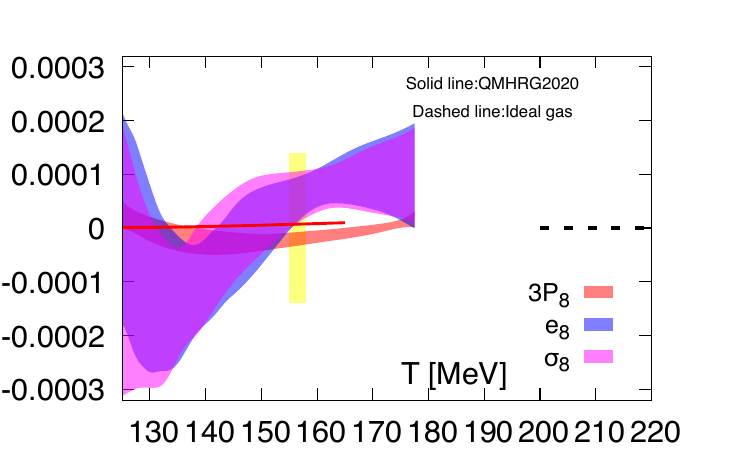}
\caption{Second-order (top, left), fourth-order (top, right), sixth-order (bottom, left) and eighth-order (bottom, right) expansion coefficients of pressure ($P_{2k}$), energy density ($\epsilon_{2k}$) and entropy density
($\sigma_{2k}$) for a isospin-symmetric, strangeness-neutral 
medium ($\mu_Q=0, n_S=0$).
Bands at high temperature show the 
$\order{g^2}$ perturbative result using a renormalization scale in the range $[4\pi T,8\pi T]$. The yellow band indicates the location of the pseudo-critical temperature at $T_{pc}$ \cite{HotQCD:2018pds}. 
}
\label{fig:pes-mu}
\end{figure*}

As discussed in Sec.~\ref{sec:Taylor}
in the case of isospin-symmetric, strangeness-neutral matter,
Taylor expansion coefficients of
net baryon-number, energy and 
entropy densities are all given in
terms of the expansion coefficients $P_{2k}$ and their temperature
derivatives $P'_{2k}$. In particular, 
the Taylor series for the $\hmu_B$-dependent
contribution to the trace anomaly $(\epsilon -3p)/T^4$ is given in terms of the 
$T$-derivatives, $P'_{2k}$, only.
Combining results for $P_{2k}$ and $P'_{2k}$,
shown in Figs.~\ref{fig:taylor8} and \ref{fig:dPdT}, we then obtain the expansion
coefficients for Taylor series of the
energy and entropy densities.
We show the 
results for these expansion 
coefficients, together with those of the pressure series\footnote{Note that the expansion coefficients for the number density are just proportional to those of the pressure series and thus are not 
shown separately in Fig.~\ref{fig:pes-mu}.},
in Fig.~\ref{fig:pes-mu}. As discussed for the expansion coefficients $P_{2k}$, also 
$\epsilon_{2k}$ and $\sigma_{2k}$
are continuum extrapolated 
for $k=1,\ 2$ and we give
spline interpolations of the 
$N_\tau=8$ results for $k=3,\ 4$.
We show a comparison with HRG model
calculations for the expansion 
coefficients of the pressure. For energy and entropy we do so only for the second-order expansion coefficients. 
As discussed in connection with Fig.~\ref{fig:dPdT},
differences between HRG model and QCD calculations are large for higher-order expansion coefficients in the entire temperature range .

At high temperatures we again show 
results from $\order{g^2}$ perturbation
theory (PT). In the case of the energy
and entropy densities this also includes
contributions arising from $T$-derivatives of the coupling $g^2(T)$,
although this might be considered as a 
part of the $\order{g^4}$ contribution
in a perturbative expansion. We thus consider our  $\order{g^2}$ PT-ansatz
as a model for describing the high-$T$
behavior of the Taylor expansion coefficients and subsequently also for
the bulk thermodynamic observables.

\begin{figure*}[t]
\includegraphics[scale=0.68,page=1]{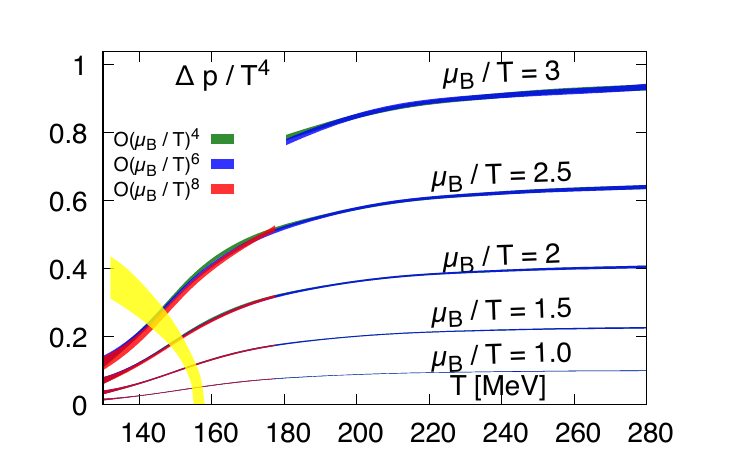}
\includegraphics[scale=0.68,page=1]{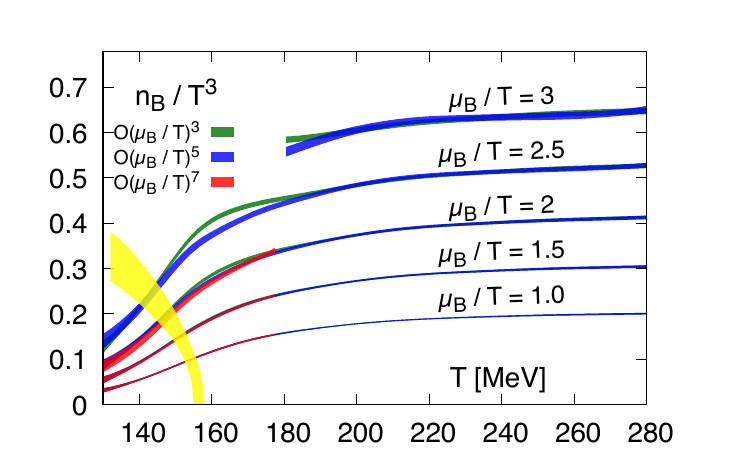}
\caption{Pressure (left) and net baryon-number density (right) versus temperature for several values of the baryon chemical potential $\hmu_B\equiv \mu_B/T$. Shown
are results obtained in different orders of the Taylor series of pressure and net baryon-number
density of isospin-symmetric, strangeness-neutral matter. 
The yellow bands highlight the
variation of $\Delta p/T^4$ and 
$n_B/T^3$ with $\hmu_B$ at $\Tpc(\hmu_B)$.
}
\label{fig:p-nB}
\end{figure*}

\section{Bulk thermodynamics of (2+1)-flavor QCD}\label{sec:bulkTherm}

We will use here the Taylor series of the pressure
and related observables to calculate thermodynamic 
observables in an isospin-symmetric, strangeness-neutral medium as function of temperature and baryon
chemical potential. Further on, we will eliminate the
baryon chemical potential in favor of other 
external control parameters like the net baryon-number
density or constant ratio of entropy over net baryon number.

\subsection{Density dependent contribution to bulk thermodynamic observables} 

In \cite{Bollweg:2022rps} we already used the new
data obtained in the temperature interval $T\in [125 {\rm MeV}:175 {\rm MeV}]$ to construct $8^{\rm th}$-order 
Taylor series for the pressure and the related
$7^{\rm th}$-order series for the net baryon-number density.
We had shown there that these expansions are reliable
for $\hmu_B\lesssim 2.5$ for the pressure and 
$\hmu_B\lesssim 2.0$ for the number density, respectively. It also 
could be shown that these expansions agree well with
corresponding \pade approximants. Here we extend
this discussion up to $T=280$~MeV using existing 
data from our previous analysis of Taylor series in QCD \cite{Bazavov:2017dus}, where 
a somewhat larger light quark mass, $m_\ell/m_s=1/20$, has been used in the
high-temperature region. Using such a 
somewhat larger light quark mass at high
temperature is of no concern as $m_\ell/T\ll 1$ and has only little influence on
thermodynamic observables in this region. 
We furthermore present results for the
energy and entropy densities, which require the 
temperature derivatives of the expansion coefficients of the pressure.

\subsubsection{EoS: Pressure versus net baryon-number density}

In Fig.~\ref{fig:p-nB} we show the  number density (right) and the density dependent ($\mu_B\ne 0$) contribution to the pressure (left) for the case
$(\mu_Q,n_S)=(0,0)$ and for several values of $\hmu_B$. Similar results for the case
$(\mu_Q,\mu_S)=(0,0)$ have been shown in
\cite{Bollweg:2022rps}.
At high temperatures, $T\gtrsim 200$~MeV, the Taylor series converges rapidly. As can be seen 
from the expansion coefficients presented in Fig.~\ref{fig:taylor8}, in this temperature range
the $4^{\rm th}$-order expansion coefficient, $P_4$, 
is more than two orders of magnitude smaller than the $2^{\rm nd}$-order coefficient, $P_2$. Furthermore, within our current statistical 
accuracy the $6^{\rm th}$-order coefficient, $P_6$, is consistent with zero. Taking into account the statistical error on $P_6$, also its magnitudes is at least 
two orders of magnitude smaller than the $4^{\rm th}$-order expansion coefficient.
For $T\gtrsim 200$~MeV we thus show results for the
pressure and net baryon-number density also for 
larger chemical potentials, e.g.
$\hmu_B\simeq 3$. Using even larger values 
for $\hmu_B$ in this high-temperature
regime seems to be possible.

For $T\le 200$~MeV contributions from higher-order expansion coefficients become important. 
The $6^{\rm th}$ and
$8^{\rm th}$-order expansion coefficients, which are clearly non-zero, show a lot 
of structure. Nonetheless, for $T\le 150$~MeV
their contribution to the Taylor series
of pressure and net baryon-number density is 
small\footnote{Note that also in a hadron resonance gas, where the $\hmu_B$-dependent contribution is
proportional to $\cosh (\hmu_B)-1$, the $\order{\hmu_B^4}$ Taylor series provides more than 90\% of the exact result. Including also the  $\order{\hmu_B^6}$ contribution
results in agreement with the exact HRG result to better than 1\% for $\hmu_B\le 2.5$.}
for $\hmu_B\le 2.5$. This is in accordance with estimates for the radius of convergence $(\mu_B/T)_{conv}$ of the Taylor series \cite{Bollweg:2022rps,Mukherjee:2019eou,Dimopoulos:2021vrk} that suggest $(\mu_B/T)_{conv}<2.5$  also in this temperature range.
For further discussion on the radius of  convergence of the Taylor series see
Refs. \cite{DElia:2016jqh,Giordano:2019slo,Giordano:2019gev}.
We also note that in this temperature range
and for $\hmu_B<2.5$ Taylor series results 
are consistent within errors with those
obtained from the $4^{\rm th}$-order series. 
For temperatures in the vicinity 
of $\Tpc$, {\it i.e.} for $150~{\rm MeV}<T<200$~MeV, contributions from higher-order Taylor expansion coefficients, however, 
need to be taken into account already for $\hmu_B>2$.

This suggests that in the three different temperature 
intervals $4^{\rm th}$ and $3^{\rm rd}$-order Taylor series 
provide a good approximation to the $8^{\rm th}$-order Taylor series
for pressure and net baryon-number density, 
as long as the baryon chemical potential is smaller than
(i) $2.5T$ for $T\le 150$~MeV, (ii) $2T$ for $150~{\rm MeV} <T< 200$~MeV 
and (iii) $3T$ for $T\ge 200$~MeV.
In these cases we may invert $p(\mu_B)$ and $n_B(T,\mu_B)$ exactly and 
provide a simple analytic expression for $p(T,n_B)$,
which provides useful insight into the structure of the 
QCD EoS in a parameter range that covers almost the entire 
parameter range accessible with the beam energy scan (BES-II) at the Relativistic Heavy Ion Collider (RHIC) in 
collider mode. In particular, at the two lowest beam energies
$\sqrt{s_{_{NN}}}=7.7$~GeV and 11.5~GeV the baryon chemical
potentials at the freeze-out temperature  $T_f(\mu_B)$ has been 
estimated to be about $\mu_B/T_f=2.8$ and $2.0$, respectively 
\cite{STAR:2017sal}.

Inverting the $\order{\hmu_B^3}$ Taylor series for the net baryon-number 
density we obtain the chemical potential as function
of $T$ and $n_B$,
\begin{widetext}
\begin{eqnarray}
    \hmu_B(T,n_B) &=&
    y(T,n_B)^{1/3} 
    -\frac{N_1^B(T)}{3N_3^B(T)}y(T,n_B)^{-1/3} 
    \; ,
\label{Yn}\\
    y(T,n_B) &=& \frac{N_1^B(T)}{N_3^B(T)}
    \left( \frac{\hn_B}{2 N_1^B(T)} +\sqrt{\frac{N_1^B(T)}{27 N_3^B(T)}+\left(\frac{\hn_B}{2 N_1^B(T)}\right)^2}\right)
  \nonumber \; ,
\end{eqnarray}
\end{widetext}
where $N_1^B(T)=2 P_2(T)$ and $N_3^B(T)=4 P_4(T)$ are the 
coefficients of $\order{\hmu_B^2}$ and $\order{\hmu_B^4}$ terms in
the net baryon-number density, which are obtained either from
the Taylor series for net baryon-number density or in $\order{g^2}$ 
high-temperature perturbation theory.

Inserting these results for the chemical potential in the $4\nth$-order 
Taylor series for the pressure we obtain a low-order approximation for
the EoS of strangeness-neutral matter,
{\it i.e.} $p(T,n_B)$,
\begin{eqnarray}
    \frac{\Delta p(T,n_B)}{T^4} &=&
    P_2(T) \hmu_B^2(T,n_B)  + P_4(T) \hmu_B^4(T,n_B) 
    \; , \nonumber \\
    && \label{PnB} \; .
\end{eqnarray}

\begin{figure}
\includegraphics[width=\linewidth]{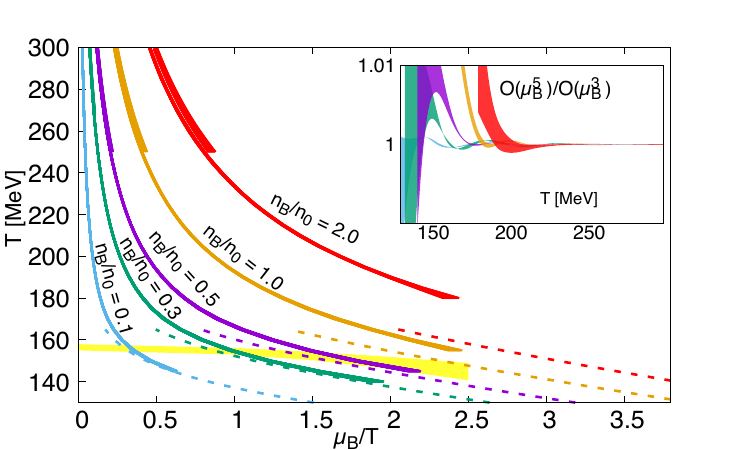}
\caption{Lines of constant net baryon-number density in the
$\hmu_B$-$T$ plane.
Solid bands indicate results obtained by numerically solving
the $\order{\hmu_B^5}$ series of $\hn_B$ for $\hmu_B$. 
Dashed lines indicate HRG results while short bands at
high temperature 
indicate the $\order{g^2}$ perturbative result with $4\leq k_T\leq8$.
The yellow band indicates $\Tpc(\hmu_B)$.
In the inset, we show 
the result from numerically inverting the Taylor series divided by the 
analytic inversion of the $\order{\hmu_B^3}$ result. 
}
\label{fig:fixednB}
\end{figure}

Going beyond the low-order approximation for the EoS we can
solve $\hn_B(T,\hmu_B)$ numerically
for $\hmu_B$ at fixed $T$. We then
obtain $\hmu_B(T,n_B)$, which 
can be inserted in the $8^{\rm th}$-order Taylor
series for $p(T,\hmu_B)$ and thus allows us to determine
$p(T,n_B)$.

In Fig.~\ref{fig:fixednB} we show results
for lines of constant net baryon-number density
in the $\hmu_B$-$T$ plane. Results are shown
for several values of $n_B/n_0$, with $n_0=0.16/{\rm fm}^{3}$ denoting 
nuclear matter density. They are
compared to QMHRG model calculations at low $T$ and $\order{g^2}$
perturbation theory at high $T$, respectively. The inset shows a comparison 
of the numerically inverted Taylor series result and the 
exact inversion of the $3^{\rm rd}$-order Taylor series for $\hn_B(\hmu_B)$
given in Eq.~\ref{Yn}. Here we show results only in $T$-intervals
where $\hmu_B$ stays below the maximal values given above.
As can be seen, the temperature 
dependence of $\hmu_B$ characterizing lines of constant
$n_B$ are well determined from
the $\order{\hmu_B^3}$ Taylor series for the net baryon number-density for $\hmu_B\le 2.5$.

\begin{figure*}[t]
\includegraphics[scale=0.68,page=1]{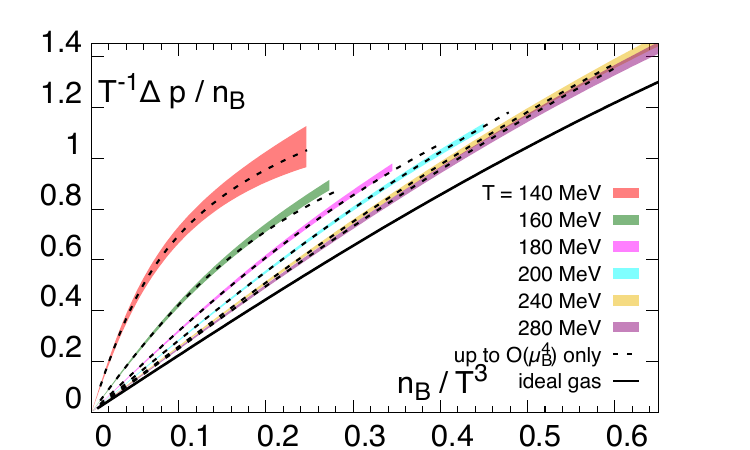}
\includegraphics[scale=0.68,page=1]{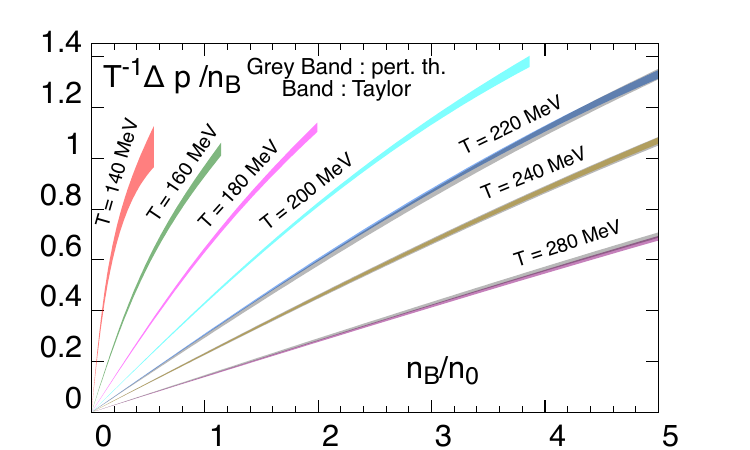}
\caption{Pressure divided by net baryon-number density versus $n_B/T^3$  (left) and $n_B/n_0$ (right), respectively.
Shown are results for strangeness-neutral, isospin-symmetric matter at several values of $T$. In the left hand figure we compare results
obtained from the full Taylor series for the
pressure with those obtained in $\order{\hmu_B^4}$ only (dashed lines). In the right
hand side the grey bands show a comparison with $\order{g^2}$ high-T perturbation theory. The bands
shown in both figure are shown up to values of $n_B/T^3$ or $n_B/n_0$ corresponding to
$\hmu_B=2.5$ for $T<200$~MeV and $\hmu_B=3$ otherwise.
}
\label{fig:p-vs-nB}
\end{figure*}

Inserting $\hmu_B(T,n_B)$ in the Taylor series for the pressure we obtain the 
contribution to the QCD EoS of strangeness-neutral, isospin-symmetric matter
that depends on non-zero net baryon number density.
Results for $\Delta p/(n_B T)$ as function of $n_B/T^3$ for several 
values of $T$ are shown in Fig.~\ref{fig:p-vs-nB}~(left). Again we only
show results for $\Delta p/(n_B T)$ for values of $n_B/T^3$ which correspond 
to $\hmu_B\le 2.5$ for $T\le 200$~MeV and $\hmu_B\le 3$ for $T > 200$~MeV. 
In Fig.~\ref{fig:p-vs-nB}~(left) we show corresponding results as function of 
$n_B/n_0$. 

For $T\gtrsim 240$~MeV it is evident that
$\Delta p(T,n_B)/(Tn_B)$, rises almost linearly in $n_B/T^3$ and is close to 
the ideal gas result. {\it I.e.}
$\Delta p(T,n_B)$  is proportional to 
$(n_B/T)^2$.
In this temperature range the density-dependent part of the
QCD EoS  is not only close to the ideal gas result, but also
agrees rather well with the $\order{g^2}$ perturbative EoS already for $T\ge 220$~MeV.
This is shown by the grey bands
in Fig.~\ref{fig:p-vs-nB}~(right) for
which we used the $\order{g^2}$ perturbative EoS for the case $(\mu_Q,n_S)=(0,0)$, given in Appendix~\ref{app:IG}, with a renormalization
scale $k_T\pi T$, using $4\le k_T\le 8$. 

From Eqs.~\ref{Yn} and \ref{PnB} we 
expect at low densities to find
$\Delta p\sim (n_B/T)^2$. The notion of low density, however, is temperature dependent. As can be seen from Eq.~\ref{Yn}, the leading order coefficient $N_1^B(T)$ sets the scale for $n_B$. As  $N_1^B(T)$ drops 
exponentially at low temperature one 
quickly enters a region of ``high density'' already at densities smaller 
than nuclear matter density, $n_0\simeq 0.16 {\rm fm}^{-3}$, whereas at high temperature the low density regime ($\Delta p\sim (n_B/T)^2$)
persists into a region well above $n_0$.

At low densities we thus expect that the 
density dependent contribution ($\Delta p$) to the total pressure is proportional to $(n_B/T)^2$
while the $n_B$-dependence weakens. 
At high densities and, within the approximation given by Eq.~\ref{PnB},
one finds $\Delta p\sim n_B^{4/3}$.

Results for $\Delta p(T,n_B)/n_B$, obtained with this $\order{\hmu_B^4}$
approximation (dashed lines), are
compared with the full Taylor series results
(bands) in Fig~\ref{fig:p-vs-nB} ~(left). The 
bands are shown up to a maximal density that can 
be reliably reached with our current $8^{\rm th}$-order Taylor series in different
temperature regions, {\it i.e.} we demand
that $\hmu_B\le 2.5$ for $T\le 200$~MeV and
$\hmu_B\le 3$ at higher temperatures. 
As can be seen the dashed lines indeed
provide a good approximation for the 
EoS at temperatures $135~{\rm MeV}< T\le 150$~MeV and
for $T\ge 200$~MeV.
Only for
$150~{\rm MeV}< T< 200~{\rm MeV}$ we find that an EoS based on the $4^{\rm th}$-order Taylor series is not sufficient. Here differences become visible in a region corresponding to $\hmu_B> 2$.

\subsubsection{Energy and entropy densities}

\begin{figure}[t]
\includegraphics[width=0.98\linewidth,page=1]{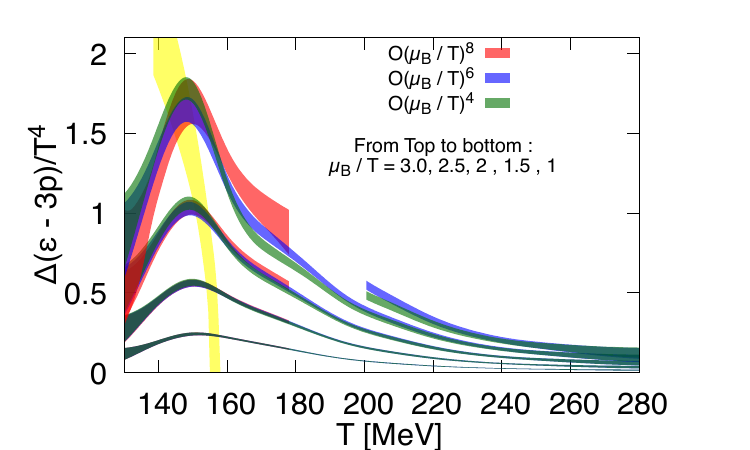}
\caption{The $\hmu_B$-dependent contribution to the 
trace anomaly in $(2+1)$-flavor QCD for several values of $\hmu_B$. The yellow band shows the
line $\Delta ((\epsilon-3p)/T^4)(\Tpc(\hmu_B))$.}
\label{fig:del_TrA}
\end{figure}

The calculation of Taylor series for the energy and entropy densities requires the 
derivatives of Taylor expansion coefficients of the pressure with respect to temperature,
$P'_{2k}\equiv T{\rm d}P_{2k}/{\rm d}T$.
These expansion coefficients are shown in
Fig.~\ref{fig:dPdT}. 
They directly give the Taylor series for the 
$\hmu_B$-dependent contribution to the trace 
anomaly, as introduced in Eq.~\ref{e3p}. In Fig.~\ref{fig:del_TrA} we show $\Delta(\epsilon-3p)/T^4$ for several values of
the baryon chemical potential. 
This should be compared to the trace 
anomaly at $\hmu_B=0$ ( Fig. 5~(right) of \cite{HotQCD:2014kol}).
We note that at $\Tpc$ and for
$\hmu_B=2.5$  the fermionic contribution to the trace anomaly
is as large as the contribution at 
$\hmu_B=0$ but rapidly drops at larger 
values of the temperature. For $T\ge 200$~MeV the 
$\hmu_B$-dependent contribution to the trace 
anomaly stays below 10\% of 
the value of the trace anomaly at $\hmu_B=0$. 

A consequence of this is that the maximum 
in the trace anomaly, which at $\hmu_B=0$ 
is reached at $T\simeq 200$~MeV \cite{HotQCD:2014kol}, gets shifted to smaller temperatures and comes closer 
to the pseudo-critical temperature. In fact,
as can be seen in Fig.~\ref{fig:del_TrA},
the location of the maximum of $\Delta (\epsilon - 3p)/T^4$ is close to the pseudo-critical
temperature
determined from chiral
observables,
\begin{equation}
    \Tpc(\hmu_B)=\Tpc^0 \left( 1-\kappa_2^{B,f}\hmu_B^2 +\order{\hmu_B^4}
    \right)    \;,
    \label{TpcmuB}
\end{equation}
with $T_{pc}^0=156.5(1.5)$~MeV and $\kappa_2^{B,f}=0.012(4)$ denoting the
curvature coefficient of the pseudo-critical 
line in strangeness-neutral matter\footnote{For a summary of different
determinations of the curvature coefficient $\kappa_2^B$, using various external constraints, see for instance \cite{DElia:2018fjp,Borsanyi:2020fev}.} \cite{HotQCD:2018pds}.
For large values of the 
chemical potential, where $\Delta (\epsilon - 3p)/T^4$ is dominated by
the $\hmu_B$-dependent contribution,
violations of the conformal relation
$\epsilon= 3p$ thus are maximal on the
pseudo-critical line.

\begin{figure*}[t]
\includegraphics[scale=0.68]{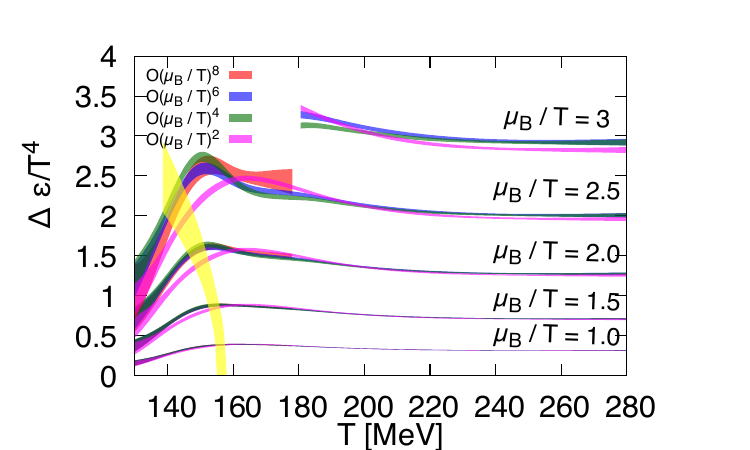}
\includegraphics[scale=0.68]{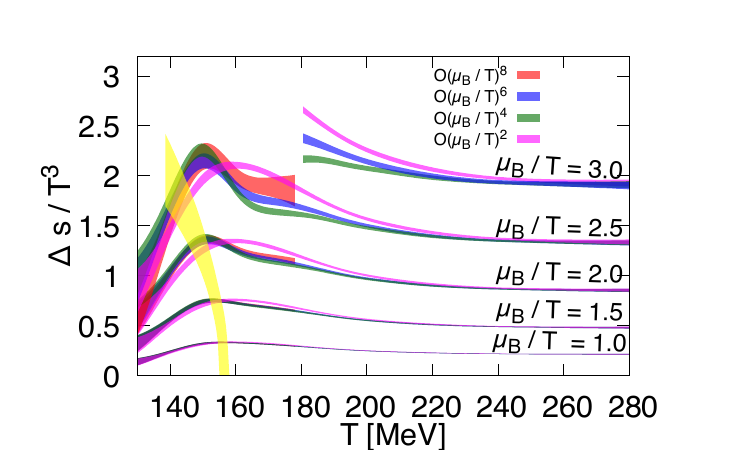}
\caption{$\hmu_B$-dependent part of the energy (left) and entropy (right) densities versus temperature. Shown are results from
$2^{\rm nd},\ 4^{\rm th},\ 6^{\rm th}$
and $8^{\rm th}$-order Taylor expansions. The yellow bands
indicate how $\Delta\epsilon/T^4$ and $\Delta s/T^3$ change with $\hmu_B$ at $\Tpc$.}
\label{fig:e-s-T}
\end{figure*}

Combining the results for $\Delta (\epsilon -3p)/T^4$ (Fig.~\ref{fig:del_TrA}) and  
$\Delta p/T^4$ (Fig.~\ref{fig:p-nB}) 
we obtain the Taylor series results for
the $\hmu_B$-dependent contributions to $\epsilon/T^4$
and $s/T^3$, respectively. These observables 
are shown in Fig.~\ref{fig:e-s-T}. 
As discussed in the case of $\Delta p/T^4$ and $n_B/T^3$, 
we also find for $\Delta (\epsilon-3p)/T^4$, as
well as for $\Delta \epsilon/T^4$ and $\Delta s/T^3$, that the
$4^{\rm th}$-order Taylor expansions provide good approximations to the full $6^{\rm th}$ and $8^{\rm th}$-order series results for $T\ge 200$~MeV
and $T\le 150$~MeV whenever 
$\hmu_B\le 2.5$ in the low-$T$
region and $\hmu_B\le 3$ at high $T$.
Higher-order corrections become
more important in the calculation
of the entropy density at high
temperature. As can be seen in Fig.~\ref{fig:e-s-T}~(right) the
$6^{\rm th}$ and $4^{\rm th}$-order correction to $s/T^3$
agree with each other in the 
region $T> 200$~MeV only up to
$\hmu_B\simeq 2.5$ but start to
differ for larger $\hmu_B$.

As can be seen in Fig.~\ref{fig:e-s-T} for large
$\hmu_B$ the $4^{\rm th}$-order Taylor expansion results for $\Delta \epsilon/T^4$ and
$\Delta s/T^3$ show a wiggly behavior in the intermediate temperature interval, 
$150~{\rm MeV}< T< 200~{\rm MeV}$, which results from the pronounced minimum of $P'_4$
in this temperature interval. This structure gets smoothed out by higher-order corrections. 
We will show in the next subsection 
that a Pad\'e-resummed Taylor series shows 
a much smoother behavior in this temperature
range.

\begin{figure*}[t]
\includegraphics[scale=0.69,page=1]{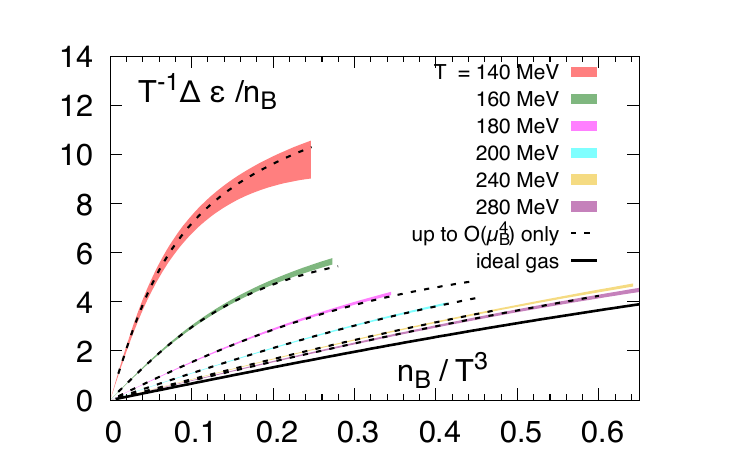}
\includegraphics[scale=0.69,page=1]{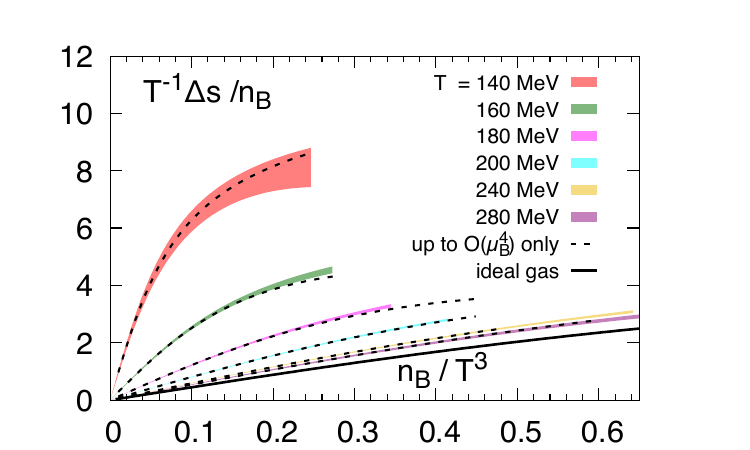}
\caption{Energy density in units of $T^4$ divided by net baryon-number density versus $n_B/T^3$  (left) and the corresponding figure for entropy density (right).
Shown are results for strangeness-neutral, isospin-symmetric matter at several values of $T$ using 
$6^{\rm th}$-order Taylor expansions for $\epsilon/T^4$ and $s/T^3$ and
a $5^{\rm th}$-order Taylor expansions for 
$n_B/T^3$. Dashed lines show a comparison
with results based on $4^{\rm th}$ and $3^{\rm rd}$-order expansions, respectively, 
using Eq.~\ref{Yn} to obtain $\epsilon (n_B)$
and $s(n_B)$.
}
\label{fig:evsnB}
\end{figure*}

We use the $6^{\rm th}$-order Taylor
expansion results for $\Delta p/T^4$, $\Delta \epsilon/T^4$ and $\Delta s/T^3$ to obtain 
the latter two as functions of $T$ and $n_B$. 
We again compare with corresponding
$4^{\rm th}$-order approximations using,
\begin{eqnarray}
\hspace{-0.5cm}    \frac{\Delta\epsilon(T,\hn_B)}{T^4}
    &=&     \epsilon_2(T) \hmu_B^2(T,n_B)  + \epsilon_4(T) \hmu_B^4(T,n_B)\; ,
    \label{enB} \\
   \hspace{-0.5cm}      \frac{\Delta s(T,\hn_B)}{T^3}
    &=&     \sigma_2(T) \hmu_B^2(T,n_B) + \sigma_4(T) \hmu_B^4(T,n_B)\; ,
\end{eqnarray}
with $\hmu_B(T,\hn_B)$ taken from Eq.~\ref{Yn}.
Results for $\Delta \epsilon(T,\hn_B)/n_B$ and
$\Delta s(T,\hn_B)/n_B$  as functions of
$\hn_B$ are shown in Fig~\ref{fig:evsnB}.

\subsubsection{Comparison of Taylor series and
their resummation using \pade approximants}

As seen already in the analysis of the Taylor
expansion for the pressure and net baryon-number density, \pade approximants
agree well with the Taylor series
themselves at low $\hmu_B$ up to the region
where we estimated the latter to provide
reliable results \cite{Bollweg:2022rps}.
We will extend this approach here to the
analysis of Taylor series for
the energy and entropy densities.

We use \pade approximants for thermodynamic
observables derived from the Taylor series
of the pressure in two ways. On the one hand
we construct \pade approximants based on the
Taylor series for a given observable, e.g. 
the energy and entropy density series given in Eqs.~\ref{energyc} and \ref{entropyc} can be resummed using
\pade approximants
similar to that of the pressure series given in Eq.~\ref{nmPade} by just
replacing the expansion coefficients $P_{2k}$
by $\epsilon_{2k}$ or $\sigma_{2k}$, respectively. On the other hand we use the
P-\pade, {\it i.e.} appropriate derivatives of the \pade approximants for the pressure, 
for the energy and entropy densities as given in Eqs.~\ref{energyPade} and \ref{entropyPade}.

\begin{figure*}
\includegraphics[scale=0.68,page=1]{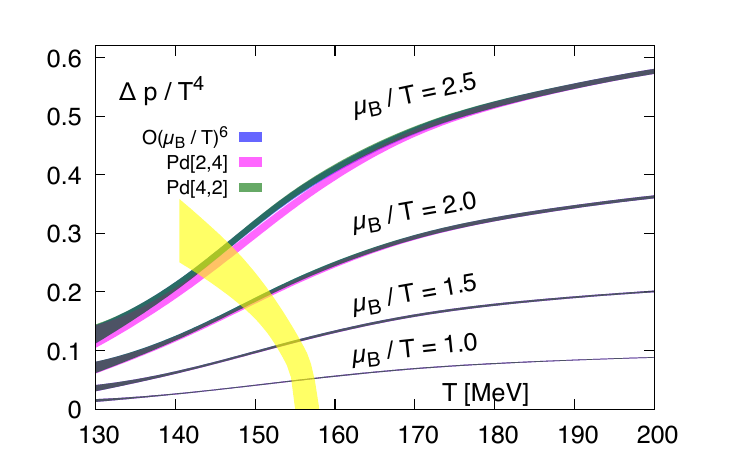}
\includegraphics[scale=0.68,page=1]{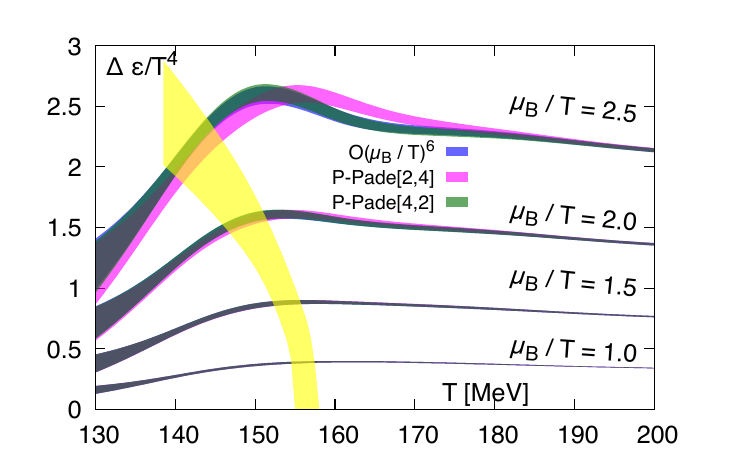}
\includegraphics[scale=0.68]{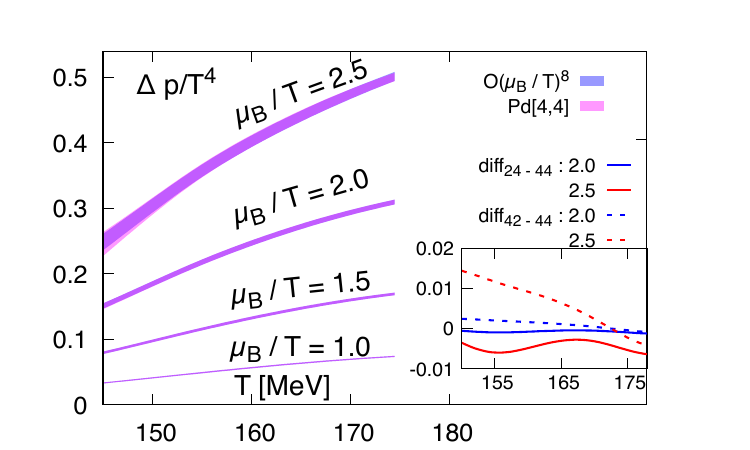}
\includegraphics[scale=0.68]{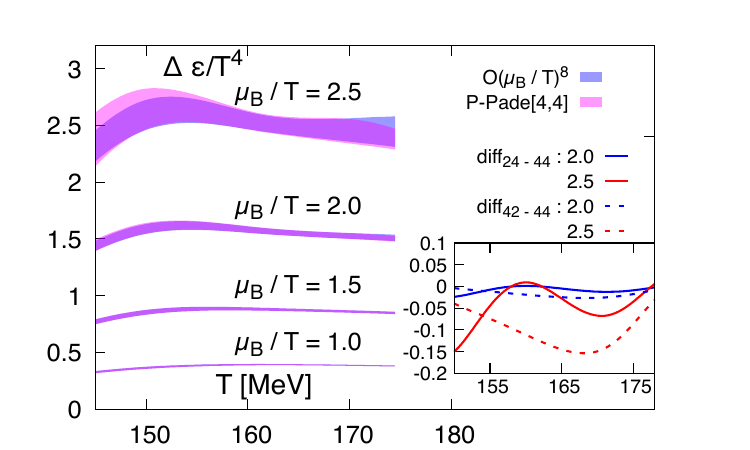}
\caption{Comparison of $6^{\rm th}$ (top)
and $8^{\rm th}$ (bottom) order Taylor expansion
results for $\Delta p/T^4$ and $\Delta \epsilon/T^4$ with corresponding [2,4] and [4,4] \pade and P-\pade approximants. 
The inset in the bottom
panels shows differences in 
the [2,4] and [4,4] as well as in 
[4,2] and [4,4] \pade approximations,
which are all based on $8^{\rm th}$-order Taylor series.
The yellow bands highlight the
variation of $\Delta p/T^4$ and 
$\Delta \epsilon/T^4$ with $\hmu_B$ at $\Tpc(\hmu_B)$.
}
\label{fig:Pe-pade}
\end{figure*}

In Fig.~\ref{fig:Pe-pade}~(left) we compare $6^{\rm th}$ 
and $8^{\rm th}$-order Taylor series for
$\Delta \hp$ with corresponding [4,2], [2,4] and [4,4]
\pade~approximants introduced in Eqs.~\ref{pd24}-\ref{nmPade}. Corresponding results for $\Delta \he$ are shown in  Fig.~\ref{fig:Pe-pade}~(right). 
In the figure for the pressure (top, left) we compare 
the $6^{\rm th}$-order Taylor expansion 
results with the two possible $[n,m]$ \pade approximants that use up to $6\nth$-order 
expansion coefficients. As can be seen the [4,2] \pade agrees with 
the Taylor series result while the
[2,4] \pade differs from these two
in the temperature interval 
$150~{\rm MeV}\lesssim T\lesssim 180~{\rm MeV}$. In the (bottom, left) figure shows
a comparison of the $8^{\rm th}$-order
Taylor expansion results with the [4,4] \pade approximant. They are in good agreement with
each other. Moreover, as can be seen from the 
inset, for large values of $\hmu_B$ the [2,4]
and [4,4] \pade approximants stay in
much better agreement with each other than the [4,2] and [4,4] \pade approximants or, equivalently, the
$6^{\rm th}$ and $8^{\rm th}$-order Taylor series. Similar conclusions can be drawn
for the energy density shown in the right hand part of the figure.

\begin{figure*}
\includegraphics[width=0.32\linewidth]{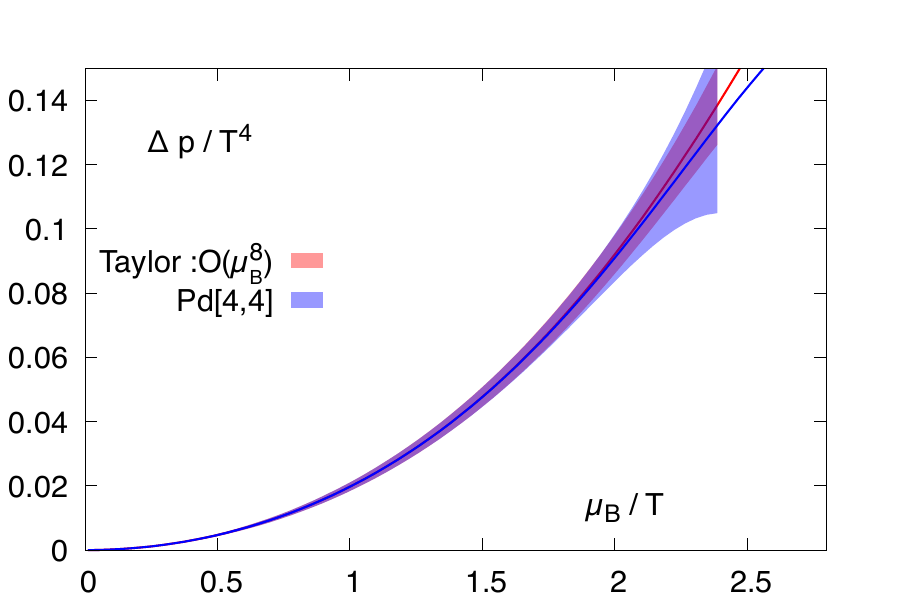}
\includegraphics[width=0.32\linewidth]{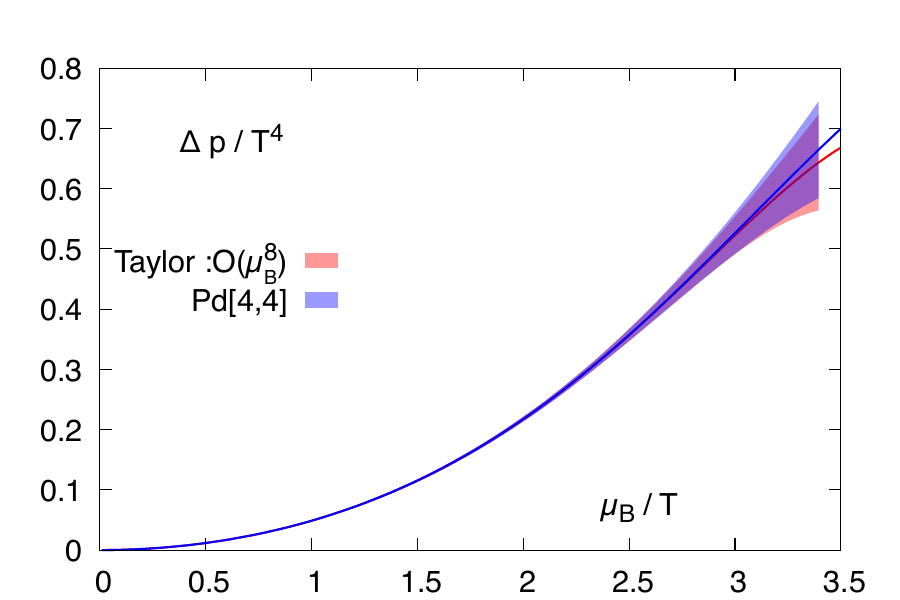}
\includegraphics[width=0.32\linewidth]{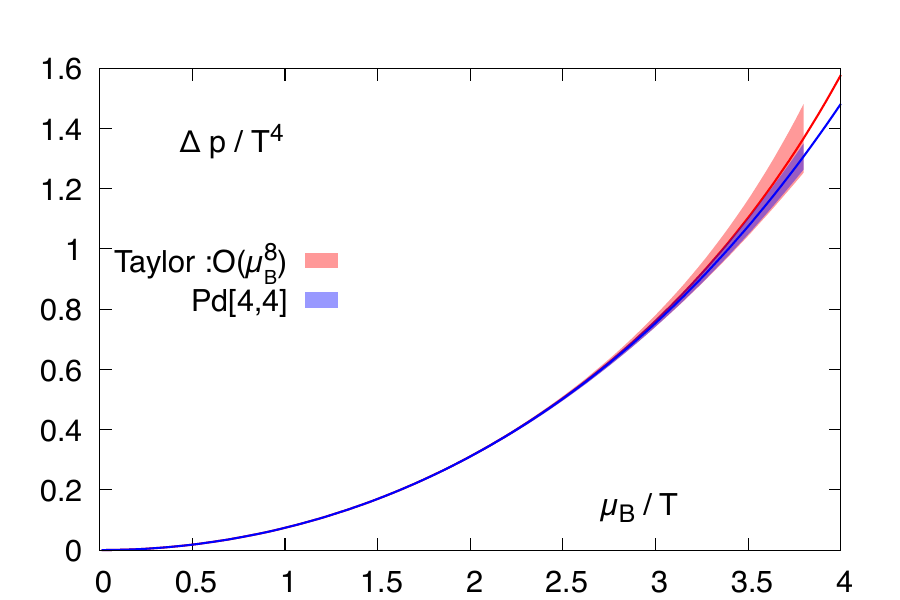}
\includegraphics[width=0.32\linewidth]{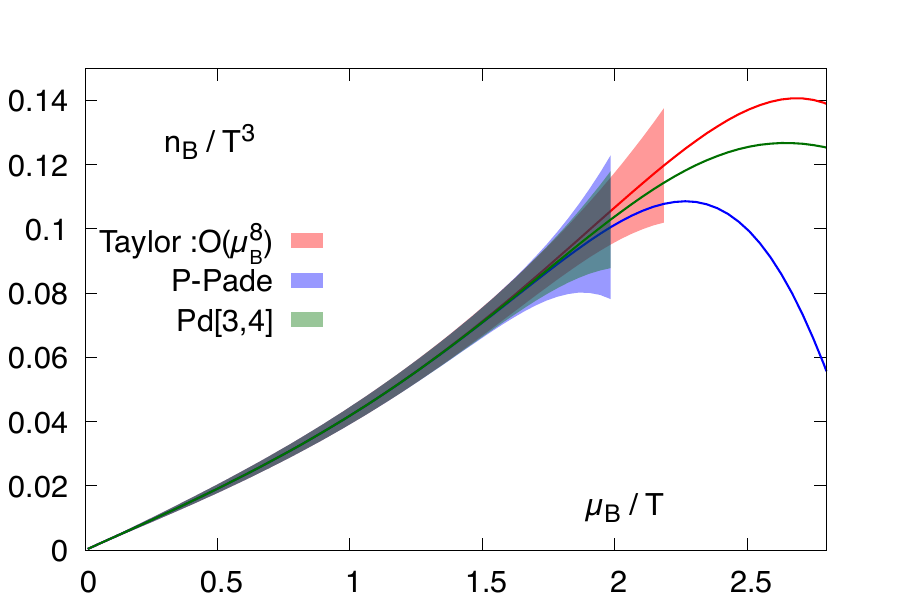}
\includegraphics[width=0.32\linewidth]{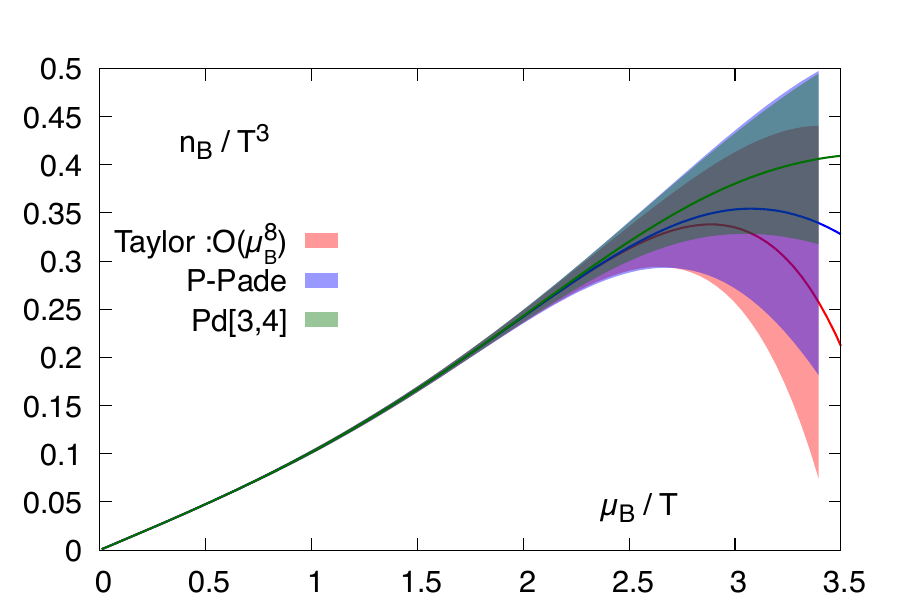}
\includegraphics[width=0.32\linewidth]{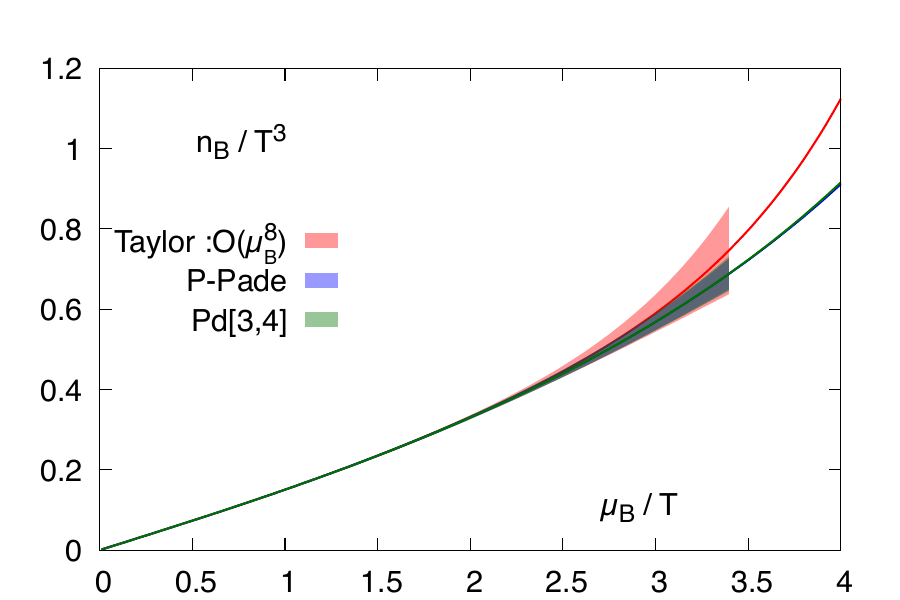}
\includegraphics[width=0.32\linewidth]{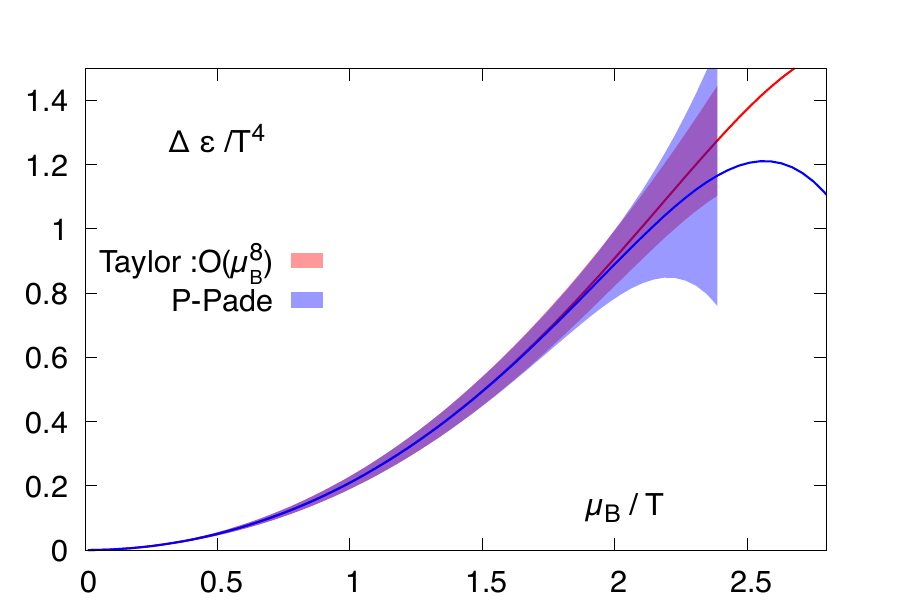}
\includegraphics[width=0.32\linewidth]{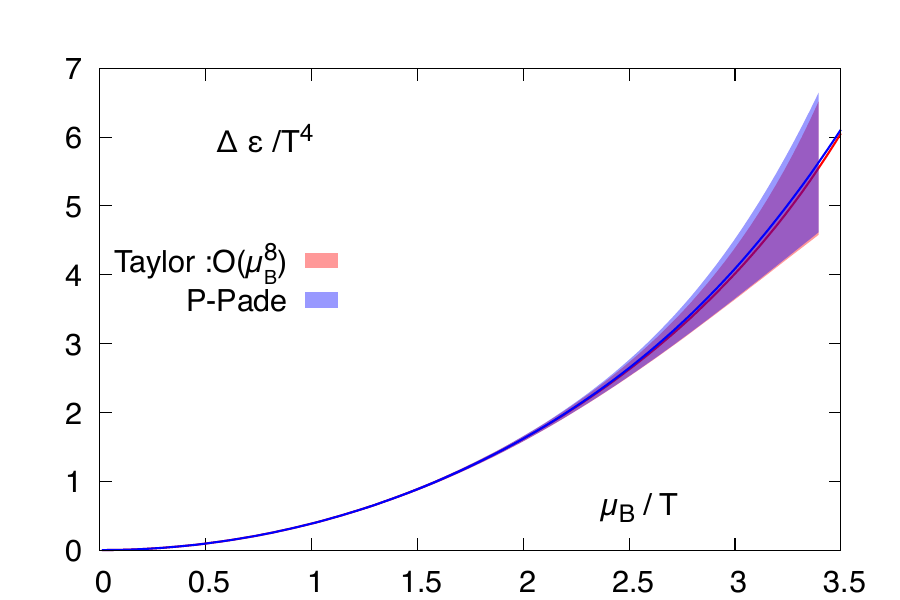}
\includegraphics[width=0.32\linewidth]{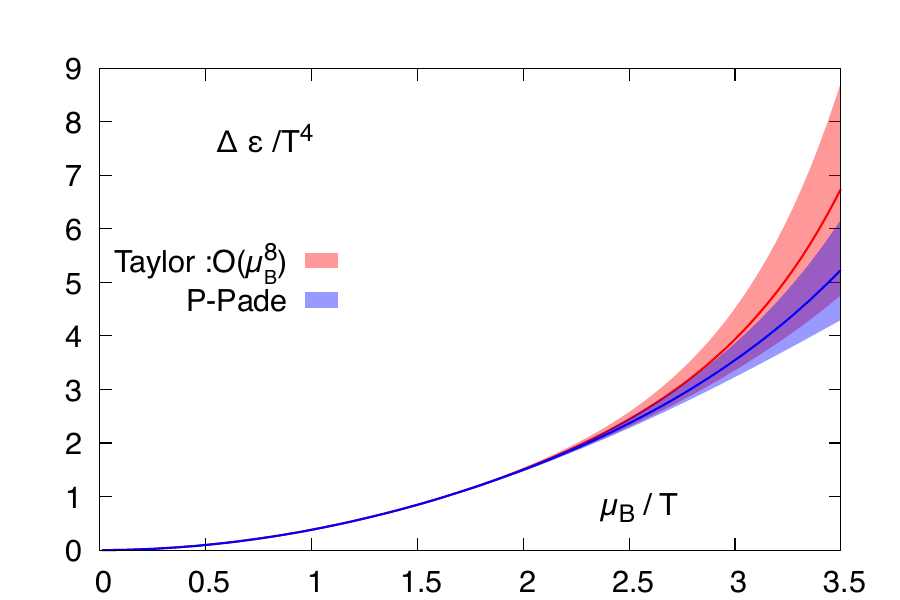}
\includegraphics[width=0.32\linewidth]{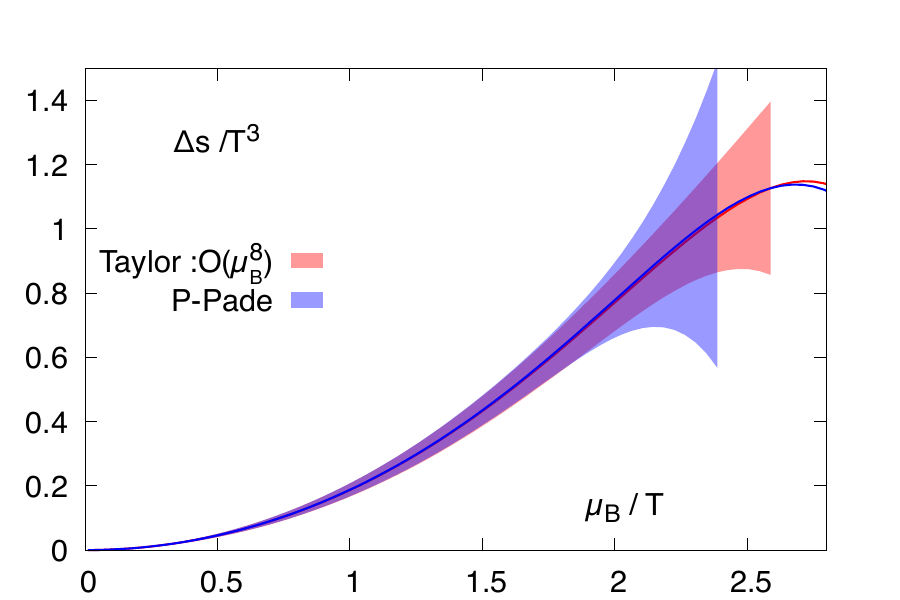}
\includegraphics[width=0.32\linewidth]{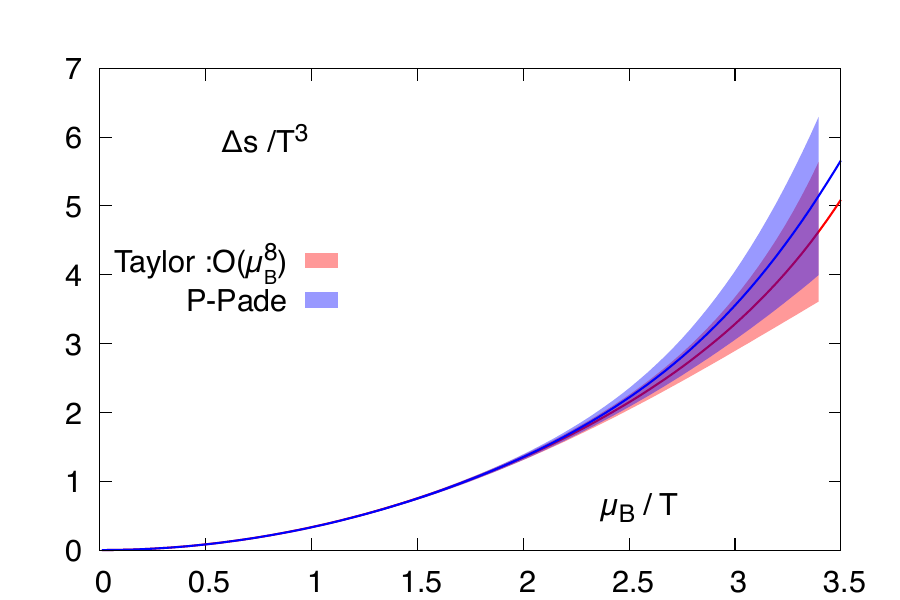}
\includegraphics[width=0.32\linewidth]{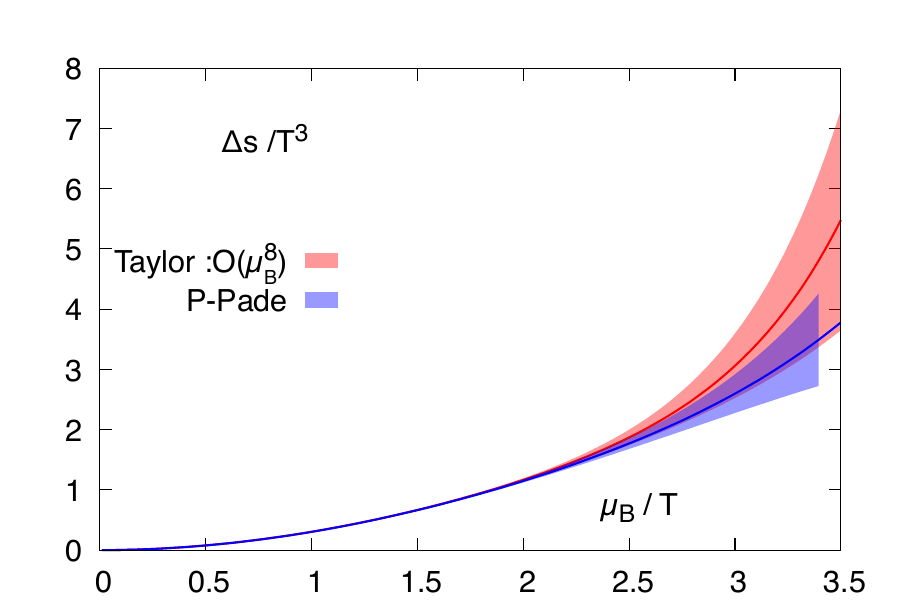}
\caption{{\it Top to bottom:} Comparison of Taylor series and
\pade approximants for pressure, net baryon-number, energy and entropy densities, at three values of the
temperature, $T=135$~MeV (left), 155 MeV (middle) and 175 MeV (right). Taylor 
expansion results are shown for $8^{\rm th}$ and $7^{\rm th}$-order (number density), respectively. The corresponding \pade approximants are [4,4] and [3,4].
Also shown are P-\pade approximants
(P-Pade) which are appropriate $T$- and $\mu_B$-derivatives of the [4,4] \pade approximant for the pressure as discussed in Sec.~\ref{sec:Pade}.
}
\label{fig:padeTaylor}
\end{figure*}

In Fig.~\ref{fig:padeTaylor} we compare at fixed
values of the temperature $8^{\rm th}$-order
Taylor expansion results with the [4,4] \pade approximant as well as the P-\pade results
for pressure, energy and entropy densities
that we have discussed above and in Sec.~\ref{sec:Pade}. We generally find that the P-\pade results for 
bulk thermodynamic observables are 
in better agreement with the Taylor
series results than the [4,4] or [3,4]
\pade approximants. This may not be
too surprising, as both approaches
are based on a thermodynamically self-consistent set of expressions.

We consider results for bulk thermodynamic observables 
based on the  $8^{\rm th}$-order Taylor series for the pressure 
reliable when the Taylor series results agree with the [4,4] \pade approximants as well as the
P-\pade approximants
shown in Fig.~\ref{fig:padeTaylor}. These
parameter ranges are similar for 
all four bulk thermodynamic observables shown in
Fig.~\ref{fig:padeTaylor} and also agree
with current estimates for a radius
of convergence of the Taylor series
given in \cite{Bollweg:2022rps,Mukherjee:2019eou,Dimopoulos:2021vrk}.
We thus conclude that
the region of reliability of current EoS results is
slightly temperature dependent, increasing from $\hmu_{B}\simeq 2.2$ at $T=135~{\rm MeV}$ to $\hmu_{B}\simeq 3.2$ at $T=180~{\rm MeV}$. The latter also holds 
for higher values of the temperature
where our results only are based on 
$6^{\rm th}$-order Taylor series expansions.

\subsection{Bulk thermodynamic observables}

Having discussed the $\mu_B$-dependent contribution
to bulk thermodynamic observables in the previous
subsection we can now combine these results with
those obtained at vanishing values of the chemical
potential, for which we use the continuum-extrapolated results 
obtained for $(2+1)$-flavor QCD by the HotQCD collaboration \cite{HotQCD:2014kol}.

\begin{figure*}
\begin{center}
\includegraphics[scale=0.48]{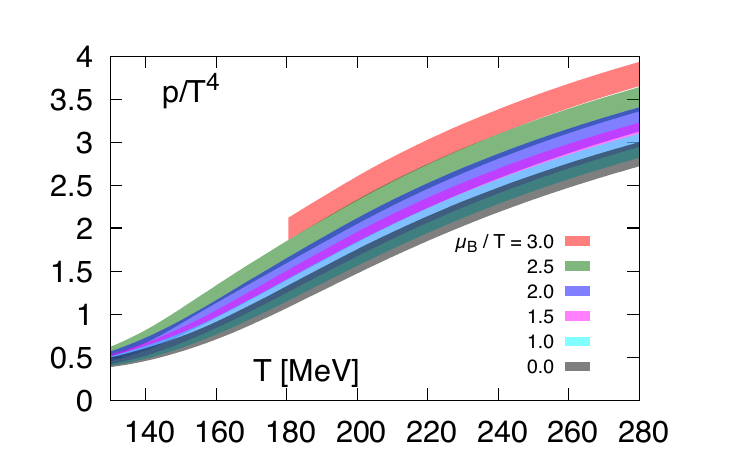}\hspace{-0.5cm}
\includegraphics[scale=0.48]{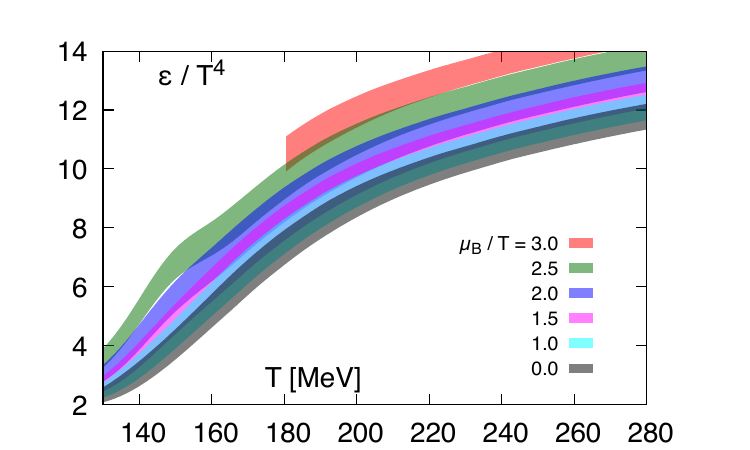}\hspace{-0.5cm}
\includegraphics[scale=0.48]{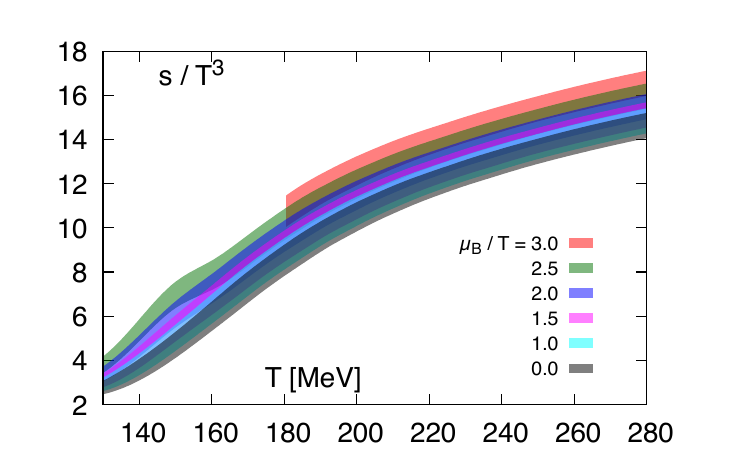}
\includegraphics[scale=0.48]{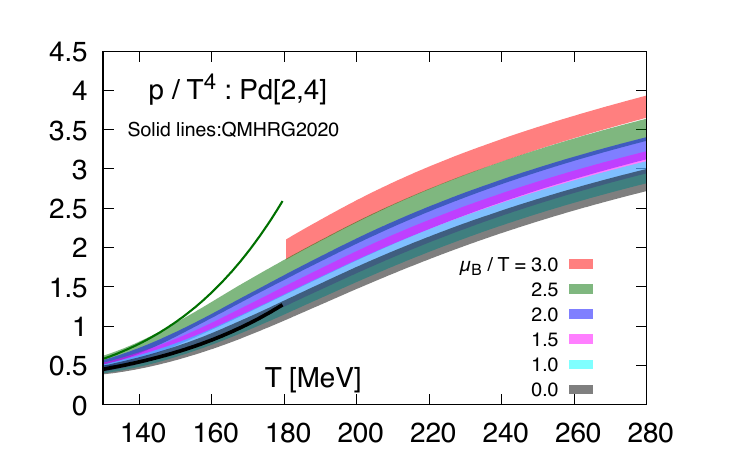}\hspace{-0.5cm}
\includegraphics[scale=0.48]{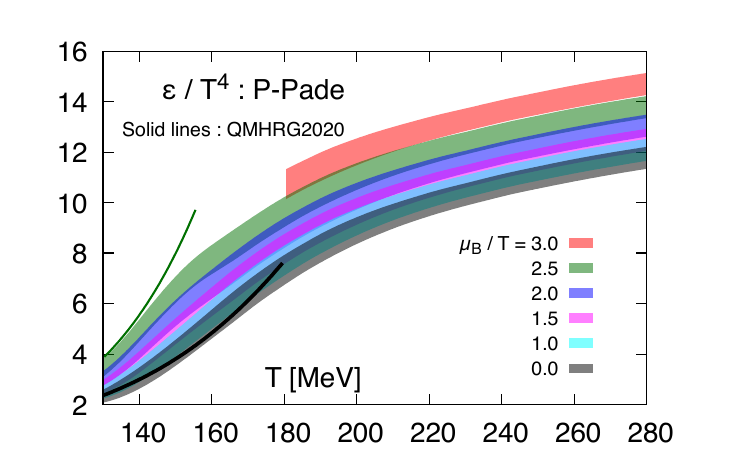}\hspace{-0.5cm}
\includegraphics[scale=0.48]{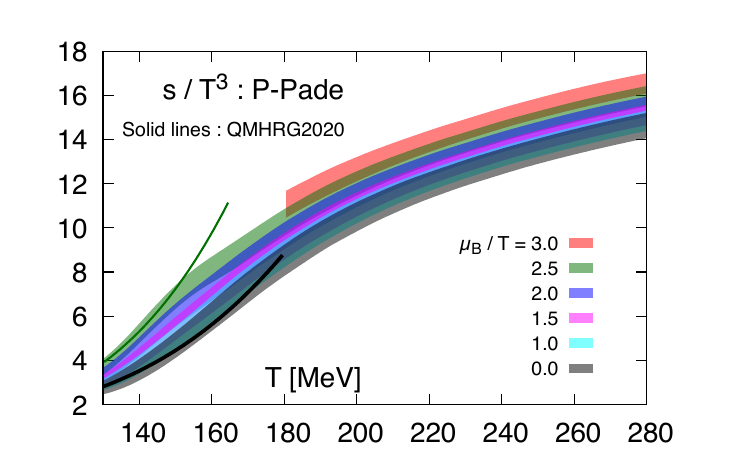}
\end{center}
\caption{Pressure (left) as well as energy (middle) and entropy (right) densities versus temperature
for several values of the baryon chemical potential. Figures 
show results for the case $n_S=0$,
$n_Q/n_B=0.5$ in the temperature interval [130 MeV:280 MeV].
Figures on the top show results 
from  $6^{\rm th}$ order Taylor series and the figures on the bottom have been used using \pade and P-\pade approximants, respectively.
The Taylor expansions are based on the continuum and spline interpolated data shown in Fig.~\ref{fig:taylor8}
for the temperature interval [130 MeV:175 MeV]. At larger temperatures we used data from our earlier 
analysis of bulk thermodynamics
in a sixth-order Taylor expansion \cite{Bazavov:2017dus}. Results for $\hmu_B=0$ have been taken from
\cite{HotQCD:2014kol}.
}
\label{fig:EoSr05}
\end{figure*}

In Fig.~\ref{fig:EoSr05} we show the total pressure (left) as well as energy (middle) and entropy (right)
densities for $\hmu_B \in [0:2.5]$ 
in the entire temperature range analyzed by us. For $\hmu_B=3$ we show results  
in the region $T\ge 180$~MeV only. 

Also shown in Fig.~\ref{fig:EoSr05} are
results from HRG model calculations
for $\hmu_B=0$ and $2.5$. 
It is apparent that the HRG model results show a stronger $\hmu_B$-dependence than
the QCD data. Already in the vicinity of $\Tpc$ differences between HRG model calculations 
and QCD increase with increasing 
values of $\hmu_B$. This is to be expected as higher-order Taylor expansion coefficients   
start to differ from HRG model results already at rather low temperatures, $T< 140$~MeV, and differences get large in the transition region (see Fig.~\ref{fig:pes-mu}).
These differences are larger for the energy and 
entropy densities than for pressure and number density. In particular,
we find that the $\order{\hmu_B^2}$ coefficients
in QCD are about 30\% smaller than in HRG model
calculations (see Fig.~\ref{fig:pes-mu}) and 
the ratio of $\order{\hmu_B^4}$ and $\order{\hmu_B^2}$ expansion
coefficients of the energy density calculated
in lattice QCD and HRG models, respectively,
differs by almost a factor 5, {\it i.e.} $\epsilon_4(\Tpc)/\epsilon_2(\Tpc) = 0.013(2)(6) $ in QCD compared to $(\epsilon_4(\Tpc)/\epsilon_2(\Tpc))_{\rm HRG} = 0.065(1)$ in the HRG model, using the QMHRG2020
hadron list. As a consequence, with increasing
$\hmu_B$ the difference between the energy density calculated in QCD
and HRG models increases on the pseudo-critical
line, $\Tpc(\hmu_B)$.

As discussed in \cite{Bazavov:2017dus}, the
curvature coefficients $\kappa_2^\epsilon$ and $\kappa_2^s$ for lines 
of constant energy  and entropy densities, respectively, are 
related through
the second-order expansion coefficients $\sigma_2$ and $s_2$,
\begin{eqnarray}
    \frac{\kappa_2^s}{\kappa_2^\epsilon}
&=& \frac{\sigma_2}{\epsilon_2} = 1-\frac{P_2}{\epsilon_2}
= 0.872(3)(5) \; .
\end{eqnarray}
Lines of constant energy density thus are flatter
than those for constant entropy density, {\it i.e.} the entropy decreases on a line of constant
energy density. Using the coefficients $\epsilon_2$ and $\sigma_2$, shown in Fig.~\ref{fig:pes-mu}, we get
\begin{eqnarray}
   \kappa_2^\epsilon &=& 0.0104(12) \;\; ,\;\;
   \kappa_2^s = 0.0091(11) \; ,
\end{eqnarray}
which should be compared to the curvature 
coefficient $\kappa_2^{B,f}$ of the 
pseudo-critical line given in Eq.~\ref{TpcmuB}.  These results thus suggest that, within current statistical errors,
energy and entropy densities stay constant on
the pseudo-critical line or drop slightly.

Using the results for $\epsilon/T^4$
shown in Fig.~\ref{fig:EoSr05}~(middle) we find for the energy density
on the pseudo-critical line,
\begin{equation}
    \epsilon(T_{\rm pc}(\hmu_B)) =
    \begin{cases}
    370(40)(30)~{\rm MeV/fm}^3\; ,\; \hmu_B=0 \\
    330(28)(53)~{\rm MeV/fm}^3 \; ,\; \hmu_B=2.5
    \end{cases}
    \label{e-QCD}
\end{equation}
Here the first error is the statistical error 
on $\epsilon$ at $\Tpc(\hmu_B)$, while the second error reflects the systematic uncertainty 
arising from the uncertainty on
$\Tpc(\hmu_B)$, which rises from
1.5~MeV at $\hmu_B=0$ to 4~MeV at $\hmu_B=2.5$. 

\begin{figure}[b]
\includegraphics[width=\linewidth]{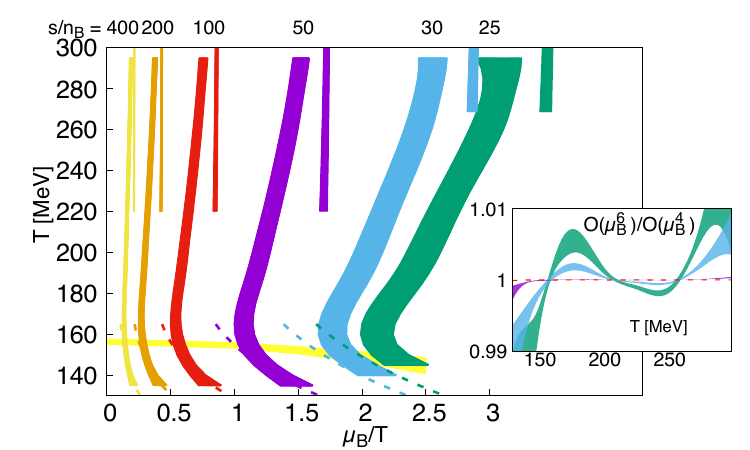}
\caption{ 
Lines of constant entropy per baryon number in the $T$-$\hmu_B$
plane. Solid bands indicate results obtained by numerically solving
$s/n_B$ derived from the $\order{\hmu_B^6}$ pressure series for $\hmu_B$. 
Dashed lines indicate HRG results while the (almost straight vertical bands 
indicate the $\order{g^2}$ perturbative result with $4\leq k_T\leq8$.
The yellow band indicates $\Tpc(\hmu_B)$.
In the inset, we show 
the result from this numerical inversion divided by the corresponding
inversion coming from the $\order{\hmu_B^4}$ pressure series.
}
\label{fig:fixedsn}
\end{figure}

\section{Thermodynamics at constant ratio of entropy to net baryon number}\label{sec:constantS}

\begin{figure*}[t]
\includegraphics[scale=0.49,page=1]{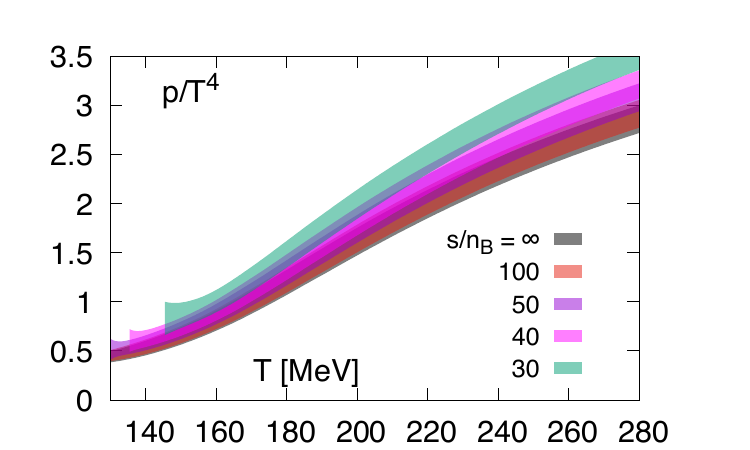}\hspace*{-0.7cm}
\includegraphics[scale=0.49,page=1]{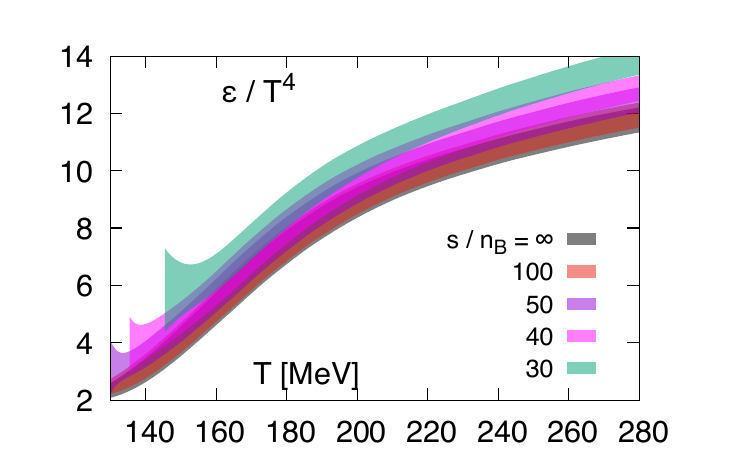}\hspace*{-0.7cm}
\includegraphics[scale=0.49,page=1]{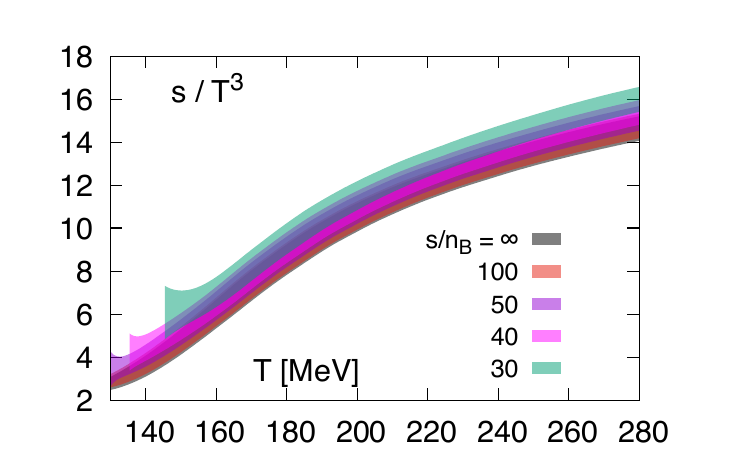}
\caption{ Total pressure (left), energy density (middle) and entropy density (right) versus temperature for several values of the ratio $s/n_B$.}
\label{fig:total}
\end{figure*}

Equilibrated strong interaction matter created, for instance
in heavy ion experiments, follows lines
of constant entropy per net baryon number
while expanding and cooling after the 
initial collision of nuclei. The
composition of the initially colliding nuclei thus fixes the conserved
charge content of the matter created
in this collisions, {\it i.e.}
$n_S=0$ and $n_Q/n_B$ fixed. 

Also for the calculation of thermodynamic
observables such as the speed of sound or
the adiabatic compressibility of matter, 
one needs to determine the lines of constant
$s/n_B$ in QCD \cite{Ejiri:2005uv,Guenther:2017hnx}. 
We have determined the  lines
of constant $s/n_B$ in a strangeness-neutral medium with fixed $n_Q/n_B$.

In Fig.~\ref{fig:fixedsn}
we show such lines in the $\hmu_B$-$T$
plane for $s/n_B$ and fixed ($n_S,n_Q/n_B)=(0,0.5)$ in the range 
$25\le s/n_B \le 400$. This roughly
corresponds to the range covered by
BES-II at RHIC in the range
of beam energies $7.7~{\rm GeV}\le \sqrt{s_{_{NN}}} \le 200~{\rm GeV}$.
Also shown in Fig.~\ref{fig:fixedsn}
is the asymptotic behavior of lines 
of constant $n_B$ at high temperature 
and the approach to the HRG model at 
low temperature. 
As can be seen, QCD results for lines of constant $s/n_B$ differ much more from 
high-$T$ perturbation theory then the lines
of constant $n_B$. This, of course, arises
from the $\mu_B=0$ contribution to the total entropy density, which approaches the ideal 
gas limit slowly. The inset in Fig.~\ref{fig:fixedsn} shows 
that these lines are already well determined using $4^{\rm th}$ order Taylor series only.

In Fig.~\ref{fig:total} we show the 
temperature dependence of pressure,
energy and entropy densities on lines 
of constant $s/n_B$. Obviously, the
$T$-dependence of these observables is
similar to that obtained for fixed
$\hmu_B$. Above
the pseudo-critical temperature, however, these observables rise somewhat faster
than at $\mu_B=0$, {\it i.e.} at $s/n_B=\infty$. This is expected as
Fig.~\ref{fig:fixedsn} shows that
the value of the chemical
potential on lines of constant
$s/n_B$ increases above $\Tpc$ with increasing temperature.

The speed of sound in strangeness-neutral matter with fixed $n_Q/n_B$ reflects the
temperature dependence of $p$ and $\epsilon$, shown in Fig.~\ref{fig:total}. As defined in Eq.~\ref{dpdesn} we have,
\begin{eqnarray}
c_s^2 &=& \left(\frac{\partial p}{\partial \epsilon} \right)_{s/n_B,n_Q/n_B,n_S}
\nonumber \\
&=& \frac{(\partial p/\partial T)_{s/n_B, n_Q/n_B,n_S}}{(\partial \epsilon /\partial T)_{s/n_B, n_Q/n_B,n_S}}
\; .
\label{speed}
\end{eqnarray}
The numerator is related to the entropy density evaluated at fixed $\vec{X}=(s/n_B,n_Q/n_B,n_S)$, while 
the denominator is a specific heat 
defined at fixed $\vec{X}$. We note, however, that the temperature derivatives taken at
fixed $\vec{X}$ differ, of course, from those taken
at fixed $\vec{\hmu}$. Some relations 
for constrained partial derivatives are 
given in Appendix~\ref{app:partial}, and in
 Appendix~\ref{app:sound} we give an 
 explicit expression for $c_s^2$.

\begin{figure*}
\includegraphics[width=0.48\linewidth]{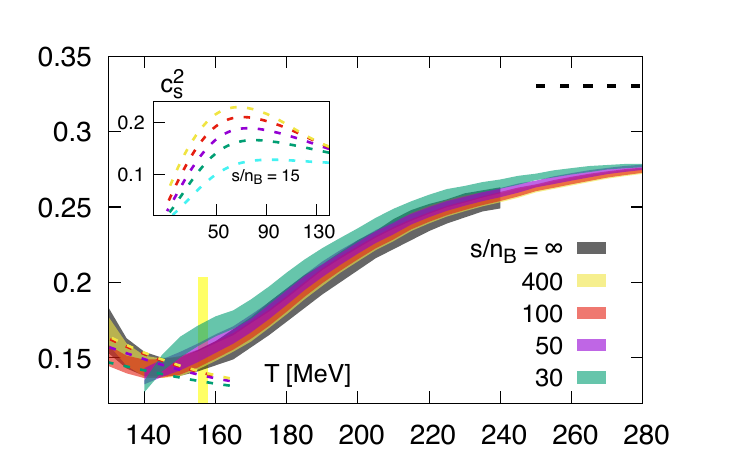}
\includegraphics[width=0.48\linewidth]{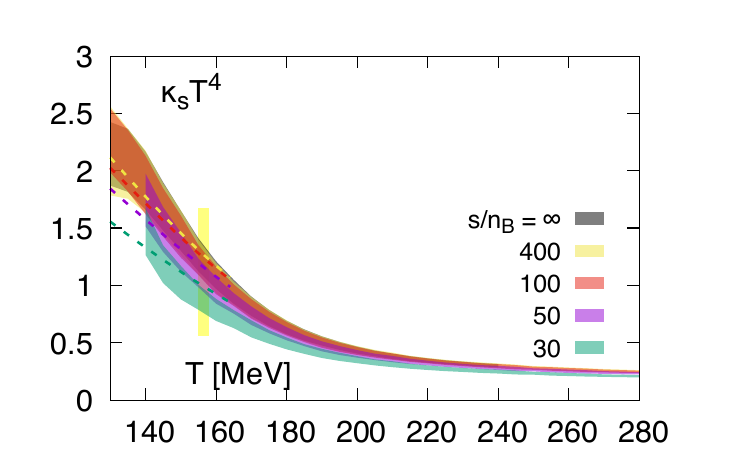}
\caption{Speed of sound (left) and adiabatic compressibility (right)
in strangeness-neutral, isospin-symmetric matter
versus temperature. Shown are results
for several values of $s/n_B$. The limit
$s/n_B=\infty$ corresponds to the case of vanishing chemical potentials.
Dashed lines at low temperatures indicate QMHRG2020 model calculations, 
at high temperatures they show the 
non-interacting quark-gluon gas  results.
In the inset HRG model calculations at lower
temperatures are shown. The yellow band indicates $\Tpc$.}
\label{fig:c2s}
\end{figure*}

For $n_S=0$ we obtain  from Eqs.~\ref{dpdesn} and \ref{kappaS} for the 
speed of sound ($c_s^2$) and the closely related
adiabatic compressibility ($\kappa_s$) 
\begin{eqnarray}
c^2_s&=&
\frac{N_{\vec{X}}}{(\epsilon+p)\ D_{\vec{X}}}\; ,\;
\nonumber \\
\kappa_s &=&\frac{1}{c_s^2(\epsilon+p)}=
\frac{D_{\vec{X}}}{N_{\vec{X}}}
\; .
\end{eqnarray}
The functions $N_{\vec{X}}(T,\vec{\mu})$ and $D_{\vec{X}}(T,\vec{\mu})$ are given in terms of second-order cumulants of the
conserved charges $(B,\ Q,\ S)$, the entropy 
density and its derivatives with respect
to $T$ as well as the three chemical potentials. All these observables themselves
are functions of $T$ and $\vec{\mu}$.
One thus may evaluate $c_s^2$ by either taking numerically  the $T$-derivatives 
appearing in Eq.~\ref{speed} or use directly
Eqs.~\ref{cX2}-\ref{bcoef}. The results 
shown in Fig.~\ref{fig:c2s} are based
on the former approach.

As can be seen in
Fig.~\ref{fig:total}~(left) and (middle) the shape of $p(T)$ and 
$\epsilon(T)$ varies little when changing
$s/n_B$ and follows a similar pattern in both quantities. In the temperature range currently
accessible to us we thus 
do not expect to observe a strong 
dependence of the speed of sound and the adiabatic compressibility on
$s/n_B$. This is indeed apparent from
the results for both observables, calculated for several
fixed ratios $s/n_B$, as shown in Fig.~\ref{fig:c2s}. At low temperatures
we show results for $c_s^2$
only down to $T$-values that can be reached at fixed $s/n_B$ for
$\hmu_B<2.5$. As can be seen in 
Fig.~\ref{fig:fixedsn} this 
allows to explore the entire available
temperature range ($T>135$~MeV) for $s/n_B\ge 30$, but 
limits the temperature range for smaller values of $s/n_B$.

At zero baryon density ($s/n_B=\infty$) the speed of sound has
a minimum at $T=145-150$ MeV \cite{HotQCD:2014kol}.
From Fig. \ref{fig:c2s}~(left) we see that
for temperatures $T>145$ MeV the
speed of sound slightly increases with decreasing $s/n_B$ and 
steadily increases with $T$. At temperatures $T<145$ MeV
the opposite trend is expected to show up in HRG model
calculations: $c_s^2$ decreases with decreasing $s/n_B$ 
and increasing $T$, see Fig. \ref{fig:c2s}~(left).
At low temperatures, the speed of sound 
is known to have a ``bump" arising
from the interplay of baryon and light 
meson contributions to $c_s^2$. This is 
apparent from the inset
in Fig.~\ref{fig:c2s}~(left), where we show
results on the speed of
sound calculated in a HRG model using
the QMHRG2020 list of hadrons for several representative values of $s/n_B$. 
The bump
in the speed of sound at low temperature
is responsible for the minimum in $c_s^2$ 
at $T=145-150$ MeV observed in lattice QCD calculations.
This bump, however, will disappear with decreasing
$s/n_B$. The HRG model calculation shown in the
inset of \ref{fig:c2s}~(left) indicate that $c_s^2$ 
will stop developing a bump at low
temperatures at $s/n_B\simeq 15$.
Therefore, we expect that the minimum in $c_s^2$
will become shallower and its position will possibly shift to
smaller temperatures with decreasing $s/n_B$.
Lattice QCD calculations for $s/n_B=100$ 
shown in Fig.~\ref{fig:c2s}~(left) seem
to confirm this expectation.
For sufficiently small $s/n_B$ the minimum in the speed of sound may
disappear altogether.

As $c_s^2$ as well as $\epsilon$ and
$p$ increase with decreasing $s/n_B$
it is clear that the adiabatic compressibility
shown in Fig.~\ref{fig:c2s}~(right) will 
decrease with decreasing $s/n_B$. Moreover we 
find that it monotonically 
decreases with increasing temperature.

\section{Conclusions}\label{sec:conclusion}

We presented an update on the 
Taylor expansion of basic thermodynamic
observable (pressure, energy and entropy densities) of strangeness-neutral, isospin-symmetric strong-interaction matter.
We showed that the $\hmu_B$-dependent parts 
of these observables approach the high-temperature perturbative regime much more 
rapidly than the contribution to these observables at $\hmu_B=0$.
The $\hmu_B$-dependent contributions are 
well described by $\order{g^2}$ perturbation
theory already for $T\gtrsim 250$~MeV.

We furthermore constructed the QCD equation of
state ($P(n_B)$, $\epsilon(n_B)$). We showed 
that the Taylor series can be resummed using
\pade approximants. The regime where Taylor 
expansions and \pade approximants agree with each other gives further confidence in the validity
range of current Taylor expansion results
varying from $\hmu_B\simeq 2.5$ at low temperatures to $\hmu_B\gtrsim 3$ at $T> 200$~MeV.
 
We showed that $\order{\hmu_B^4}$ Taylor 
series results for the pressure provide a good
description of the EoS in  almost the entire parameter
range currently accessible in the beam energy
scan performed at RHIC in collider mode. This
allows for an analytic ansatz for the 
$\hmu_B$-dependent part of the EoS of strangeness-neutral, isospin-symmetric matter,
valid at
temperatures $T\gtrsim 135$~MeV and $\mu_B/T\lesssim 2.5$, which requires as input
only the two leading-order Taylor expansion
coefficients ($P_2(T),\ P_4(T))$ of the pressure series.
We note, however, that this conclusion is based on lattice calculation with $N_{\tau}=8$ lattices only.

For the first time, we presented results for the
speed of sound in strangeness-neutral matter
at non-zero net baryon-number density. 
We 
showed that the minimum in $c_s^2$
becomes shallower and its position shifts toward smaller temperatures
for $s/n_B \le 100$. 
It is possible that this minimum will disappear for sufficiently
small $s/n_B$. All data presented in the figures of this paper can be found in Ref.\cite{epubdata:2980715}.

\vspace{0.5cm}
\emph{Acknowledgments.---} 
This work was supported by: (i) The U.S. Department of Energy, Office of
Science, Office of Nuclear Physics through the Contract No. DE-SC0012704;
(ii) The U.S. Department of Energy, Office of Science, Office of Nuclear
Physics within the framework of Scientific Discovery through Advance Computing (SciDAC) award
\textit{Fundamental Nuclear Physics at the Exascale and Beyond};
(iii) The Deutsche Forschungsgemeinschaft (DFG, German Research Foundation) - Project numbers 315477589-TRR 211 and 460248186 (PUNCH4NFDI);
(iv) The grant 05P2018 (ErUM-FSP T01) of the German Bundesministerium f\"ur Bildung und Forschung;
(v) The grant 283286 of the European Union.
D.B. was supported by the Intel Corporation.

This research used awards of computer time provided by:
(i) The INCITE program at Oak Ridge Leadership Computing Facility, a DOE
Office of Science User Facility operated under Contract No. DE-AC05-00OR22725;
(ii) The ALCC program at National Energy Research Scientific Computing Center,
a U.S. Department of Energy Office of Science User Facility operated under
Contract No. DE-AC02-05CH11231;
(iii) The INCITE program at Argonne Leadership Computing Facility, a U.S.
Department of Energy Office of Science User Facility operated under Contract
No. DE-AC02-06CH11357;
(iv) The USQCD resources at the Thomas Jefferson National Accelerator Facility.

This research also used computing resources made available through:
(i) a  PRACE grant at CINECA, Italy;
(ii) the Gauss Center at NIC-J\"ulich, Germany;
(iii) the GPU-cluster at Bielefeld University, Germany.

We thank the Bielefeld HPC.NRW team for their support and Dietrich 
Boedeker for very helpful discussions.
\vspace{0.2cm}

\appendix

\section{Constrained partial derivatives}
\label{app:partial}
We summarize here relations for partial
derivatives of thermodynamic observables
with respect to temperature, keeping specific
external conditions $(x,y,z)$ fixed, 

For any thermodynamic function
$f(T,\mu_B,\mu_Q,\mu_S)$ we have
\begin{eqnarray}
    \left(\frac{\partial f}{\partial T} \right)_{(x,y,z)}
    &=& \left(\frac{\partial f}{\partial T} \right)_{(\mu_B,\mu_Q,\mu_S)} 
    \label{dfdT} \\
    &&+
    \left(\frac{\partial f}{\partial \mu_B} \right)_{(T,\mu_Q,\mu_S)}
    \left(\frac{\partial \mu_B}{\partial T} \right)_{(x,y,z)}
    \nonumber \\
    &&+
        \left(\frac{\partial f}{\partial \mu_Q} \right)_{(T,\mu_B,\mu_S)}
    \left(\frac{\partial \mu_Q}{\partial T} \right)_{(x,y,z)}
    \nonumber \\
    &&+
    \left(\frac{\partial f}{\partial \mu_S} \right)_{(T,\mu_B,\mu_Q)}
    \left(\frac{\partial \mu_S}{\partial T} \right)_{(x,y,z)} \; .
\nonumber
\end{eqnarray}
Similarly one has for two 
thermodynamic functions 
$f(T,\mu_B,\mu_Q,\mu_S)$ and $g(T,\mu_B,\mu_Q,\mu_S)$ the relation
\begin{equation}
    \left( \frac{\partial f}{ \partial g} \right)_{(x,y,z)}
    = \frac{\left( \partial f / \partial T \right)_{(x,y,z)}}{\left( \partial g / \partial T \right)_{(x,y,z)}}
    \label{dfdTdgdT}
\end{equation}
In Eqs.~\ref{dfdT} and \ref{dfdTdgdT} the derivatives of
the chemical potentials are taken
on lines of constant $x(T,\mu_B,\mu_Q,\mu_S)$,
$y(T,\mu_B,\mu_Q,\mu_S)$ and $z(T,\mu_B,\mu_Q,\mu_S)$ in the 
space of external parameters $(T,\mu_B,\mu_Q,\mu_S)$.
In the lattice QCD context
we usually work in the parameter space ($T,\hmu\equiv \mu/T$). Moreover, we conveniently
work with reduced, i.e. dimensionless,  thermodynamic observables, i.e. we want to replace e.g.
$\he=\epsilon/T^4$, etc. 

Changing the partial derivatives $\partial_{\mu_B}$
to $\partial_{\mu_B/T}$ and introducing reduced 
observables is straightforward, as these derivatives 
are taken at fixed $T$. We have for an observable 
that has dimension of $T^n$ the relation,
\begin{equation}
\left.    \frac{\partial f}{\partial \mu_B}\right|_T =
\left.    T^{n-1} \frac{\partial \hat{f}}{\partial \hmu_B}\right|_T
\; .
\end{equation}
Rewriting the temperature derivatives one has to be  
a bit more careful,
\begin{eqnarray}
\left.    \frac{\partial f}{\partial T}\right|_{\vec{\mu}} &=&
\left.  \frac{\partial f}{\partial T}\right|_{\vec{\hmu}} - \sum_{i=B,Q,S}\frac{\mu_i}{T^2}
\left.  \frac{\partial f}{\partial \mu_i/T}\right|_T
\nonumber \\
&=& \left.  \frac{\partial T^n \hat{f}}{\partial T}\right|_{\vec{\hmu}} - \sum_i \frac{\mu_i}{T^2} T^{n}
\left.  \frac{\partial \hat{f}}{\partial \hmu_i}\right|_T
 \\
&=& T^{n-1}\left( n \hat{f} + T \frac{\partial \hat{f}}{\partial T}\Bigg|_{\vec{\hmu}}
- \sum_i \hmu_i\frac{\partial \hat{f}}{\partial \hmu_i}\Bigg|_T \right)
\; .
\nonumber
\end{eqnarray}

\section{Ideal gas and high-temperature perturbation theory
for isospin-symmetric, strangeness-neutral matter}
\label{app:IG}

At large values of the temperature, the pressure approaches that of a massless
ideal gas of quarks and gluons. 
In the case of massless quarks, the leading order, ideal gas result
as well as the $\order{g^2}$ 
correction are second-order 
polynomials in $\hmu_f^2$ \cite{Shuryak:1977ut,Chin:1978gj},
\begin{equation}
    p/T^4= \hp_{id} - g^2 \hp_2\; ,
\end{equation}
with the ideal gas term
\begin{eqnarray}
\hp_{id}  &=&  \frac{8 \pi^2}{45}\left(1 +
\frac{21 n_f }{32}\right)
\nonumber \\
 &&\hspace{0.5cm}+\sum_{f=u,d,s}
 \left[\frac{1}{2}  \left(\frac{\mu_f}{T}\right)^2 
+ \frac{1}{4 \pi^2} \left(\frac{\mu_f}{T}\right)^4 
\right] \; ,
\label{free}
\end{eqnarray}
and the $\order{g^2}$ correction
\begin{eqnarray}
\hp_{2}  &=&  \frac{1}{6}\left(1+\frac{5n_f}{12}\right)  
\nonumber \\
&&\hspace{0.1cm}+ \frac{1}{2\pi^2} \sum_{f=u,d,s} \left[
\frac{1}{2}  \left(\frac{\mu_f}{T}\right)^2 
+ \frac{1}{4\pi^2} \left(\frac{\mu_f}{T}\right)^4 
\right] \; ,
\label{PTg2}
\end{eqnarray}
Expressing the flavor chemical potentials
in terms of conserved charge chemical potentials,
\begin{eqnarray}
    \hmu_u &=& (\hmu_B+2 \hmu_Q)/3\; ,
    \nonumber \\
    \hmu_d &=& (\hmu_B - \hmu_Q)/3 \; ,
    \nonumber \\
    \hmu_s &=& (\hmu_B - \hmu_Q)/3 -\hmu_S \; ,
    \label{mu-uds}
\end{eqnarray}
the strangeness number density can be written as
\begin{eqnarray}
   \hn_S&=&  -\frac{1-q_1 -3 s_1}{3}
   \left(1+\frac{1}{2\pi^2}g^2\right) 
   \nonumber \\
  && \times\left( \hmu_B +
        -\frac{(1-q_1 -3 s_1)^2}{18 \pi^2}  \hmu_B^3\right)
        \;
        \label{nSid}
\end{eqnarray}
Here we introduced the ratio of chemical potentials, $s_1=\hmu_S/\hmu_B$ and $q_1=\hmu_Q/\hmu_B$. From Eq.~\ref{nSid} we 
find the constraint for a
strangeness-neutral ideal gas,
\begin{equation}
    s_1 = \frac{1-q_1}{3} \; .
\end{equation}
Inserting this constraint in  Eq.~\ref{mu-uds} one finds that up 
to $\order{g^2}$ in perturbation theory
the strange quark chemical potential
vanishes in strangeness-neutral matter, $\mu_s=0$. Moreover, the electric charge chemical potential vanishes ($\mu_Q=0$)
in the isospin-symmetric case $n_u=n_d$.
Up to $\order{g^2}$ we thus may
write for the $\hmu_B$-dependent
contribution to the pressure of an 
isospin-symmetric medium and the two
\begin{figure*}
\includegraphics[width=0.43\linewidth]{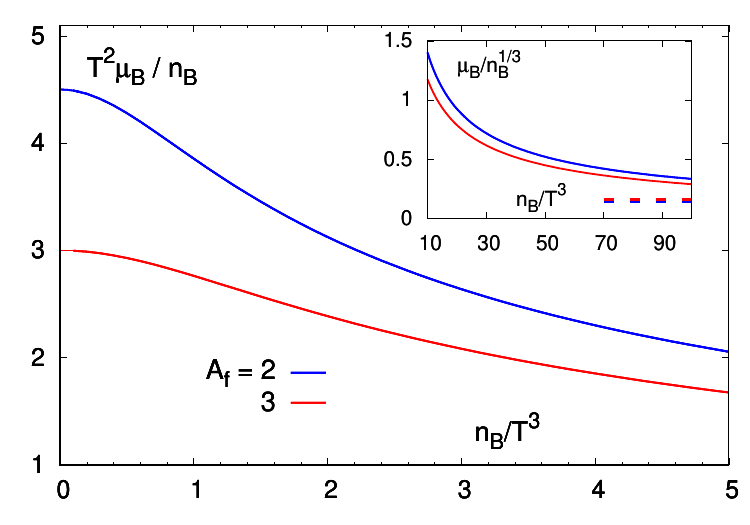}
\includegraphics[width=0.43\linewidth]{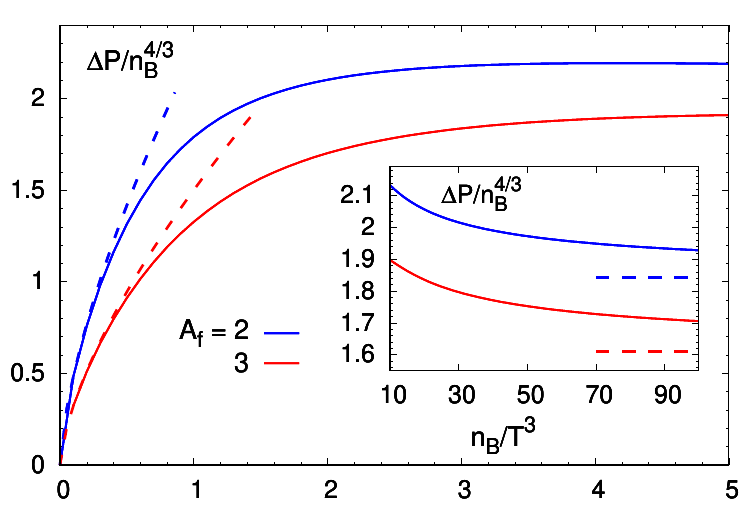}
\caption{Baryon chemical potential (left) and the $\mu_B$-dependent part of the pressure (right)
of a 3-flavor, ideal gas of massless quarks versus net 
baryon-number density ($n_B$). Shown are results
for the cases $(\mu_Q=0,n_S=0)$ ($A_f=2)$ 
and $(\mu_Q=0,\mu_S=0)$ ($A_f=3)$.
Results are rescaled with powers of $n_B$ appropriate in the 
low density (main panel) and high density (insets) limits,
respectively. Dashed lines show the respective low and high density limits.
}
\label{fig:EoSid}
\end{figure*}
cases\footnote{
Because of isospin symmetry and
strangeness-neutrality conditions that we impose, one has $\mu_s=0$ at high
temperature. For this reason, strange quarks
do not contribute to the density dependent part, in particular the net baryon-number 
density of the ideal gas and LO perturbative limits. This is reflected in the parameter 
$A_f$.} 
of interest to us (i) $\mu_S=0$
or (ii) $n_S=0$,
\begin{eqnarray}
    \Delta p/T^4 &=&
    \frac{A_f}{18}\left(1-\frac{g^2}{2\pi^2}\right) \left( \hmu_B^2 
    +\frac{1}{18\pi^2}\hmu_B^4\right) \; ,
    \label{pid} \\
    n_B/T^3 &=& \frac{A_f}{9}\left(1-\frac{g^2}{2\pi^2}\right) \left( \hmu_B 
    +\frac{1}{9\pi^2}\hmu_B^3\right) \; .
    \label{nBid}
\end{eqnarray}
where $A_f=3$ for $\mu_S=0$ and $A_f=2$
for $n_S=0$. 

For the coupling $g^2$ we use the 
two-loop running coupling, allowing for a
free parameter in the renormalization 
scale, $k_T \pi T$, as it also is
used in more refined, resummed approaches,
e.g. hard thermal loop (HTL) calculations
\cite{Mogliacci:2013mca},
\begin{eqnarray}
  \beta_0 &=& \frac{11 -2 n_f/3}{4\pi} \nonumber \\
  \beta_1 &=& \frac{102 - 38 n_f/3}{16\pi^2}  \nonumber \\
    g^2_{1loop}&=&
    \frac{2\pi}{\beta_0\ln\left(k_T\pi T/\Lambda_{\overline{MS}}\right)}
\; , 
\label{LORG}
\end{eqnarray}
\begin{eqnarray}
    g^2 &=& g^2_{1loop}\left[1 - \frac{\beta_1}{\beta_0^2}\frac{\ln(2\ln (k_T\pi T/\Lambda_{\overline{MS}}))}{2\ln (k_T\pi T/\Lambda_{\overline{MS}})}\right]
\; , 
\label{LORG-2loop}
\end{eqnarray}
with $n_f=3$.  
As a scale factor we use the average of recent values summarized in the last FLAG report
$\Lambda_{\overline{MS}}=339(12)$~MeV \cite{Aoki:2021kgd}.

From Eqs.~\ref{pid} and \ref{nBid} we 
easily find the low and high density limits
of the EoS of a massless, strangeness-neutral, isospin-symmetric quark gas to
$\order{g^2}$ in high-$T$ perturbation
theory,
\begin{equation}
    \frac{\Delta p}{T^4} =
    \begin{cases}
    \frac{9}{2 A_f(1-g^2/2\pi^2)} \left(\frac{n_B}{T^3} \right)^2 &\; , \; \frac{n_B}{T^3} \ll 1\; , \\
       \frac{1}{4}\left(\frac{81 \pi^2}{A_f(1-g^2/2\pi^2)}\right)^{1/3} \left(\frac{n_B}{T^3} \right)^{4/3} &\; , \; \frac{n_B}{T^3} \gg 1\; .
    \end{cases}
\end{equation}

For arbitrary net baryon-number densities 
we invert Eq.~\ref{nBid} to obtain the baryon
chemical potential as function of $n_B/T^3$
in $\order{g^2}$ perturbation theory,
\begin{eqnarray}\label{hmuBpert}
    \hmu_B (\hn_B) &=& y^{1/3} - \frac{3 \pi^2}{y^{1/3}} 
 \; ,
\end{eqnarray}
with 
\begin{eqnarray}
y &=& \frac{3^4 \pi^2}{2 A_f(1-g^2/2\pi^2)}
    \\
    &&\hspace{0.4cm}\times\left( \hn_B+\sqrt{\hn_B^2+(2\pi A_f (1-g^2/2\pi^2))^2/3^5} \right)\;\; .\;\; 
\label{yid} \nonumber
\end{eqnarray}

Inserting Eq.~\ref{hmuBpert} in Eq.~\ref{pid} we obtain
the equation of state of strangeness-neutral,
isospin-symmetric matter in the high-temperature limit. The result in the 
ideal gas limit ($g^2=0$) is shown in Fig.~\ref{fig:EoSid}. Comparing with 
Fig.~\ref{fig:pes-mu} one sees that 
the density range covered at high
temperature in current Taylor series
expansions with $\hmu_B\le 2.5$, {\it i.e.}
$n_B/T^3 < 0.6$ corresponds to the 
low density limit of an ideal gas,
$\mu_B/T \sim n_B/T^3$. 
The transition between the low and
high density regions, however, starts to
set in for $n_B/T^3 \gtrsim 0.6$.

\section{Speed of sound}
\label{app:sound}

The speed of sound is defined as
\begin{equation}
    c_{\vec{X}}^2 = \left( \frac{\partial p}{\partial \epsilon} \right)_{\vec{X}} \; ,
\end{equation}
where $\vec{X}$ defines the external conditions, e.g. constant entropy per particle number. We 
generalize this here to the case of QCD, where three conserved currents need to be specified.
We do so by choosing $\vec{X}=(s/n_B,n_Q/n_B,n_S)$
With this we obtain 

\begin{eqnarray}
c^2_{\vec{X}} &=&
\frac{N_{\vec{X}}}{(\epsilon+p-\mu_S n_S)\ D_{\vec{X}}}\; ,\;
    \label{cX2}
    \end{eqnarray}
with
\begin{eqnarray}
N_{\vec{X}}&=&
s^2 a_{s2}+ s' a_{s}+ 
 s\ (s_B  a_{B}+  s_Q  a_{Q}+
 s_S a_{S})
\nonumber \\
&& + s_B^2 a_{B2} 
+  s_Q^2 a_{Q2}+s_S^2 a_{S2}
+ s_B s_Q  a_{BQ}
\nonumber \\
&&
 +s_B s_S  a_{BS}+s_Q s_S  a_{QS}
\label{cs2BQS} \\
D_{\vec{X}}&=&
 s' b_{s}+s_B^2 b_{B2}+ s_Q^2 b_{Q2} +s_S^2 b_{S2}
\nonumber \\
&& +s_B s_Q  b_{BQ}+s_B s_S
 b_{BS}+s_Q s_S  b_{QS}
\nonumber
\end{eqnarray}
Here we use the abbreviation $s_i$
for the derivatives of the entropy
with respect to the chemical potential $\mu_i$, {\it i.e.}
$s_i=\partial s/\partial \mu_i$.

Using $n_Q/n_B=r$, the various coefficients
appearing in Eq.~\ref{cs2BQS} are
given by

\begin{eqnarray}
a_{s2}&=& -(\chi_{11}^{BS})^2 \chi_2^Q + 2 \chi_{11}^{BQ} \chi_{11}^{BS} \chi_{11}^{QS} - \chi_2^B (\chi_{11}^{QS})^2 
\nonumber \\
&&
- (\chi_{11}^{BQ})^2 \chi_2^S + \chi_2^B \chi_2^Q \chi_2^S
\nonumber \\
a_{s}&=& -(r^2 b_{Q2} + r b_{BQ} +b_{B2}) n_B^2
\nonumber \\
&&
- \frac{1}{2}(r b_{QS} + b_{BS}) n_B n_S
\nonumber \\
a_{B}&=&
 (2 b_{B2} + r b_{BQ}
 ) n_B 
+ \frac{1}{2} b_{BS} n_S
\nonumber \\
a_{Q}&=&  ( b_{BQ} +2 r b_{Q2} )n_B  
+ \frac{1}{2} b_{QS} n_S   
\nonumber \\
a_{S}&=& 
 ( b_{BS}+ r b_{QS}) n_B + b_{S2} n_S
\nonumber 
\end{eqnarray}
\begin{eqnarray}
a_{B2}&=&  - r^2 \chi_2^S  n_B^2+r \chi_{11}^{QS} n_B n_S
\nonumber \\
a_{Q2}&=& - \chi_2^S n_B^2+\chi_{11}^{BS} n_B n_S
 \label{acoef}\\
a_{S2}&=&(2r \chi_{11}^{BQ}  - r^2\chi_2^B - \chi_2^Q) n_B^2
\nonumber \\
a_{BQ}&=&
 2r \chi_2^S n_B^2
 - (r\chi_{11}^{BS}  + \chi_{11}^{QS}) n_B n_S 
\nonumber \\
a_{BS}&=&
-2r (\chi_{11}^{QS} - r \chi_{11}^{BS}) n_B^2
+ (\chi_2^Q - r \chi_{11}^{BQ}) n_B n_S
\nonumber \\
a_{QS}&=&
 2 (\chi_{11}^{QS} - r \chi_{11}^{BS}) n_B^2
+ (r \chi_2^B - \chi_{11}^{BQ}) n_B n_S
\nonumber
\end{eqnarray}

\begin{eqnarray}
b_{s}&=& a_{s2}
\nonumber \\
b_{B2}&=& (\chi_{11}^{QS})^2 - \chi_2^Q \chi_2^S
\nonumber \\
b_{Q2}&=& (\chi_{11}^{BS})^2 - \chi_2^B \chi_2^S
\nonumber \\
b_{S2}&=& (\chi_{11}^{BQ})^2 - \chi_2^B \chi_2^Q
\nonumber \\
b_{BQ}&=& 2(\chi_{11}^{BQ}  \chi_2^S- \chi_{11}^{BS} \chi_{11}^{QS} )
\nonumber \\
b_{BS}&=& 2 (\chi_{11}^{BS} \chi_2^Q -  \chi_{11}^{BQ} \chi_{11}^{QS})
\nonumber \\
b_{QS}&=& 2(  \chi_{11}^{QS} \chi_2^B - \chi_{11}^{BQ} \chi_{11}^{BS} )
\label{bcoef}
\end{eqnarray}
We note that all cumulants $\chi_{11}^{ab}$ and $\chi_2^c$, with $a,\ b= B,\ Q,\ S$,
appearing in Eqs.~\ref{acoef}
and \ref{bcoef} are defined as $\mu_i$, $i=a,b$, derivatives of $P$ and as such are functions of the chemical
potentials, {\it i.e} $\chi_{11}^{ab}\equiv \chi_{11}^{ab}(\mu_B,\mu_Q,\mu_S)$. They need
to be chosen such that the constraint $\vec{X}$
is satisfied.

For $\vec{\mu}=0$ Eq.~\ref{cs2BQS}
reduces to 
$N_{\vec{\mu}=0}=s^2 a_{s2}$ and
$D_{\vec{\mu}=0}= s' b_{s}$. All other coefficients appearing in Es.~\ref{bcoef} and
\ref{acoef} are proportional to $\mu^2$.

As discussed in Appendix~\ref{app:partial}
some care has to be taken when  rewriting the 
derivatives of e.g. entropy density with respect
to $T$,
\begin{eqnarray}
  s'\equiv  \frac{\partial s}{\partial T}\Bigg|_{\vec{\mu}} 
&=&  T^{2}\left( 3 \hat{s} + T \frac{\partial \hat{s}}{\partial T}\Bigg|_{\vec{\hmu}}
- \sum_i \hmu_i
  \frac{\partial \hat{s}}{\partial \hmu_i}\Bigg|_T \right).
\label{fixedmuT}
\end{eqnarray}

For $\vec{\mu}=0$ we obtain the well known
expression
\begin{equation}
  \left.  c^2_{s}\right|_{\vec{\mu}=0}=\left. \frac{s}{T s'}\right|_{\vec{\mu}=0} \; .
\end{equation}
Furthermore, for an ideal gas Eq.~\ref{cX2} gives, of course, 
$c_s^2=1/3$ for arbitrary $(\mu_B, \mu_Q, \mu_S)$.

We are interested here in the speed of sound in
strangeness-neutral matter. In that case all
terms proportional to $n_S$ appearing in Eqs.~\ref{cX2} and \ref{acoef} can be set to 
zero.

\vfill
\bibliography{HICbib}

\begin{thebibliography}{51}%
\makeatletter
\providecommand \@ifxundefined [1]{%
 \@ifx{#1\undefined}
}%
\providecommand \@ifnum [1]{%
 \ifnum #1\expandafter \@firstoftwo
 \else \expandafter \@secondoftwo
 \fi
}%
\providecommand \@ifx [1]{%
 \ifx #1\expandafter \@firstoftwo
 \else \expandafter \@secondoftwo
 \fi
}%
\providecommand \natexlab [1]{#1}%
\providecommand \enquote  [1]{``#1''}%
\providecommand \bibnamefont  [1]{#1}%
\providecommand \bibfnamefont [1]{#1}%
\providecommand \citenamefont [1]{#1}%
\providecommand \href@noop [0]{\@secondoftwo}%
\providecommand \href [0]{\begingroup \@sanitize@url \@href}%
\providecommand \@href[1]{\@@startlink{#1}\@@href}%
\providecommand \@@href[1]{\endgroup#1\@@endlink}%
\providecommand \@sanitize@url [0]{\catcode `\\12\catcode `\$12\catcode
  `\&12\catcode `\#12\catcode `\^12\catcode `\_12\catcode `\%12\relax}%
\providecommand \@@startlink[1]{}%
\providecommand \@@endlink[0]{}%
\providecommand \url  [0]{\begingroup\@sanitize@url \@url }%
\providecommand \@url [1]{\endgroup\@href {#1}{\urlprefix }}%
\providecommand \urlprefix  [0]{URL }%
\providecommand \Eprint [0]{\href }%
\providecommand \doibase [0]{https://doi.org/}%
\providecommand \selectlanguage [0]{\@gobble}%
\providecommand \bibinfo  [0]{\@secondoftwo}%
\providecommand \bibfield  [0]{\@secondoftwo}%
\providecommand \translation [1]{[#1]}%
\providecommand \BibitemOpen [0]{}%
\providecommand \bibitemStop [0]{}%
\providecommand \bibitemNoStop [0]{.\EOS\space}%
\providecommand \EOS [0]{\spacefactor3000\relax}%
\providecommand \BibitemShut  [1]{\csname bibitem#1\endcsname}%
\let\auto@bib@innerbib\@empty
%</preamble>
\bibitem [{\citenamefont {Braun-Munzinger}\ \emph {et~al.}(2016)\citenamefont
  {Braun-Munzinger}, \citenamefont {Koch}, \citenamefont {Sch\"afer},\ and\
  \citenamefont {Stachel}}]{Braun-Munzinger:2015hba}%
  \BibitemOpen
  \bibfield  {author} {\bibinfo {author} {\bibfnamefont {P.}~\bibnamefont
  {Braun-Munzinger}}, \bibinfo {author} {\bibfnamefont {V.}~\bibnamefont
  {Koch}}, \bibinfo {author} {\bibfnamefont {T.}~\bibnamefont {Sch\"afer}},\
  and\ \bibinfo {author} {\bibfnamefont {J.}~\bibnamefont {Stachel}},\
  }\bibfield  {title} {\bibinfo {title} {{Properties of hot and dense matter
  from relativistic heavy ion collisions}},\ }\href
  {https://doi.org/10.1016/j.physrep.2015.12.003} {\bibfield  {journal}
  {\bibinfo  {journal} {Phys. Rept.}\ }\textbf {\bibinfo {volume} {621}},\
  \bibinfo {pages} {76} (\bibinfo {year} {2016})},\ \Eprint
  {https://arxiv.org/abs/1510.00442} {arXiv:1510.00442 [nucl-th]} \BibitemShut
  {NoStop}%
\bibitem [{\citenamefont {Middeldorf-Wygas}\ \emph {et~al.}(2022)\citenamefont
  {Middeldorf-Wygas}, \citenamefont {Oldengott}, \citenamefont {B\"odeker},\
  and\ \citenamefont {Schwarz}}]{Middeldorf-Wygas:2020glx}%
  \BibitemOpen
  \bibfield  {author} {\bibinfo {author} {\bibfnamefont {M.~M.}\ \bibnamefont
  {Middeldorf-Wygas}}, \bibinfo {author} {\bibfnamefont {I.~M.}\ \bibnamefont
  {Oldengott}}, \bibinfo {author} {\bibfnamefont {D.}~\bibnamefont
  {B\"odeker}},\ and\ \bibinfo {author} {\bibfnamefont {D.~J.}\ \bibnamefont
  {Schwarz}},\ }\bibfield  {title} {\bibinfo {title} {{Cosmic QCD transition
  for large lepton flavor asymmetries}},\ }\href
  {https://doi.org/10.1103/PhysRevD.105.123533} {\bibfield  {journal} {\bibinfo
   {journal} {Phys. Rev. D}\ }\textbf {\bibinfo {volume} {105}},\ \bibinfo
  {pages} {123533} (\bibinfo {year} {2022})},\ \Eprint
  {https://arxiv.org/abs/2009.00036} {arXiv:2009.00036 [hep-ph]} \BibitemShut
  {NoStop}%
\bibitem [{\citenamefont {Karsch}\ \emph {et~al.}(2000)\citenamefont {Karsch},
  \citenamefont {Laermann},\ and\ \citenamefont {Peikert}}]{Karsch:2000ps}%
  \BibitemOpen
  \bibfield  {author} {\bibinfo {author} {\bibfnamefont {F.}~\bibnamefont
  {Karsch}}, \bibinfo {author} {\bibfnamefont {E.}~\bibnamefont {Laermann}},\
  and\ \bibinfo {author} {\bibfnamefont {A.}~\bibnamefont {Peikert}},\
  }\bibfield  {title} {\bibinfo {title} {{The Pressure in two flavor,
  (2+1)-flavor and three flavor QCD}},\ }\href
  {https://doi.org/10.1016/S0370-2693(00)00292-6} {\bibfield  {journal}
  {\bibinfo  {journal} {Phys. Lett. B}\ }\textbf {\bibinfo {volume} {478}},\
  \bibinfo {pages} {447} (\bibinfo {year} {2000})},\ \Eprint
  {https://arxiv.org/abs/hep-lat/0002003} {arXiv:hep-lat/0002003} \BibitemShut
  {NoStop}%
\bibitem [{\citenamefont {Borsanyi}\ \emph {et~al.}(2014)\citenamefont
  {Borsanyi}, \citenamefont {Fodor}, \citenamefont {Hoelbling}, \citenamefont
  {Katz}, \citenamefont {Krieg},\ and\ \citenamefont
  {Szabo}}]{Borsanyi:2013bia}%
  \BibitemOpen
  \bibfield  {author} {\bibinfo {author} {\bibfnamefont {S.}~\bibnamefont
  {Borsanyi}}, \bibinfo {author} {\bibfnamefont {Z.}~\bibnamefont {Fodor}},
  \bibinfo {author} {\bibfnamefont {C.}~\bibnamefont {Hoelbling}}, \bibinfo
  {author} {\bibfnamefont {S.~D.}\ \bibnamefont {Katz}}, \bibinfo {author}
  {\bibfnamefont {S.}~\bibnamefont {Krieg}},\ and\ \bibinfo {author}
  {\bibfnamefont {K.~K.}\ \bibnamefont {Szabo}},\ }\bibfield  {title} {\bibinfo
  {title} {{Full result for the QCD equation of state with 2+1 flavors}},\
  }\href {https://doi.org/10.1016/j.physletb.2014.01.007} {\bibfield  {journal}
  {\bibinfo  {journal} {Phys. Lett. B}\ }\textbf {\bibinfo {volume} {730}},\
  \bibinfo {pages} {99} (\bibinfo {year} {2014})},\ \Eprint
  {https://arxiv.org/abs/1309.5258} {arXiv:1309.5258 [hep-lat]} \BibitemShut
  {NoStop}%
\bibitem [{\citenamefont {Bazavov}\ \emph {et~al.}(2014)\citenamefont {Bazavov}
  \emph {et~al.}}]{HotQCD:2014kol}%
  \BibitemOpen
  \bibfield  {author} {\bibinfo {author} {\bibfnamefont {A.}~\bibnamefont
  {Bazavov}} \emph {et~al.} (\bibinfo {collaboration} {HotQCD}),\ }\bibfield
  {title} {\bibinfo {title} {{Equation of state in (2+1)-flavor QCD}},\ }\href
  {https://doi.org/10.1103/PhysRevD.90.094503} {\bibfield  {journal} {\bibinfo
  {journal} {Phys. Rev. D}\ }\textbf {\bibinfo {volume} {90}},\ \bibinfo
  {pages} {094503} (\bibinfo {year} {2014})},\ \Eprint
  {https://arxiv.org/abs/1407.6387} {arXiv:1407.6387 [hep-lat]} \BibitemShut
  {NoStop}%
\bibitem [{\citenamefont {Gavai}\ and\ \citenamefont
  {Gupta}(2001)}]{Gavai:2001fr}%
  \BibitemOpen
  \bibfield  {author} {\bibinfo {author} {\bibfnamefont {R.~V.}\ \bibnamefont
  {Gavai}}\ and\ \bibinfo {author} {\bibfnamefont {S.}~\bibnamefont {Gupta}},\
  }\bibfield  {title} {\bibinfo {title} {{Quark number susceptibilities,
  strangeness and dynamical confinement}},\ }\href
  {https://doi.org/10.1103/PhysRevD.64.074506} {\bibfield  {journal} {\bibinfo
  {journal} {Phys. Rev. D}\ }\textbf {\bibinfo {volume} {64}},\ \bibinfo
  {pages} {074506} (\bibinfo {year} {2001})},\ \Eprint
  {https://arxiv.org/abs/hep-lat/0103013} {arXiv:hep-lat/0103013} \BibitemShut
  {NoStop}%
\bibitem [{\citenamefont {Allton}\ \emph {et~al.}(2002)\citenamefont {Allton},
  \citenamefont {Ejiri}, \citenamefont {Hands}, \citenamefont {Kaczmarek},
  \citenamefont {Karsch}, \citenamefont {Laermann}, \citenamefont {Schmidt},\
  and\ \citenamefont {Scorzato}}]{Allton:2002zi}%
  \BibitemOpen
  \bibfield  {author} {\bibinfo {author} {\bibfnamefont {C.~R.}\ \bibnamefont
  {Allton}}, \bibinfo {author} {\bibfnamefont {S.}~\bibnamefont {Ejiri}},
  \bibinfo {author} {\bibfnamefont {S.~J.}\ \bibnamefont {Hands}}, \bibinfo
  {author} {\bibfnamefont {O.}~\bibnamefont {Kaczmarek}}, \bibinfo {author}
  {\bibfnamefont {F.}~\bibnamefont {Karsch}}, \bibinfo {author} {\bibfnamefont
  {E.}~\bibnamefont {Laermann}}, \bibinfo {author} {\bibfnamefont
  {C.}~\bibnamefont {Schmidt}},\ and\ \bibinfo {author} {\bibfnamefont
  {L.}~\bibnamefont {Scorzato}},\ }\bibfield  {title} {\bibinfo {title} {{The
  QCD thermal phase transition in the presence of a small chemical
  potential}},\ }\href {https://doi.org/10.1103/PhysRevD.66.074507} {\bibfield
  {journal} {\bibinfo  {journal} {Phys. Rev. D}\ }\textbf {\bibinfo {volume}
  {66}},\ \bibinfo {pages} {074507} (\bibinfo {year} {2002})},\ \Eprint
  {https://arxiv.org/abs/hep-lat/0204010} {arXiv:hep-lat/0204010} \BibitemShut
  {NoStop}%
\bibitem [{\citenamefont {D'Elia}\ and\ \citenamefont
  {Lombardo}(2003)}]{DElia:2002tig}%
  \BibitemOpen
  \bibfield  {author} {\bibinfo {author} {\bibfnamefont {M.}~\bibnamefont
  {D'Elia}}\ and\ \bibinfo {author} {\bibfnamefont {M.-P.}\ \bibnamefont
  {Lombardo}},\ }\bibfield  {title} {\bibinfo {title} {{Finite density QCD via
  imaginary chemical potential}},\ }\href
  {https://doi.org/10.1103/PhysRevD.67.014505} {\bibfield  {journal} {\bibinfo
  {journal} {Phys. Rev. D}\ }\textbf {\bibinfo {volume} {67}},\ \bibinfo
  {pages} {014505} (\bibinfo {year} {2003})},\ \Eprint
  {https://arxiv.org/abs/hep-lat/0209146} {arXiv:hep-lat/0209146} \BibitemShut
  {NoStop}%
\bibitem [{\citenamefont {de~Forcrand}\ and\ \citenamefont
  {Philipsen}(2002)}]{deForcrand:2002hgr}%
  \BibitemOpen
  \bibfield  {author} {\bibinfo {author} {\bibfnamefont {P.}~\bibnamefont
  {de~Forcrand}}\ and\ \bibinfo {author} {\bibfnamefont {O.}~\bibnamefont
  {Philipsen}},\ }\bibfield  {title} {\bibinfo {title} {{The QCD phase diagram
  for small densities from imaginary chemical potential}},\ }\href
  {https://doi.org/10.1016/S0550-3213(02)00626-0} {\bibfield  {journal}
  {\bibinfo  {journal} {Nucl. Phys. B}\ }\textbf {\bibinfo {volume} {642}},\
  \bibinfo {pages} {290} (\bibinfo {year} {2002})},\ \Eprint
  {https://arxiv.org/abs/hep-lat/0205016} {arXiv:hep-lat/0205016} \BibitemShut
  {NoStop}%
\bibitem [{\citenamefont {Borsanyi}\ \emph {et~al.}(2012)\citenamefont
  {Borsanyi}, \citenamefont {Endrodi}, \citenamefont {Fodor}, \citenamefont
  {Katz}, \citenamefont {Krieg}, \citenamefont {Ratti},\ and\ \citenamefont
  {Szabo}}]{Borsanyi:2012cr}%
  \BibitemOpen
  \bibfield  {author} {\bibinfo {author} {\bibfnamefont {S.}~\bibnamefont
  {Borsanyi}}, \bibinfo {author} {\bibfnamefont {G.}~\bibnamefont {Endrodi}},
  \bibinfo {author} {\bibfnamefont {Z.}~\bibnamefont {Fodor}}, \bibinfo
  {author} {\bibfnamefont {S.~D.}\ \bibnamefont {Katz}}, \bibinfo {author}
  {\bibfnamefont {S.}~\bibnamefont {Krieg}}, \bibinfo {author} {\bibfnamefont
  {C.}~\bibnamefont {Ratti}},\ and\ \bibinfo {author} {\bibfnamefont {K.~K.}\
  \bibnamefont {Szabo}},\ }\bibfield  {title} {\bibinfo {title} {{QCD equation
  of state at nonzero chemical potential: continuum results with physical quark
  masses at order $\mu^2$}},\ }\href {https://doi.org/10.1007/JHEP08(2012)053}
  {\bibfield  {journal} {\bibinfo  {journal} {JHEP}\ }\textbf {\bibinfo
  {volume} {08}},\ \bibinfo {pages} {053}},\ \Eprint
  {https://arxiv.org/abs/1204.6710} {arXiv:1204.6710 [hep-lat]} \BibitemShut
  {NoStop}%
\bibitem [{\citenamefont {Guenther}\ \emph {et~al.}(2017)\citenamefont
  {Guenther}, \citenamefont {Bellwied}, \citenamefont {Borsanyi}, \citenamefont
  {Fodor}, \citenamefont {Katz}, \citenamefont {Pasztor}, \citenamefont
  {Ratti},\ and\ \citenamefont {Szab\'o}}]{Guenther:2017hnx}%
  \BibitemOpen
  \bibfield  {author} {\bibinfo {author} {\bibfnamefont {J.~N.}\ \bibnamefont
  {Guenther}}, \bibinfo {author} {\bibfnamefont {R.}~\bibnamefont {Bellwied}},
  \bibinfo {author} {\bibfnamefont {S.}~\bibnamefont {Borsanyi}}, \bibinfo
  {author} {\bibfnamefont {Z.}~\bibnamefont {Fodor}}, \bibinfo {author}
  {\bibfnamefont {S.~D.}\ \bibnamefont {Katz}}, \bibinfo {author}
  {\bibfnamefont {A.}~\bibnamefont {Pasztor}}, \bibinfo {author} {\bibfnamefont
  {C.}~\bibnamefont {Ratti}},\ and\ \bibinfo {author} {\bibfnamefont {K.~K.}\
  \bibnamefont {Szab\'o}},\ }\bibfield  {title} {\bibinfo {title} {{The QCD
  equation of state at finite density from analytical continuation}},\ }\href
  {https://doi.org/10.1016/j.nuclphysa.2017.05.044} {\bibfield  {journal}
  {\bibinfo  {journal} {Nucl. Phys. A}\ }\textbf {\bibinfo {volume} {967}},\
  \bibinfo {pages} {720} (\bibinfo {year} {2017})},\ \Eprint
  {https://arxiv.org/abs/1607.02493} {arXiv:1607.02493 [hep-lat]} \BibitemShut
  {NoStop}%
\bibitem [{\citenamefont {Borsanyi}\ \emph {et~al.}(2022)\citenamefont
  {Borsanyi}, \citenamefont {Guenther}, \citenamefont {Kara}, \citenamefont
  {Fodor}, \citenamefont {Parotto}, \citenamefont {Pasztor}, \citenamefont
  {Ratti},\ and\ \citenamefont {Szabo}}]{Borsanyi:2022qlh}%
  \BibitemOpen
  \bibfield  {author} {\bibinfo {author} {\bibfnamefont {S.}~\bibnamefont
  {Borsanyi}}, \bibinfo {author} {\bibfnamefont {J.~N.}\ \bibnamefont
  {Guenther}}, \bibinfo {author} {\bibfnamefont {R.}~\bibnamefont {Kara}},
  \bibinfo {author} {\bibfnamefont {Z.}~\bibnamefont {Fodor}}, \bibinfo
  {author} {\bibfnamefont {P.}~\bibnamefont {Parotto}}, \bibinfo {author}
  {\bibfnamefont {A.}~\bibnamefont {Pasztor}}, \bibinfo {author} {\bibfnamefont
  {C.}~\bibnamefont {Ratti}},\ and\ \bibinfo {author} {\bibfnamefont {K.~K.}\
  \bibnamefont {Szabo}},\ }\bibfield  {title} {\bibinfo {title} {{Resummed
  lattice QCD equation of state at finite baryon density: Strangeness
  neutrality and beyond}},\ }\href
  {https://doi.org/10.1103/PhysRevD.105.114504} {\bibfield  {journal} {\bibinfo
   {journal} {Phys. Rev. D}\ }\textbf {\bibinfo {volume} {105}},\ \bibinfo
  {pages} {114504} (\bibinfo {year} {2022})},\ \Eprint
  {https://arxiv.org/abs/2202.05574} {arXiv:2202.05574 [hep-lat]} \BibitemShut
  {NoStop}%
\bibitem [{\citenamefont {Ejiri}\ \emph {et~al.}(2006)\citenamefont {Ejiri},
  \citenamefont {Karsch}, \citenamefont {Laermann},\ and\ \citenamefont
  {Schmidt}}]{Ejiri:2005uv}%
  \BibitemOpen
  \bibfield  {author} {\bibinfo {author} {\bibfnamefont {S.}~\bibnamefont
  {Ejiri}}, \bibinfo {author} {\bibfnamefont {F.}~\bibnamefont {Karsch}},
  \bibinfo {author} {\bibfnamefont {E.}~\bibnamefont {Laermann}},\ and\
  \bibinfo {author} {\bibfnamefont {C.}~\bibnamefont {Schmidt}},\ }\bibfield
  {title} {\bibinfo {title} {{The Isentropic equation of state of 2-flavor
  QCD}},\ }\href {https://doi.org/10.1103/PhysRevD.73.054506} {\bibfield
  {journal} {\bibinfo  {journal} {Phys. Rev. D}\ }\textbf {\bibinfo {volume}
  {73}},\ \bibinfo {pages} {054506} (\bibinfo {year} {2006})},\ \Eprint
  {https://arxiv.org/abs/hep-lat/0512040} {arXiv:hep-lat/0512040} \BibitemShut
  {NoStop}%
\bibitem [{\citenamefont {Bazavov}\ \emph {et~al.}(2017)\citenamefont {Bazavov}
  \emph {et~al.}}]{Bazavov:2017dus}%
  \BibitemOpen
  \bibfield  {author} {\bibinfo {author} {\bibfnamefont {A.}~\bibnamefont
  {Bazavov}} \emph {et~al.},\ }\bibfield  {title} {\bibinfo {title} {{The QCD
  Equation of State to $\mathcal{O}(\mu_B^6)$ from Lattice QCD}},\ }\href
  {https://doi.org/10.1103/PhysRevD.95.054504} {\bibfield  {journal} {\bibinfo
  {journal} {Phys. Rev. D}\ }\textbf {\bibinfo {volume} {95}},\ \bibinfo
  {pages} {054504} (\bibinfo {year} {2017})},\ \Eprint
  {https://arxiv.org/abs/1701.04325} {arXiv:1701.04325 [hep-lat]} \BibitemShut
  {NoStop}%
\bibitem [{\citenamefont {Bollweg}\ \emph
  {et~al.}(2022{\natexlab{a}})\citenamefont {Bollweg}, \citenamefont {Goswami},
  \citenamefont {Kaczmarek}, \citenamefont {Karsch}, \citenamefont {Mukherjee},
  \citenamefont {Petreczky}, \citenamefont {Schmidt},\ and\ \citenamefont
  {Scior}}]{Bollweg:2022rps}%
  \BibitemOpen
  \bibfield  {author} {\bibinfo {author} {\bibfnamefont {D.}~\bibnamefont
  {Bollweg}}, \bibinfo {author} {\bibfnamefont {J.}~\bibnamefont {Goswami}},
  \bibinfo {author} {\bibfnamefont {O.}~\bibnamefont {Kaczmarek}}, \bibinfo
  {author} {\bibfnamefont {F.}~\bibnamefont {Karsch}}, \bibinfo {author}
  {\bibfnamefont {S.}~\bibnamefont {Mukherjee}}, \bibinfo {author}
  {\bibfnamefont {P.}~\bibnamefont {Petreczky}}, \bibinfo {author}
  {\bibfnamefont {C.}~\bibnamefont {Schmidt}},\ and\ \bibinfo {author}
  {\bibfnamefont {P.}~\bibnamefont {Scior}} (\bibinfo {collaboration}
  {HotQCD}),\ }\bibfield  {title} {\bibinfo {title} {{Taylor expansions and
  Pad\'e approximants for cumulants of conserved charge fluctuations at
  nonvanishing chemical potentials}},\ }\href
  {https://doi.org/10.1103/PhysRevD.105.074511} {\bibfield  {journal} {\bibinfo
   {journal} {Phys. Rev. D}\ }\textbf {\bibinfo {volume} {105}},\ \bibinfo
  {pages} {074511} (\bibinfo {year} {2022}{\natexlab{a}})},\ \Eprint
  {https://arxiv.org/abs/2202.09184} {arXiv:2202.09184 [hep-lat]} \BibitemShut
  {NoStop}%
\bibitem [{\citenamefont {Gavai}\ \emph {et~al.}(2005)\citenamefont {Gavai},
  \citenamefont {Gupta},\ and\ \citenamefont {Mukherjee}}]{Gavai:2004se}%
  \BibitemOpen
  \bibfield  {author} {\bibinfo {author} {\bibfnamefont {R.~V.}\ \bibnamefont
  {Gavai}}, \bibinfo {author} {\bibfnamefont {S.}~\bibnamefont {Gupta}},\ and\
  \bibinfo {author} {\bibfnamefont {S.}~\bibnamefont {Mukherjee}},\ }\bibfield
  {title} {\bibinfo {title} {{The Speed of sound and specific heat in the QCD
  plasma: Hydrodynamics, fluctuations and conformal symmetry}},\ }\href
  {https://doi.org/10.1103/PhysRevD.71.074013} {\bibfield  {journal} {\bibinfo
  {journal} {Phys. Rev. D}\ }\textbf {\bibinfo {volume} {71}},\ \bibinfo
  {pages} {074013} (\bibinfo {year} {2005})},\ \Eprint
  {https://arxiv.org/abs/hep-lat/0412036} {arXiv:hep-lat/0412036} \BibitemShut
  {NoStop}%
\bibitem [{\citenamefont {Floerchinger}\ and\ \citenamefont
  {Martinez}(2015)}]{Floerchinger:2015efa}%
  \BibitemOpen
  \bibfield  {author} {\bibinfo {author} {\bibfnamefont {S.}~\bibnamefont
  {Floerchinger}}\ and\ \bibinfo {author} {\bibfnamefont {M.}~\bibnamefont
  {Martinez}},\ }\bibfield  {title} {\bibinfo {title} {{Fluid dynamic
  propagation of initial baryon number perturbations on a Bjorken flow
  background}},\ }\href {https://doi.org/10.1103/PhysRevC.92.064906} {\bibfield
   {journal} {\bibinfo  {journal} {Phys. Rev. C}\ }\textbf {\bibinfo {volume}
  {92}},\ \bibinfo {pages} {064906} (\bibinfo {year} {2015})},\ \Eprint
  {https://arxiv.org/abs/1507.05569} {arXiv:1507.05569 [nucl-th]} \BibitemShut
  {NoStop}%
\bibitem [{\citenamefont {Bazavov}\ \emph {et~al.}(2020)\citenamefont {Bazavov}
  \emph {et~al.}}]{Bazavov:2020bjn}%
  \BibitemOpen
  \bibfield  {author} {\bibinfo {author} {\bibfnamefont {A.}~\bibnamefont
  {Bazavov}} \emph {et~al.},\ }\bibfield  {title} {\bibinfo {title} {{Skewness,
  kurtosis, and the fifth and sixth order cumulants of net baryon-number
  distributions from lattice QCD confront high-statistics STAR data}},\ }\href
  {https://doi.org/10.1103/PhysRevD.101.074502} {\bibfield  {journal} {\bibinfo
   {journal} {Phys. Rev. D}\ }\textbf {\bibinfo {volume} {101}},\ \bibinfo
  {pages} {074502} (\bibinfo {year} {2020})},\ \Eprint
  {https://arxiv.org/abs/2001.08530} {arXiv:2001.08530 [hep-lat]} \BibitemShut
  {NoStop}%
\bibitem [{\citenamefont {Dimopoulos}\ \emph {et~al.}(2022)\citenamefont
  {Dimopoulos}, \citenamefont {Dini}, \citenamefont {Di~Renzo}, \citenamefont
  {Goswami}, \citenamefont {Nicotra}, \citenamefont {Schmidt}, \citenamefont
  {Singh}, \citenamefont {Zambello},\ and\ \citenamefont
  {Ziesch\'e}}]{Dimopoulos:2021vrk}%
  \BibitemOpen
  \bibfield  {author} {\bibinfo {author} {\bibfnamefont {P.}~\bibnamefont
  {Dimopoulos}}, \bibinfo {author} {\bibfnamefont {L.}~\bibnamefont {Dini}},
  \bibinfo {author} {\bibfnamefont {F.}~\bibnamefont {Di~Renzo}}, \bibinfo
  {author} {\bibfnamefont {J.}~\bibnamefont {Goswami}}, \bibinfo {author}
  {\bibfnamefont {G.}~\bibnamefont {Nicotra}}, \bibinfo {author} {\bibfnamefont
  {C.}~\bibnamefont {Schmidt}}, \bibinfo {author} {\bibfnamefont
  {S.}~\bibnamefont {Singh}}, \bibinfo {author} {\bibfnamefont
  {K.}~\bibnamefont {Zambello}},\ and\ \bibinfo {author} {\bibfnamefont
  {F.}~\bibnamefont {Ziesch\'e}},\ }\bibfield  {title} {\bibinfo {title}
  {{Contribution to understanding the phase structure of strong interaction
  matter: Lee-Yang edge singularities from lattice QCD}},\ }\href
  {https://doi.org/10.1103/PhysRevD.105.034513} {\bibfield  {journal} {\bibinfo
   {journal} {Phys. Rev. D}\ }\textbf {\bibinfo {volume} {105}},\ \bibinfo
  {pages} {034513} (\bibinfo {year} {2022})},\ \Eprint
  {https://arxiv.org/abs/2110.15933} {arXiv:2110.15933 [hep-lat]} \BibitemShut
  {NoStop}%
\bibitem [{\citenamefont {Bazavov}\ \emph {et~al.}(2019)\citenamefont {Bazavov}
  \emph {et~al.}}]{HotQCD:2018pds}%
  \BibitemOpen
  \bibfield  {author} {\bibinfo {author} {\bibfnamefont {A.}~\bibnamefont
  {Bazavov}} \emph {et~al.} (\bibinfo {collaboration} {HotQCD}),\ }\bibfield
  {title} {\bibinfo {title} {{Chiral crossover in QCD at zero and non-zero
  chemical potentials}},\ }\href
  {https://doi.org/10.1016/j.physletb.2019.05.013} {\bibfield  {journal}
  {\bibinfo  {journal} {Phys. Lett. B}\ }\textbf {\bibinfo {volume} {795}},\
  \bibinfo {pages} {15} (\bibinfo {year} {2019})},\ \Eprint
  {https://arxiv.org/abs/1812.08235} {arXiv:1812.08235 [hep-lat]} \BibitemShut
  {NoStop}%
\bibitem [{\citenamefont {Mazur}(2021)}]{mazur2021}%
  \BibitemOpen
  \bibfield  {author} {\bibinfo {author} {\bibfnamefont {L.}~\bibnamefont
  {Mazur}},\ }\bibfield  {title} {\bibinfo {title} {{Topological Aspects in
  Lattice QCD}},\ }\bibfield  {journal} {\bibinfo  {journal} {Ph.D. thesis,
  Bielefeld University}\ }\href {https://doi.org/10.4119/unibi/2956493}
  {10.4119/unibi/2956493} (\bibinfo {year} {2021})\BibitemShut {NoStop}%
\bibitem [{\citenamefont {Bollweg}\ \emph
  {et~al.}(2022{\natexlab{b}})\citenamefont {Bollweg}, \citenamefont
  {Altenkort}, \citenamefont {Clarke}, \citenamefont {Kaczmarek}, \citenamefont
  {Mazur}, \citenamefont {Schmidt}, \citenamefont {Scior},\ and\ \citenamefont
  {Shu}}]{Bollweg:2021cvl}%
  \BibitemOpen
  \bibfield  {author} {\bibinfo {author} {\bibfnamefont {D.}~\bibnamefont
  {Bollweg}}, \bibinfo {author} {\bibfnamefont {L.}~\bibnamefont {Altenkort}},
  \bibinfo {author} {\bibfnamefont {D.~A.}\ \bibnamefont {Clarke}}, \bibinfo
  {author} {\bibfnamefont {O.}~\bibnamefont {Kaczmarek}}, \bibinfo {author}
  {\bibfnamefont {L.}~\bibnamefont {Mazur}}, \bibinfo {author} {\bibfnamefont
  {C.}~\bibnamefont {Schmidt}}, \bibinfo {author} {\bibfnamefont
  {P.}~\bibnamefont {Scior}},\ and\ \bibinfo {author} {\bibfnamefont {H.-T.}\
  \bibnamefont {Shu}},\ }\bibfield  {title} {\bibinfo {title} {{HotQCD on
  multi-GPU Systems}},\ }\href {https://doi.org/10.22323/1.396.0196} {\bibfield
   {journal} {\bibinfo  {journal} {PoS}\ }\textbf {\bibinfo {volume}
  {LATTICE2021}},\ \bibinfo {pages} {196} (\bibinfo {year}
  {2022}{\natexlab{b}})},\ \Eprint {https://arxiv.org/abs/2111.10354}
  {arXiv:2111.10354 [hep-lat]} \BibitemShut {NoStop}%
\bibitem [{\citenamefont {Mazur}\ \emph {et~al.}(2023)\citenamefont {Mazur}
  \emph {et~al.}}]{Mazur:2023lvn}%
  \BibitemOpen
  \bibfield  {author} {\bibinfo {author} {\bibfnamefont {L.}~\bibnamefont
  {Mazur}} \emph {et~al.},\ }\bibfield  {title} {\bibinfo {title}
  {{SIMULATeQCD: A simple multi-GPU lattice code for QCD calculations}},\
  }\href@noop {} {\  (\bibinfo {year} {2023})},\ \Eprint
  {https://arxiv.org/abs/2306.01098} {arXiv:2306.01098 [hep-lat]} \BibitemShut
  {NoStop}%
\bibitem [{\citenamefont {Bollweg}\ \emph {et~al.}(2021)\citenamefont
  {Bollweg}, \citenamefont {Goswami}, \citenamefont {Kaczmarek}, \citenamefont
  {Karsch}, \citenamefont {Mukherjee}, \citenamefont {Petreczky}, \citenamefont
  {Schmidt},\ and\ \citenamefont {Scior}}]{Bollweg:2021vqf}%
  \BibitemOpen
  \bibfield  {author} {\bibinfo {author} {\bibfnamefont {D.}~\bibnamefont
  {Bollweg}}, \bibinfo {author} {\bibfnamefont {J.}~\bibnamefont {Goswami}},
  \bibinfo {author} {\bibfnamefont {O.}~\bibnamefont {Kaczmarek}}, \bibinfo
  {author} {\bibfnamefont {F.}~\bibnamefont {Karsch}}, \bibinfo {author}
  {\bibfnamefont {S.}~\bibnamefont {Mukherjee}}, \bibinfo {author}
  {\bibfnamefont {P.}~\bibnamefont {Petreczky}}, \bibinfo {author}
  {\bibfnamefont {C.}~\bibnamefont {Schmidt}},\ and\ \bibinfo {author}
  {\bibfnamefont {P.}~\bibnamefont {Scior}} (\bibinfo {collaboration}
  {HotQCD}),\ }\bibfield  {title} {\bibinfo {title} {{Second order cumulants of
  conserved charge fluctuations revisited: Vanishing chemical potentials}},\
  }\bibfield  {journal} {\bibinfo  {journal} {Phys. Rev. D}\ }\textbf {\bibinfo
  {volume} {104}},\ \href {https://doi.org/10.1103/PhysRevD.104.074512}
  {10.1103/PhysRevD.104.074512} (\bibinfo {year} {2021}),\ \Eprint
  {https://arxiv.org/abs/2107.10011} {arXiv:2107.10011 [hep-lat]} \BibitemShut
  {NoStop}%
\bibitem [{too()}]{toolbox}%
  \BibitemOpen
  \href@noop {} {\bibinfo {title} {{AnalysisToolbox: A set of Python tools for
  analyzing physics data, in particular targeting lattice QCD}}},\ \bibinfo
  {howpublished}
  {\url{https://github.com/LatticeQCD/AnalysisToolbox}}\BibitemShut {NoStop}%
\bibitem [{\citenamefont {Virtanen}\ \emph {et~al.}(2020)\citenamefont
  {Virtanen} \emph {et~al.}}]{2020SciPy-NMeth}%
  \BibitemOpen
  \bibfield  {author} {\bibinfo {author} {\bibfnamefont {P.}~\bibnamefont
  {Virtanen}} \emph {et~al.},\ }\bibfield  {title} {\bibinfo {title} {{{SciPy}
  1.0: Fundamental Algorithms for Scientific Computing in Python}},\ }\href
  {https://doi.org/10.1038/s41592-019-0686-2} {\bibfield  {journal} {\bibinfo
  {journal} {Nature Methods}\ }\textbf {\bibinfo {volume} {17}},\ \bibinfo
  {pages} {261} (\bibinfo {year} {2020})}\BibitemShut {NoStop}%
\bibitem [{\citenamefont {D'Elia}\ \emph {et~al.}(2017)\citenamefont {D'Elia},
  \citenamefont {Gagliardi},\ and\ \citenamefont {Sanfilippo}}]{DElia:2016jqh}%
  \BibitemOpen
  \bibfield  {author} {\bibinfo {author} {\bibfnamefont {M.}~\bibnamefont
  {D'Elia}}, \bibinfo {author} {\bibfnamefont {G.}~\bibnamefont {Gagliardi}},\
  and\ \bibinfo {author} {\bibfnamefont {F.}~\bibnamefont {Sanfilippo}},\
  }\bibfield  {title} {\bibinfo {title} {{Higher order quark number
  fluctuations via imaginary chemical potentials in $N_f=2+1$ QCD}},\ }\href
  {https://doi.org/10.1103/PhysRevD.95.094503} {\bibfield  {journal} {\bibinfo
  {journal} {Phys. Rev. D}\ }\textbf {\bibinfo {volume} {95}},\ \bibinfo
  {pages} {094503} (\bibinfo {year} {2017})},\ \Eprint
  {https://arxiv.org/abs/1611.08285} {arXiv:1611.08285 [hep-lat]} \BibitemShut
  {NoStop}%
\bibitem [{\citenamefont {Borsanyi}\ \emph {et~al.}(2018)\citenamefont
  {Borsanyi}, \citenamefont {Fodor}, \citenamefont {Guenther}, \citenamefont
  {Katz}, \citenamefont {Szabo}, \citenamefont {Pasztor}, \citenamefont
  {Portillo},\ and\ \citenamefont {Ratti}}]{Borsanyi:2018grb}%
  \BibitemOpen
  \bibfield  {author} {\bibinfo {author} {\bibfnamefont {S.}~\bibnamefont
  {Borsanyi}}, \bibinfo {author} {\bibfnamefont {Z.}~\bibnamefont {Fodor}},
  \bibinfo {author} {\bibfnamefont {J.~N.}\ \bibnamefont {Guenther}}, \bibinfo
  {author} {\bibfnamefont {S.~K.}\ \bibnamefont {Katz}}, \bibinfo {author}
  {\bibfnamefont {K.~K.}\ \bibnamefont {Szabo}}, \bibinfo {author}
  {\bibfnamefont {A.}~\bibnamefont {Pasztor}}, \bibinfo {author} {\bibfnamefont
  {I.}~\bibnamefont {Portillo}},\ and\ \bibinfo {author} {\bibfnamefont
  {C.}~\bibnamefont {Ratti}},\ }\bibfield  {title} {\bibinfo {title} {{Higher
  order fluctuations and correlations of conserved charges from lattice QCD}},\
  }\href {https://doi.org/10.1007/JHEP10(2018)205} {\bibfield  {journal}
  {\bibinfo  {journal} {JHEP}\ }\textbf {\bibinfo {volume} {10}},\ \bibinfo
  {pages} {205}},\ \Eprint {https://arxiv.org/abs/1805.04445} {arXiv:1805.04445
  [hep-lat]} \BibitemShut {NoStop}%
\bibitem [{\citenamefont {Haque}\ \emph {et~al.}(2014)\citenamefont {Haque},
  \citenamefont {Bandyopadhyay}, \citenamefont {Andersen}, \citenamefont
  {Mustafa}, \citenamefont {Strickland},\ and\ \citenamefont
  {Su}}]{Haque:2014rua}%
  \BibitemOpen
  \bibfield  {author} {\bibinfo {author} {\bibfnamefont {N.}~\bibnamefont
  {Haque}}, \bibinfo {author} {\bibfnamefont {A.}~\bibnamefont
  {Bandyopadhyay}}, \bibinfo {author} {\bibfnamefont {J.~O.}\ \bibnamefont
  {Andersen}}, \bibinfo {author} {\bibfnamefont {M.~G.}\ \bibnamefont
  {Mustafa}}, \bibinfo {author} {\bibfnamefont {M.}~\bibnamefont
  {Strickland}},\ and\ \bibinfo {author} {\bibfnamefont {N.}~\bibnamefont
  {Su}},\ }\bibfield  {title} {\bibinfo {title} {{Three-loop HTLpt
  thermodynamics at finite temperature and chemical potential}},\ }\href
  {https://doi.org/10.1007/JHEP05(2014)027} {\bibfield  {journal} {\bibinfo
  {journal} {JHEP}\ }\textbf {\bibinfo {volume} {05}},\ \bibinfo {pages}
  {027}},\ \Eprint {https://arxiv.org/abs/1402.6907} {arXiv:1402.6907 [hep-ph]}
  \BibitemShut {NoStop}%
\bibitem [{\citenamefont {Mogliacci}\ \emph {et~al.}(2013)\citenamefont
  {Mogliacci}, \citenamefont {Andersen}, \citenamefont {Strickland},
  \citenamefont {Su},\ and\ \citenamefont {Vuorinen}}]{Mogliacci:2013mca}%
  \BibitemOpen
  \bibfield  {author} {\bibinfo {author} {\bibfnamefont {S.}~\bibnamefont
  {Mogliacci}}, \bibinfo {author} {\bibfnamefont {J.~O.}\ \bibnamefont
  {Andersen}}, \bibinfo {author} {\bibfnamefont {M.}~\bibnamefont
  {Strickland}}, \bibinfo {author} {\bibfnamefont {N.}~\bibnamefont {Su}},\
  and\ \bibinfo {author} {\bibfnamefont {A.}~\bibnamefont {Vuorinen}},\
  }\bibfield  {title} {\bibinfo {title} {{Equation of State of hot and dense
  QCD: Resummed perturbation theory confronts lattice data}},\ }\href
  {https://doi.org/10.1007/JHEP12(2013)055} {\bibfield  {journal} {\bibinfo
  {journal} {JHEP}\ }\textbf {\bibinfo {volume} {12}},\ \bibinfo {pages}
  {055}},\ \Eprint {https://arxiv.org/abs/1307.8098} {arXiv:1307.8098 [hep-ph]}
  \BibitemShut {NoStop}%
\bibitem [{\citenamefont {Bazavov}\ \emph {et~al.}(2013)\citenamefont
  {Bazavov}, \citenamefont {Ding}, \citenamefont {Hegde}, \citenamefont
  {Karsch}, \citenamefont {Miao}, \citenamefont {Mukherjee}, \citenamefont
  {Petreczky}, \citenamefont {Schmidt},\ and\ \citenamefont
  {Velytsky}}]{Bazavov:2013uja}%
  \BibitemOpen
  \bibfield  {author} {\bibinfo {author} {\bibfnamefont {A.}~\bibnamefont
  {Bazavov}}, \bibinfo {author} {\bibfnamefont {H.~T.}\ \bibnamefont {Ding}},
  \bibinfo {author} {\bibfnamefont {P.}~\bibnamefont {Hegde}}, \bibinfo
  {author} {\bibfnamefont {F.}~\bibnamefont {Karsch}}, \bibinfo {author}
  {\bibfnamefont {C.}~\bibnamefont {Miao}}, \bibinfo {author} {\bibfnamefont
  {S.}~\bibnamefont {Mukherjee}}, \bibinfo {author} {\bibfnamefont
  {P.}~\bibnamefont {Petreczky}}, \bibinfo {author} {\bibfnamefont
  {C.}~\bibnamefont {Schmidt}},\ and\ \bibinfo {author} {\bibfnamefont
  {A.}~\bibnamefont {Velytsky}},\ }\bibfield  {title} {\bibinfo {title} {{Quark
  number susceptibilities at high temperatures}},\ }\href
  {https://doi.org/10.1103/PhysRevD.88.094021} {\bibfield  {journal} {\bibinfo
  {journal} {Phys. Rev. D}\ }\textbf {\bibinfo {volume} {88}},\ \bibinfo
  {pages} {094021} (\bibinfo {year} {2013})},\ \Eprint
  {https://arxiv.org/abs/1309.2317} {arXiv:1309.2317 [hep-lat]} \BibitemShut
  {NoStop}%
\bibitem [{\citenamefont {Ding}\ \emph {et~al.}(2015)\citenamefont {Ding},
  \citenamefont {Mukherjee}, \citenamefont {Ohno}, \citenamefont {Petreczky},\
  and\ \citenamefont {Schadler}}]{Ding:2015fca}%
  \BibitemOpen
  \bibfield  {author} {\bibinfo {author} {\bibfnamefont {H.~T.}\ \bibnamefont
  {Ding}}, \bibinfo {author} {\bibfnamefont {S.}~\bibnamefont {Mukherjee}},
  \bibinfo {author} {\bibfnamefont {H.}~\bibnamefont {Ohno}}, \bibinfo {author}
  {\bibfnamefont {P.}~\bibnamefont {Petreczky}},\ and\ \bibinfo {author}
  {\bibfnamefont {H.~P.}\ \bibnamefont {Schadler}},\ }\bibfield  {title}
  {\bibinfo {title} {{Diagonal and off-diagonal quark number susceptibilities
  at high temperatures}},\ }\href {https://doi.org/10.1103/PhysRevD.92.074043}
  {\bibfield  {journal} {\bibinfo  {journal} {Phys. Rev. D}\ }\textbf {\bibinfo
  {volume} {92}},\ \bibinfo {pages} {074043} (\bibinfo {year} {2015})},\
  \Eprint {https://arxiv.org/abs/1507.06637} {arXiv:1507.06637 [hep-lat]}
  \BibitemShut {NoStop}%
\bibitem [{\citenamefont {Bellwied}\ \emph
  {et~al.}(2015{\natexlab{a}})\citenamefont {Bellwied}, \citenamefont
  {Borsanyi}, \citenamefont {Fodor}, \citenamefont {Katz}, \citenamefont
  {Pasztor}, \citenamefont {Ratti},\ and\ \citenamefont
  {Szabo}}]{Bellwied:2015lba}%
  \BibitemOpen
  \bibfield  {author} {\bibinfo {author} {\bibfnamefont {R.}~\bibnamefont
  {Bellwied}}, \bibinfo {author} {\bibfnamefont {S.}~\bibnamefont {Borsanyi}},
  \bibinfo {author} {\bibfnamefont {Z.}~\bibnamefont {Fodor}}, \bibinfo
  {author} {\bibfnamefont {S.~D.}\ \bibnamefont {Katz}}, \bibinfo {author}
  {\bibfnamefont {A.}~\bibnamefont {Pasztor}}, \bibinfo {author} {\bibfnamefont
  {C.}~\bibnamefont {Ratti}},\ and\ \bibinfo {author} {\bibfnamefont {K.~K.}\
  \bibnamefont {Szabo}},\ }\bibfield  {title} {\bibinfo {title} {{Fluctuations
  and correlations in high temperature QCD}},\ }\href
  {https://doi.org/10.1103/PhysRevD.92.114505} {\bibfield  {journal} {\bibinfo
  {journal} {Phys. Rev. D}\ }\textbf {\bibinfo {volume} {92}},\ \bibinfo
  {pages} {114505} (\bibinfo {year} {2015}{\natexlab{a}})},\ \Eprint
  {https://arxiv.org/abs/1507.04627} {arXiv:1507.04627 [hep-lat]} \BibitemShut
  {NoStop}%
\bibitem [{\citenamefont {Prosperi}\ \emph {et~al.}(2007)\citenamefont
  {Prosperi}, \citenamefont {Raciti},\ and\ \citenamefont
  {Simolo}}]{Prosperi:2006hx}%
  \BibitemOpen
  \bibfield  {author} {\bibinfo {author} {\bibfnamefont {G.~M.}\ \bibnamefont
  {Prosperi}}, \bibinfo {author} {\bibfnamefont {M.}~\bibnamefont {Raciti}},\
  and\ \bibinfo {author} {\bibfnamefont {C.}~\bibnamefont {Simolo}},\
  }\bibfield  {title} {\bibinfo {title} {{On the running coupling constant in
  QCD}},\ }\href {https://doi.org/10.1016/j.ppnp.2006.09.001} {\bibfield
  {journal} {\bibinfo  {journal} {Prog. Part. Nucl. Phys.}\ }\textbf {\bibinfo
  {volume} {58}},\ \bibinfo {pages} {387} (\bibinfo {year} {2007})},\ \Eprint
  {https://arxiv.org/abs/hep-ph/0607209} {arXiv:hep-ph/0607209} \BibitemShut
  {NoStop}%
\bibitem [{\citenamefont {Clarke}\ \emph {et~al.}(2021)\citenamefont {Clarke},
  \citenamefont {Kaczmarek}, \citenamefont {Karsch}, \citenamefont {Lahiri},\
  and\ \citenamefont {Sarkar}}]{Clarke:2020htu}%
  \BibitemOpen
  \bibfield  {author} {\bibinfo {author} {\bibfnamefont {D.~A.}\ \bibnamefont
  {Clarke}}, \bibinfo {author} {\bibfnamefont {O.}~\bibnamefont {Kaczmarek}},
  \bibinfo {author} {\bibfnamefont {F.}~\bibnamefont {Karsch}}, \bibinfo
  {author} {\bibfnamefont {A.}~\bibnamefont {Lahiri}},\ and\ \bibinfo {author}
  {\bibfnamefont {M.}~\bibnamefont {Sarkar}},\ }\bibfield  {title} {\bibinfo
  {title} {{Sensitivity of the Polyakov loop and related observables to chiral
  symmetry restoration}},\ }\href
  {https://doi.org/10.1103/PhysRevD.103.L011501} {\bibfield  {journal}
  {\bibinfo  {journal} {Phys. Rev. D}\ }\textbf {\bibinfo {volume} {103}},\
  \bibinfo {pages} {L011501} (\bibinfo {year} {2021})},\ \Eprint
  {https://arxiv.org/abs/2008.11678} {arXiv:2008.11678 [hep-lat]} \BibitemShut
  {NoStop}%
\bibitem [{\citenamefont {Borsanyi}\ \emph {et~al.}(2020)\citenamefont
  {Borsanyi}, \citenamefont {Fodor}, \citenamefont {Guenther}, \citenamefont
  {Kara}, \citenamefont {Katz}, \citenamefont {Parotto}, \citenamefont
  {Pasztor}, \citenamefont {Ratti},\ and\ \citenamefont
  {Szabo}}]{Borsanyi:2020fev}%
  \BibitemOpen
  \bibfield  {author} {\bibinfo {author} {\bibfnamefont {S.}~\bibnamefont
  {Borsanyi}}, \bibinfo {author} {\bibfnamefont {Z.}~\bibnamefont {Fodor}},
  \bibinfo {author} {\bibfnamefont {J.~N.}\ \bibnamefont {Guenther}}, \bibinfo
  {author} {\bibfnamefont {R.}~\bibnamefont {Kara}}, \bibinfo {author}
  {\bibfnamefont {S.~D.}\ \bibnamefont {Katz}}, \bibinfo {author}
  {\bibfnamefont {P.}~\bibnamefont {Parotto}}, \bibinfo {author} {\bibfnamefont
  {A.}~\bibnamefont {Pasztor}}, \bibinfo {author} {\bibfnamefont
  {C.}~\bibnamefont {Ratti}},\ and\ \bibinfo {author} {\bibfnamefont {K.~K.}\
  \bibnamefont {Szabo}},\ }\bibfield  {title} {\bibinfo {title} {{QCD Crossover
  at Finite Chemical Potential from Lattice Simulations}},\ }\href
  {https://doi.org/10.1103/PhysRevLett.125.052001} {\bibfield  {journal}
  {\bibinfo  {journal} {Phys. Rev. Lett.}\ }\textbf {\bibinfo {volume} {125}},\
  \bibinfo {pages} {052001} (\bibinfo {year} {2020})},\ \Eprint
  {https://arxiv.org/abs/2002.02821} {arXiv:2002.02821 [hep-lat]} \BibitemShut
  {NoStop}%
\bibitem [{\citenamefont {Kotov}\ \emph {et~al.}(2021)\citenamefont {Kotov},
  \citenamefont {Lombardo},\ and\ \citenamefont {Trunin}}]{Kotov:2021rah}%
  \BibitemOpen
  \bibfield  {author} {\bibinfo {author} {\bibfnamefont {A.~Y.}\ \bibnamefont
  {Kotov}}, \bibinfo {author} {\bibfnamefont {M.~P.}\ \bibnamefont
  {Lombardo}},\ and\ \bibinfo {author} {\bibfnamefont {A.}~\bibnamefont
  {Trunin}},\ }\bibfield  {title} {\bibinfo {title} {{QCD transition at the
  physical point, and its scaling window from twisted mass Wilson fermions}},\
  }\href {https://doi.org/10.1016/j.physletb.2021.136749} {\bibfield  {journal}
  {\bibinfo  {journal} {Phys. Lett. B}\ }\textbf {\bibinfo {volume} {823}},\
  \bibinfo {pages} {136749} (\bibinfo {year} {2021})},\ \Eprint
  {https://arxiv.org/abs/2105.09842} {arXiv:2105.09842 [hep-lat]} \BibitemShut
  {NoStop}%
\bibitem [{\citenamefont {Pelissetto}\ and\ \citenamefont
  {Vicari}(2013)}]{Pelissetto:2013hqa}%
  \BibitemOpen
  \bibfield  {author} {\bibinfo {author} {\bibfnamefont {A.}~\bibnamefont
  {Pelissetto}}\ and\ \bibinfo {author} {\bibfnamefont {E.}~\bibnamefont
  {Vicari}},\ }\bibfield  {title} {\bibinfo {title} {{Relevance of the axial
  anomaly at the finite-temperature chiral transition in QCD}},\ }\href
  {https://doi.org/10.1103/PhysRevD.88.105018} {\bibfield  {journal} {\bibinfo
  {journal} {Phys. Rev. D}\ }\textbf {\bibinfo {volume} {88}},\ \bibinfo
  {pages} {105018} (\bibinfo {year} {2013})},\ \Eprint
  {https://arxiv.org/abs/1309.5446} {arXiv:1309.5446 [hep-lat]} \BibitemShut
  {NoStop}%
\bibitem [{\citenamefont {Kaczmarek}\ \emph {et~al.}(2011)\citenamefont
  {Kaczmarek}, \citenamefont {Karsch}, \citenamefont {Laermann}, \citenamefont
  {Miao}, \citenamefont {Mukherjee}, \citenamefont {Petreczky}, \citenamefont
  {Schmidt}, \citenamefont {Soeldner},\ and\ \citenamefont
  {Unger}}]{Kaczmarek:2011zz}%
  \BibitemOpen
  \bibfield  {author} {\bibinfo {author} {\bibfnamefont {O.}~\bibnamefont
  {Kaczmarek}}, \bibinfo {author} {\bibfnamefont {F.}~\bibnamefont {Karsch}},
  \bibinfo {author} {\bibfnamefont {E.}~\bibnamefont {Laermann}}, \bibinfo
  {author} {\bibfnamefont {C.}~\bibnamefont {Miao}}, \bibinfo {author}
  {\bibfnamefont {S.}~\bibnamefont {Mukherjee}}, \bibinfo {author}
  {\bibfnamefont {P.}~\bibnamefont {Petreczky}}, \bibinfo {author}
  {\bibfnamefont {C.}~\bibnamefont {Schmidt}}, \bibinfo {author} {\bibfnamefont
  {W.}~\bibnamefont {Soeldner}},\ and\ \bibinfo {author} {\bibfnamefont
  {W.}~\bibnamefont {Unger}},\ }\bibfield  {title} {\bibinfo {title} {{Phase
  boundary for the chiral transition in (2+1) -flavor QCD at small values of
  the chemical potential}},\ }\href
  {https://doi.org/10.1103/PhysRevD.83.014504} {\bibfield  {journal} {\bibinfo
  {journal} {Phys. Rev. D}\ }\textbf {\bibinfo {volume} {83}},\ \bibinfo
  {pages} {014504} (\bibinfo {year} {2011})},\ \Eprint
  {https://arxiv.org/abs/1011.3130} {arXiv:1011.3130 [hep-lat]} \BibitemShut
  {NoStop}%
\bibitem [{\citenamefont {Karsch}(2019)}]{Karsch:2019mbv}%
  \BibitemOpen
  \bibfield  {author} {\bibinfo {author} {\bibfnamefont {F.}~\bibnamefont
  {Karsch}},\ }\bibfield  {title} {\bibinfo {title} {{Critical behavior and
  net-charge fluctuations from lattice QCD}},\ }\href
  {https://doi.org/10.22323/1.347.0163} {\bibfield  {journal} {\bibinfo
  {journal} {PoS}\ }\textbf {\bibinfo {volume} {CORFU2018}},\ \bibinfo {pages}
  {163} (\bibinfo {year} {2019})},\ \Eprint {https://arxiv.org/abs/1905.03936}
  {arXiv:1905.03936 [hep-lat]} \BibitemShut {NoStop}%
\bibitem [{\citenamefont {Bellwied}\ \emph
  {et~al.}(2015{\natexlab{b}})\citenamefont {Bellwied}, \citenamefont
  {Borsanyi}, \citenamefont {Fodor}, \citenamefont {G\"unther}, \citenamefont
  {Katz}, \citenamefont {Ratti},\ and\ \citenamefont
  {Szabo}}]{Bellwied:2015rza}%
  \BibitemOpen
  \bibfield  {author} {\bibinfo {author} {\bibfnamefont {R.}~\bibnamefont
  {Bellwied}}, \bibinfo {author} {\bibfnamefont {S.}~\bibnamefont {Borsanyi}},
  \bibinfo {author} {\bibfnamefont {Z.}~\bibnamefont {Fodor}}, \bibinfo
  {author} {\bibfnamefont {J.}~\bibnamefont {G\"unther}}, \bibinfo {author}
  {\bibfnamefont {S.~D.}\ \bibnamefont {Katz}}, \bibinfo {author}
  {\bibfnamefont {C.}~\bibnamefont {Ratti}},\ and\ \bibinfo {author}
  {\bibfnamefont {K.~K.}\ \bibnamefont {Szabo}},\ }\bibfield  {title} {\bibinfo
  {title} {{The QCD phase diagram from analytic continuation}},\ }\href
  {https://doi.org/10.1016/j.physletb.2015.11.011} {\bibfield  {journal}
  {\bibinfo  {journal} {Phys. Lett. B}\ }\textbf {\bibinfo {volume} {751}},\
  \bibinfo {pages} {559} (\bibinfo {year} {2015}{\natexlab{b}})},\ \Eprint
  {https://arxiv.org/abs/1507.07510} {arXiv:1507.07510 [hep-lat]} \BibitemShut
  {NoStop}%
\bibitem [{\citenamefont {Bonati}\ \emph {et~al.}(2018)\citenamefont {Bonati},
  \citenamefont {D'Elia}, \citenamefont {Negro}, \citenamefont {Sanfilippo},\
  and\ \citenamefont {Zambello}}]{Bonati:2018nut}%
  \BibitemOpen
  \bibfield  {author} {\bibinfo {author} {\bibfnamefont {C.}~\bibnamefont
  {Bonati}}, \bibinfo {author} {\bibfnamefont {M.}~\bibnamefont {D'Elia}},
  \bibinfo {author} {\bibfnamefont {F.}~\bibnamefont {Negro}}, \bibinfo
  {author} {\bibfnamefont {F.}~\bibnamefont {Sanfilippo}},\ and\ \bibinfo
  {author} {\bibfnamefont {K.}~\bibnamefont {Zambello}},\ }\bibfield  {title}
  {\bibinfo {title} {{Curvature of the pseudocritical line in QCD: Taylor
  expansion matches analytic continuation}},\ }\href
  {https://doi.org/10.1103/PhysRevD.98.054510} {\bibfield  {journal} {\bibinfo
  {journal} {Phys. Rev. D}\ }\textbf {\bibinfo {volume} {98}},\ \bibinfo
  {pages} {054510} (\bibinfo {year} {2018})},\ \Eprint
  {https://arxiv.org/abs/1805.02960} {arXiv:1805.02960 [hep-lat]} \BibitemShut
  {NoStop}%
\bibitem [{\citenamefont {Mukherjee}\ and\ \citenamefont
  {Skokov}(2021)}]{Mukherjee:2019eou}%
  \BibitemOpen
  \bibfield  {author} {\bibinfo {author} {\bibfnamefont {S.}~\bibnamefont
  {Mukherjee}}\ and\ \bibinfo {author} {\bibfnamefont {V.}~\bibnamefont
  {Skokov}},\ }\bibfield  {title} {\bibinfo {title} {{Universality driven
  analytic structure of the QCD crossover: radius of convergence in the baryon
  chemical potential}},\ }\href {https://doi.org/10.1103/PhysRevD.103.L071501}
  {\bibfield  {journal} {\bibinfo  {journal} {Phys. Rev. D}\ }\textbf {\bibinfo
  {volume} {103}},\ \bibinfo {pages} {L071501} (\bibinfo {year} {2021})},\
  \Eprint {https://arxiv.org/abs/1909.04639} {arXiv:1909.04639 [hep-ph]}
  \BibitemShut {NoStop}%
\bibitem [{\citenamefont {Giordano}\ and\ \citenamefont
  {P\'asztor}(2019)}]{Giordano:2019slo}%
  \BibitemOpen
  \bibfield  {author} {\bibinfo {author} {\bibfnamefont {M.}~\bibnamefont
  {Giordano}}\ and\ \bibinfo {author} {\bibfnamefont {A.}~\bibnamefont
  {P\'asztor}},\ }\bibfield  {title} {\bibinfo {title} {{Reliable estimation of
  the radius of convergence in finite density QCD}},\ }\href
  {https://doi.org/10.1103/PhysRevD.99.114510} {\bibfield  {journal} {\bibinfo
  {journal} {Phys. Rev. D}\ }\textbf {\bibinfo {volume} {99}},\ \bibinfo
  {pages} {114510} (\bibinfo {year} {2019})},\ \Eprint
  {https://arxiv.org/abs/1904.01974} {arXiv:1904.01974 [hep-lat]} \BibitemShut
  {NoStop}%
\bibitem [{\citenamefont {Giordano}\ \emph {et~al.}(2020)\citenamefont
  {Giordano}, \citenamefont {Kapas}, \citenamefont {Katz}, \citenamefont
  {Nogradi},\ and\ \citenamefont {Pasztor}}]{Giordano:2019gev}%
  \BibitemOpen
  \bibfield  {author} {\bibinfo {author} {\bibfnamefont {M.}~\bibnamefont
  {Giordano}}, \bibinfo {author} {\bibfnamefont {K.}~\bibnamefont {Kapas}},
  \bibinfo {author} {\bibfnamefont {S.~D.}\ \bibnamefont {Katz}}, \bibinfo
  {author} {\bibfnamefont {D.}~\bibnamefont {Nogradi}},\ and\ \bibinfo {author}
  {\bibfnamefont {A.}~\bibnamefont {Pasztor}},\ }\bibfield  {title} {\bibinfo
  {title} {{Radius of convergence in lattice QCD at finite $\mu_B$ with rooted
  staggered fermions}},\ }\href {https://doi.org/10.1103/PhysRevD.101.074511}
  {\bibfield  {journal} {\bibinfo  {journal} {Phys. Rev. D}\ }\textbf {\bibinfo
  {volume} {101}},\ \bibinfo {pages} {074511} (\bibinfo {year} {2020})},\
  \bibinfo {note} {[Erratum: Phys.Rev.D 104, 119901 (2021)]},\ \Eprint
  {https://arxiv.org/abs/1911.00043} {arXiv:1911.00043 [hep-lat]} \BibitemShut
  {NoStop}%
\bibitem [{\citenamefont {Adamczyk}\ \emph {et~al.}(2017)\citenamefont
  {Adamczyk} \emph {et~al.}}]{STAR:2017sal}%
  \BibitemOpen
  \bibfield  {author} {\bibinfo {author} {\bibfnamefont {L.}~\bibnamefont
  {Adamczyk}} \emph {et~al.} (\bibinfo {collaboration} {STAR}),\ }\bibfield
  {title} {\bibinfo {title} {{Bulk Properties of the Medium Produced in
  Relativistic Heavy-Ion Collisions from the Beam Energy Scan Program}},\
  }\href {https://doi.org/10.1103/PhysRevC.96.044904} {\bibfield  {journal}
  {\bibinfo  {journal} {Phys. Rev. C}\ }\textbf {\bibinfo {volume} {96}},\
  \bibinfo {pages} {044904} (\bibinfo {year} {2017})},\ \Eprint
  {https://arxiv.org/abs/1701.07065} {arXiv:1701.07065 [nucl-ex]} \BibitemShut
  {NoStop}%
\bibitem [{\citenamefont {D'Elia}(2019)}]{DElia:2018fjp}%
  \BibitemOpen
  \bibfield  {author} {\bibinfo {author} {\bibfnamefont {M.}~\bibnamefont
  {D'Elia}},\ }\bibfield  {title} {\bibinfo {title} {{High-Temperature QCD:
  theory overview}},\ }\href {https://doi.org/10.1016/j.nuclphysa.2018.10.042}
  {\bibfield  {journal} {\bibinfo  {journal} {Nucl. Phys. A}\ }\textbf
  {\bibinfo {volume} {982}},\ \bibinfo {pages} {99} (\bibinfo {year} {2019})},\
  \Eprint {https://arxiv.org/abs/1809.10660} {arXiv:1809.10660 [hep-lat]}
  \BibitemShut {NoStop}%
\bibitem [{\citenamefont {Bollweg}\ \emph {et~al.}(2023)\citenamefont
  {Bollweg}, \citenamefont {Clarke}, \citenamefont {Goswami}, \citenamefont
  {Kaczmarek}, \citenamefont {Karsch}, \citenamefont {Mukherjee}, \citenamefont
  {Petreczky}, \citenamefont {Schmidt},\ and\ \citenamefont
  {Sharma}}]{epubdata:2980715}%
  \BibitemOpen
  \bibfield  {author} {\bibinfo {author} {\bibfnamefont {D.}~\bibnamefont
  {Bollweg}}, \bibinfo {author} {\bibfnamefont {D.~A.}\ \bibnamefont {Clarke}},
  \bibinfo {author} {\bibfnamefont {J.}~\bibnamefont {Goswami}}, \bibinfo
  {author} {\bibfnamefont {O.}~\bibnamefont {Kaczmarek}}, \bibinfo {author}
  {\bibfnamefont {F.}~\bibnamefont {Karsch}}, \bibinfo {author} {\bibfnamefont
  {S.}~\bibnamefont {Mukherjee}}, \bibinfo {author} {\bibfnamefont
  {P.}~\bibnamefont {Petreczky}}, \bibinfo {author} {\bibfnamefont
  {C.}~\bibnamefont {Schmidt}},\ and\ \bibinfo {author} {\bibfnamefont
  {S.}~\bibnamefont {Sharma}},\ }\href {https://doi.org/10.4119/unibi/2980715}
  {\bibinfo {title} {{Dataset for ''Equation of state and speed of sound of
  (2+1)-flavor QCD in strangeness-neutral matter at non-vanishing net
  baryon-number density''}}} (\bibinfo {year} {2023})\BibitemShut {NoStop}%
\bibitem [{\citenamefont {Shuryak}(1978)}]{Shuryak:1977ut}%
  \BibitemOpen
  \bibfield  {author} {\bibinfo {author} {\bibfnamefont {E.~V.}\ \bibnamefont
  {Shuryak}},\ }\bibfield  {title} {\bibinfo {title} {{Theory of Hadronic
  Plasma}},\ }\href@noop {} {\bibfield  {journal} {\bibinfo  {journal} {Sov.
  Phys. JETP}\ }\textbf {\bibinfo {volume} {47}},\ \bibinfo {pages} {212}
  (\bibinfo {year} {1978})}\BibitemShut {NoStop}%
\bibitem [{\citenamefont {Chin}(1978)}]{Chin:1978gj}%
  \BibitemOpen
  \bibfield  {author} {\bibinfo {author} {\bibfnamefont {S.~A.}\ \bibnamefont
  {Chin}},\ }\bibfield  {title} {\bibinfo {title} {{Transition to Hot Quark
  Matter in Relativistic Heavy Ion Collision}},\ }\href
  {https://doi.org/10.1016/0370-2693(78)90637-8} {\bibfield  {journal}
  {\bibinfo  {journal} {Phys. Lett. B}\ }\textbf {\bibinfo {volume} {78}},\
  \bibinfo {pages} {552} (\bibinfo {year} {1978})}\BibitemShut {NoStop}%
\bibitem [{\citenamefont {Aoki}\ \emph {et~al.}(2022)\citenamefont {Aoki} \emph
  {et~al.}}]{Aoki:2021kgd}%
  \BibitemOpen
  \bibfield  {author} {\bibinfo {author} {\bibfnamefont {Y.}~\bibnamefont
  {Aoki}} \emph {et~al.} (\bibinfo {collaboration} {Flavour Lattice Averaging
  Group (FLAG)}),\ }\bibfield  {title} {\bibinfo {title} {{FLAG Review 2021}},\
  }\href {https://doi.org/10.1140/epjc/s10052-022-10536-1} {\bibfield
  {journal} {\bibinfo  {journal} {Eur. Phys. J. C}\ }\textbf {\bibinfo {volume}
  {82}},\ \bibinfo {pages} {869} (\bibinfo {year} {2022})},\ \Eprint
  {https://arxiv.org/abs/2111.09849} {arXiv:2111.09849 [hep-lat]} \BibitemShut
  {NoStop}%
\end{thebibliography}%

\end{document}